\documentclass[a4paper,11pt]{article}
\usepackage{graphicx} % Permite inserir gr?ficos no documento.
\usepackage{geometry} % Configura??o das dimens?es da p?gina
\usepackage{setspace}  %\doublespacing
\usepackage{amsmath,amsthm,amsfonts,amssymb,amscd,amsxtra}
\usepackage{soul}
\usepackage{pgf,tikz}
\usetikzlibrary{arrows}
\usepackage[]{algorithm2e}
\usepackage{algpseudocode}
\usepackage{color}
\usepackage{cases}
\usepackage{authblk}
\usepackage{hyperref}
\usepackage{bbm}
\usepackage{caption}
\usepackage{subcaption}

\newtheorem{lemma}{Lemma}[section]

\newtheorem{remark}[lemma]{Remark}

\theoremstyle{definition}

\newcommand{\XX}{\mathbf{X}}	
\newcommand{\VV}{\mathbf{V}}
\newcommand{\RR}{\mathbb{R}}

\newcommand{\LL}{\mathcal{L}}

\newcommand{\llambda}{\boldsymbol{\lambda}}

\newcommand{\bXX}{\bar{\mathbf{X}}}

\newcommand{\bllambda}{\bar{\boldsymbol{\lambda}}}

\newcommand{\mC}{\mathcal{C}}
\numberwithin{equation}{section}

\title{Studying ballistic aggregation phenomena through efficient Time Stepping approaches}
\author[1]{P. Degond}
\author[2]{G. Dimarco}
\author[3]{M. A. Ferreira}
\author[4]{S. Hecht}
\affil[1]{Institut de Math{\'e}matiques de Toulouse ; UMR5219
	Universit{\'e} de Toulouse ; CNRS UPS, F-31062 Toulouse Cedex 9, France}
\affil[2]{University of Ferrara, Department of Mathematics and Computer Science \& Center for Modeling, Computing and Statistics, CMCS, Ferrara, Italy}
\affil[3]{CMUC, Department of Mathematics, University of Coimbra, 3000-413 Coimbra, Portugal}
\affil[4]{Sorbonne Universit{\'e}, CNRS, Universit\'{e} de Paris, Laboratoire Jacques-Louis Lions UMR7598, F-75005 Paris, France}

\date{\today}

\begin{document}

\maketitle

\begin{abstract}
	This paper deals with the problem of simulating dense dispersed systems composed by large numbers of particles undergoing ballistic aggregation. The most classical approaches for dealing with such problems are represented by the so-called event-driven methods. Despite being more accurate, these methods become computationally very expensive as the number of particles increases. Typically, their computational cost is proportional to the square of the number of particles and thus they become extremely demanding as soon as this number becomes sufficiently large. An alternative approach, called time-stepping, consists in evolving the problem over small time-intervals and to handle all collisions occurring during each time interval simultaneously. In this work, we follow this second direction and we introduce a new time stepping method which recasts the problem of the multiple collisions in a minimization framework. The objective of this work is twofold, first to show that the statistical description of the resulting aggregates obtained with this new time stepping method is sufficiently close to that of the event driven methods. The second goal consists in showing that, in terms of computational performances, this new approach {is competitive and may} outperform standard methodologies especially when the number of particles becomes sufficiently large. Numerical results obtained in the case of spherical particles moving in a two dimensional box show that these two properties are indeed satisfied.
\end{abstract}

\medskip

\noindent{\bf Keywords:} Ballistic aggregation, fractal aggregates, time-stepping methods, event driven methods, optimization.

\medskip

\noindent{\bf Mathematics Subject Classification:} 70F35, 74G65, 65K10

\section{Introduction}
Aggregation is an ubiquitous phenomenon in nature. It can be met in many different fields of physics: nucleation and crystallization phenomena, aerosols, raindrops formation, sprays, polymers, formation of planets and galaxies~\cite{Elimelech1995BOOK, Vicsek1992}. In biology, cells aggregate to form tissues or to repair injuries (blood coagulation), bacteria aggregate to form biofilms~\cite{Kreft, Wimpenny}. Aggregation phenomena can be observed and studied at different scales. For instance, at the microscopic level, particle based models describe the evolution of the mass, position and velocity of each particle~\cite{dey2011lattice, frachebourg1999exact, frachebourg2000ballistic, grzegorczyk2004ballistic, majumdar2009statistical, martin1994one, pathak2014inhomogeneous}. %or the Marcus-Lushnikov microscopic model~\cite{Aldous1999, Fournier2009}
{At the mesoscopic level, one typically studies the evolution of the statistical distribution of sizes, positions and velocities of the aggregates.
	Within this description, one approach is given by the inhomogeneous Smoluchowski coagulation equation~\cite{broazat, escobedo2010scalings, HammondRezak, jiang1994kinetic}, a population balance model describing the time evolution of the distribution of clusters over size, space and velocity as they diffuse and coagulate to form larger clusters. On the opposite side of the spectrum, at the macroscopic level, one instead describes averages over velocity and space leading to the widely studied Smoluchowski coagulation equations~\cite{book_Laurencot} in which the distribution function only depends on the mass variable.}
Despite being much easier to treat, both computationally and analytically, the mesoscopic and macroscopic models do not incorporate all information about position or velocity of the clusters. Moreover, this upscaling procedure often assumes that clusters are spherical entities. A particle based model instead is able to give more detailed spatial information about the shape of the aggregates and it allows the construction of more realistic models~\cite{dirkse1995modified, grzegorczyk2004ballistic}. In particular, it can be used to study fractal growth phenomena, which has been an object of study in mathematical physics motivated also by applications \cite{Vicsek1992}. Additionally, particle methods provide an appropriate framework to study the randomness arising from the initial conditions~\cite{frachebourg2000ballistic, majumdar2009statistical} and the statistical properties of the growing aggregates related to the evolution of the shape~\cite{amitrano1986growth, kardar1986dynamic, plischke1984active}. The structure of aggregates obtained in the long-time limit can also be studied within this microscopic approach~\cite{martin1994one}. For the above reasons, being interested in the microscopic properties of the aggregates, in this work we choose the particle approach and we develop a new technique which permits to deal with a very large number of particles.

Specifically, we focus on a particular type of aggregation phenomenon, called ballistic aggregation, in which hard-spheres move in straight trajectories until they collide. When a collision event takes place, particles stick together and form growing and moving aggregates~\cite{Elimelech1995BOOK, Vicsek1992}. In our model, we consider that collisions are inelastic and frictionless and that, during a collision event, the total mass and momentum are conserved while kinetic energy is dissipated. In this first attempt to develop a new methodology to efficiently describe aggregation with large number of agents, we do not incorporate other aspects of the phenomenon such as angular momentum conservation or chemical reactions into the model. Consequently, particles do not rotate around their center of mass and their internal structure does not change as effect of collisions. The resulting model is among the simplest description one can use for treating the interaction of rigid particles. This contributed to its strong popularity in the past (see for instance the recent book~\cite{Pelliccione2008BOOK} or \cite{Vicsek1992}. In particular, this model has been used to describe evaporated target foils~\cite{stoner1989properties}, silica-sol gel films~\cite{grzegorczyk2005modeling}, granular gases~\cite{pathak2014inhomogeneous}, cosmic dust~\cite{martin1996aggregation}, protoplanetary disks~\cite{okuzumi2009numerical}, among many other different applications. However, despite its simplicity, this model becomes extremely difficult to handle once the number of particles and the rate of collisions increase. Since, complexity is one of the main challenges in the simulation of particles interactions such as in molecular dynamics \cite{Rapaport}, the ballistic aggregation model we study can be considered as a toy model for testing and comparing numerical methods which successively can be adapted to more realistic problems.

Numerical methods to simulate hard-spheres dynamics may be divided into two main classes (see ~\cite{brogliato2002overviewNonsmooth, Rapaport} for overviews). The first are the so-called event-driven method (ED)~\cite{frachebourg1999exact, frachebourg2000ballistic}. In this approach, given the position of every particle, the time of the next collision event is exactly computed. Then, the system evolves until that time and the single collision is performed exactly. The main difficulty of such approach is to determine the sequence of collisions efficiently, which makes ED schemes computationally very expensive when the number of particles grows. A crude estimate gives the complexity of the order of $\mathcal O(N^2)$ with $N$ the number of particles, even if remedy to this complexity has been designed \cite{DEMICHELE,Donev}. The second class of methods is the so-called time-driven or time-stepping (TS) method~\cite{Haile}. In this approach, the time is discretized into equal intervals and all collisions occurring within a time-interval are solved simultaneously. This approach in general allows to reduce the computational cost, the price to pay being the handling of multiple collision situations. In fact, time stepping methods are in general less accurate than ED methods, however, they may provide a very efficient alternative to ED methods for very large systems. The main idea, in a nutshell, is to get at each time-step the positions and velocities of the particles by integrating Newton's second law and successively to correct possible overlaps between particles, usually by projecting onto an admissible space of non-overlapping configurations. This projection can be formulated as a minimization problem with non-overlapping constraints, which may be imposed on the velocities or on the positions of the particles, leading to Moreau-Jean-type schemes~\cite{moreau1988unilateral, saussine2006ballast, Maury2006}  or Schatzman-Paoli-type schemes~\cite{paoli2002impactI, paoli2002impactII}, respectively. We also recall that, when the constraints are imposed onto the velocities, the computational cost of the minimization problem may strongly depend on the initial configuration and this effect becomes worse as the number of particles $N$ increases~\cite{degond2017damped}. Aside from the complexity problem, one of the main question that arises when TS methods are used to replace event driven ones is if this approach is able to describe the same type of dynamics of the ED methods. 

In this work we follow the path of the TS methods and we propose a new algorithm in which the projection step  over a non-overlapping configuration, occurring in the time-stepping methodology, is substituted by the local minimization of an energy associated to the adhesive links between the colliding spheres. In this proposed approach, the minimization is done subject to non-overlapping constraints on the positions of the spheres. Thanks to this method, we ensure that the colliding spheres remain close to each other at all times. The minimization method is based on a modified version of the  Arrow-Hurwicz algorithm \cite{Uzawa58} in which we include a damping term. This method has been recently proposed to tackle minimization problems with non-convex and non-overlapping constraints~\cite{degond2017damped} in a different setting. The main idea consists of the introduction of a new term in the original equation with the scope of accelerating the convergence towards an equilibrium state. In particular, in ~\cite{degond2017damped}, it has been shown that better performances with respect to the original method can be obtained in the context of sphere packing, even if only a relatively small number of particles were considered. Here, we deal with much larger sets of spheres in the framework of ballistic aggregation. In particular, we study the behavior of a damped Arrow-Hurwicz method in which we let the numerical coefficients evolve over the simulation in such a way that an important step forward in the direction of computational efficiency is achieved. One of the main challenge faced in the present work consists also in optimizing the cluster/particle localization to avoid excessive computation at each time step of the procedure. This part often represents a second bottleneck for the simulation of particle interactions \cite{Has}.

In the sequel, we perform a careful study of our TS approach by following two main directions. The first one is about complexity. Thus, we look for the best possible version of our algorithm in such a way that the computation is as efficient as possible. We then compare this new method to standard Event-Driven methodologies showing that we are able to outperform standard approaches based on the exact computation of the collision events. The second direction we follow is about the analysis and the comparisons of the Time-Stepping and the Event-Driven method in terms of their ability to describe the ballistic aggregation phenomenon. In detail, by considering as a benchmark the Event-Driven method, we statistically compare the shape of the final clusters and the computational time by introducing different indicators. Indeed, the results show that a TS method based on a minimization procedurev%% to which we refer in the rest of the paper to as the Time-Stepping-Minimization (TSM) method, 
can be successfully used to describe aggregation dynamics and that, when large size clusters have to be considered, it is faster than traditional Event-Driven approaches while maintaining a sufficient precision in the physical description of the phenomena of interest.

The plan of the article is the following. In Section \ref{Sec:Cont_mod}, we introduce the ballistic aggregation model. In Section \ref{sec:TS}, the TS method is described in detail together with the minimization algorithm, characterizing the main part of the scheme. In Section~\ref{sec:test}, we define the experimental set up which will be used to study the performances of the TS scheme and we compare it to the ED method in terms of capability of capturing the correct physics and in terms of computational performances. In a final section, we draw some conclusions and suggest future investigations. The supplementary material (SM) contains the details of the classical Event-Driven method which will be used to produce the benchmark results in terms of shape of the clusters and computational efforts in Section \ref{sec:ED}. It also contains in Section \ref{app_quant} additional information about the shape analysis of the clusters we first discuss in Section \ref{sec:numerical}. Finally the supplementary material contains the description of several videos (\href{https://figshare.com/articles/media/Modeling_ballistic_aggregation_by_time_stepping_approaches/24081027}{figshare.com/articles/media/Modeling\_ballistic\_aggregation\_by\_time\_stepping\_approaches/24081027})\\ related to the aggregation phenomenon obtained with respectively the TS and the ED methods in \ref{sec:videos}.

\section{A simple model of ballistic aggregation}
\label{Sec:Cont_mod}
We start by describing the model of ballistic aggregation used in our investigation. We stress that at this stage, as already stated in the Introduction, we deal with a model containing the basic features of the clustering dynamics. This choice is due to the fact that we are interested in deriving a new resolution approach for cluster formation and consequently {we want to focus only on the key physical mechanisms which are responsible for the description of that phenomenon. Once the new method is validated extensions to more general settings can be set into place without modification of the core aspects of the approach described in the sequel.}  

\begin{figure}
	\centering
	\includegraphics[scale=0.3]{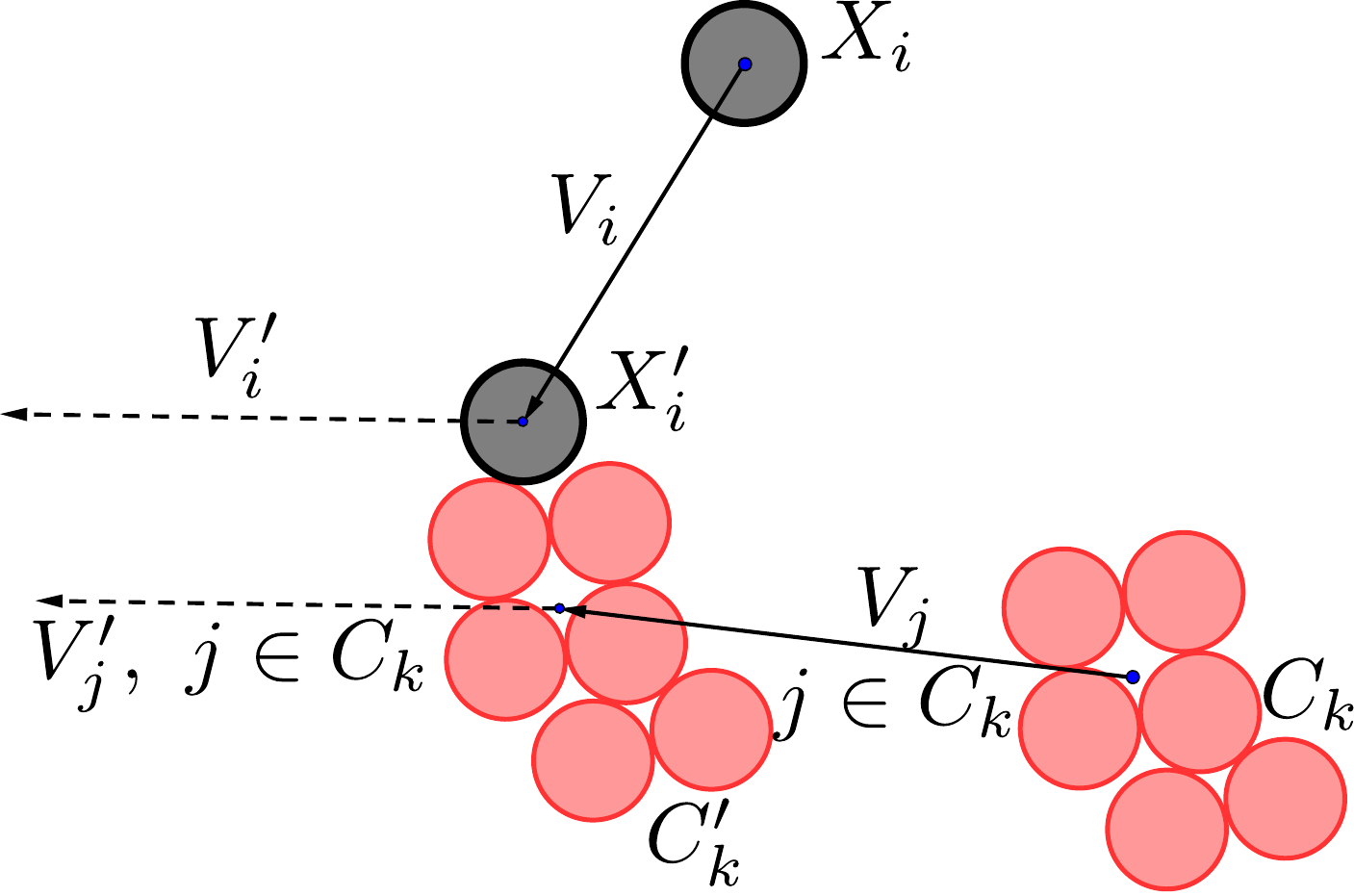}
	\caption{Example in two space dimensions of a collision between a cluster formed by a single sphere $i$ and a cluster $k$. The successive aggregation phenomenon is also shown. The pre and post collisional velocities are represented by solid and dashed lines, respectively.}
	\label{fig:coll2}
\end{figure}

We consider a system of $N$ self-propelled hard spheres on a $d-$dimensional box $[0,L]^d$. Specifically, even if not strictly necessary, we focus on the case $d=2$ in the sequel, thus we consider a dynamics on a plane. The spheres are characterized by their radius $R_i>0$, their position $X_i=(x_{i_1},\ldots,x_{i_d}),$ their velocity $V_i=(v_{i_1},\ldots,v_{i_d})$ and their mass $m_{i}>0,\ i=1,\ldots,N$. They interact via inelastic frictionless collisions. Thus, once a collision event happens, a link is created between the colliding particles which avoids successive fragmentation. 
At the same time, when a particle touches the boundary, it obeys to a specular reflection rule. No other forces are exerted on the spheres. Thus, between two consecutive collisions, particles travel in straight trajectories and at a constant speed. Given the times $t_1$ and $t_2$ of two consecutive events, i.e. collisions among spheres or with the boundaries, the velocities $\VV(t)=\{V_i(t)\}_{i=1,\ldots,N}$ and positions $\XX(t)=\{X_i(t)\}_{i=1,\ldots,N}$ of the particles for $t\in [t_1, t_2]$ satisfy the following equations on the box
\begin{equation}\label{eq:continuous_velocity}
	\frac{dX_i}{dt}(t)=V_i(t),\quad \frac{dV_i}{dt}(t)=0, \quad t  \in [t_1,t_2],\quad i=1,\ldots,N.
\end{equation}
Concerning the collisions with the boundaries occurring at time $t^*$, the rule is given by
\begin{equation}\label{boundcoll}
	V'_i(t^*)=V_i(t^*)-2n_b(n_b\cdot V_i(t^*))
\end{equation}
with $n_b$ the unit normal to the boundary and where $V'_i(t^*)$ indicates the instantaneous post-collision velocity. When, instead, the event corresponds to a collision between two particles $i$ and $j$, we start the description of such event by hypothesizing a dissipative mechanism in which the normal relative velocity is lost. In general, for dissipative hard sphere dynamics, one can then write \cite{toscani,villani}
\begin{equation}\label{velnormcol}
	(V'_i(t^*)-V'_j(t^*))\cdot n_c=-\epsilon(V'_i(t^*)-V'_j(t^*))\cdot n_c
\end{equation}
where $\epsilon \in [0,1]$ is the so-called restitution coefficient, $\epsilon=1$ in the case of elastic collisions, and $n_c$ represents the unit vector directed along the line joining the center of the spheres, i.e.
\begin{equation}
	n_c=\frac{(X_i-X_j)}{|X_i-X_j|}
\end{equation} 
with $|\cdot|$ denoting the standard Euclidean norm. Assuming now conservation of momentum 
\begin{equation}\label{velcol}
	m_iV_i(t^*)+m_jV_j(t^*)=m_iV'_i(t^*)+m_jV'_j(t^*)
\end{equation}
one finds the following rule for collision
\begin{equation}\label{velcol1}
	\begin{cases}
		&	V'_i(t^*)=V_i(t^*)-2\alpha(1-\beta)[(V_i(t^*)-V_j(t^*))\cdot n_c]n_c
		\\
		&V'_j(t^*)=V_j(t^*)+2(1-\alpha)(1-\beta)[(V_i(t^*)-V_j(t^*))\cdot n_c]n_c	\end{cases}
\end{equation}
where the mass ratio $\alpha$ and the inelasticity parameter $\beta$ reads
{\begin{equation}
		\alpha=\frac{m_j}{m_i+m_j},\quad \beta=\frac{1-\epsilon}{2}
\end{equation}}
and where $0<\alpha<1$ and $0\leq\beta\leq 1/2$ with $\beta=0$ representing the standard hard sphere dynamics giving rise to the well known Boltzmann equation in the limit $N\to \infty$ \cite{cer}, while the case $\beta=1/2$ corresponds to the perfect inelastic situation, typically encountered in granular gas models \cite{toscani}. Let us also observe that, within this framework, the kinetic energy of the system is dissipated in time:
\begin{equation}\label{kinen}
	m_i|V'_i|^2+m_j|V'_j|^2-(m_i|V_i|^2+m_j|V_j|^2)=-4\frac{m_im_j}{m_i+m_j}\beta(1-\beta)|(V_i-V_j)\cdot n_c|^2\leq 0,
\end{equation}
where in the last equation we omitted the collision time $t^*$.
With the chosen rules, particles preserve their energy during collisions with the boundary while when they meet each other their relative velocity is set to zero in the case $\beta=1/2$. However, this is not enough to make them stick together since the velocity component in the orthogonal direction with respect to the line joining their center of masses can be different causing two particles to slip one with respect to the other. {Thus, in order to impose the particles to hold together after a collision event, we introduce an additional hypothesis: namely after the contact two particles stick together and they assume the same speed. More precisely all the particles which belong to the new formed cluster assume the same speed and once a cluster is formed, fragmentation is not allowed.}
%a non smooth adhesion force acting between the colliding objects such that
%\[
%F_{ad}(X_i-X_j) = 0 \, ,  \quad \forall\, X_i, \, X_j \quad \mbox{ s.t.} \quad
%|X_i-X_j|>R_i+R_j,\]
%and
%\[
%F_{ad}(X_i-X_j) = -(X_i-X_j) \, ,  \quad \forall\, X_i, \, X_j \quad \mbox{ s.t.} \quad
%|X_i-X_j|=R_i+R_j,\]
%which in addition avoids the fragmentation of the clusters once formed. 
%The effect of this force is to average the velocity in the direction $\tau_c$ orthogonal to $n_c$. 
{We then assume} the following post collision rule
\begin{equation}\nonumber
	V'_i=V'_j=V'_c = \frac{m_iV_i+m_jV_j}{m_i+m_j}
\end{equation}
where $V'_c$ indicates the weighted (by the particle masses) average between the pre-collision velocities and it represents the velocity of a new moving cluster of size $2$ which we denote by $\mC_k$ with
corresponding center of mass $$X_{c_k} := (m_iX_i+m_jX_j)/(m_i+m_j).$$ 
{ In Figure \ref{fig:coll2} it is depicted an example of aggregation between two different clusters in the two space dimensional case. The pre and post collisional velocities are also shown.}

In addition to this straight line movement, in general each cluster spins around its center of mass with an angular velocity $\Omega' $ given by 
$$\Omega' = \frac{M_c}{J}$$
with $M_c$ the angular momentum given by $$M_c = m_i(X_c-X_i) \times (V_i-V_i\cdot(X_c-X_i)n_{ci}) + m_j (X_c-X_j) \times  (V_j-V_j\cdot(X_c-X_j)n_{cj})$$
where $\times$ denotes the cross product, $n_{ci},n_{cj}$ the unit vector directed along the line joining the center of the cluster and the center of the sphere $i$ or $j$ and $J$ the moment of inertia with respect to the axis passing through the center of mass of the cluster and directed orthogonally to the plane. In our model we omit this additional mechanism and we simply suppose that clusters translate without rotating.

We give now details about the cluster formation in the case of sizes greater than two, generalizing the above described dynamics. First, since we suppose no other forces act on the particles, the spheres remain attached for the rest of the time and fragmentation does not take place. Now, during the time evolution of the system, other collisions occur and growing aggregates of multiple spheres are formed. In this situation, when a cluster $\mC_k$ collides with a cluster  $\mC_\ell$ they give rise to a larger new cluster $\mC_{k,\ell}$ with new center of mass $$ X_{c_{k,\ell}} = \sum_{i\in \mC_{k,\ell}} m_iX_i/ (\sum_{i\in \mC_{k,\ell}} m_i).$$ 
The post event velocity $V_{c_{k,\ell}}'$ of the center of mass of this new object is then a weighted (by the masses of the two clusters) average between the pre-collision velocities $V_i, i\in (\mC_k\cup\mC_\ell)=\mC_{k,\ell}$ of the spheres that constitute the two colliding clusters:
\begin{equation}\label{eq:new_velocity}
	V_{c_{k,\ell}}'=\frac{\sum_{i\in  \mC_{k,\ell}} m_iV_i}{\sum_{i\in  \mC_{k,\ell}} m_i}, \quad V'_i=V'_j=V_{c_{k,\ell}}', \quad i,j\in(\mC_k\cup\mC_\ell).
\end{equation}
In general, the new aggregate shows also a rotational movement around the new center of mass $X_{c_{k,\ell}}$. This is omitted as for the case of a two particle cluster. Consequently, the dynamics considered do not conserve the angular momentum. 
In fact, for what concerns the comparison between event driven methods and time stepping algorithms, we do not expect the introduction of the rotation dynamics into the system to bring a key contribution but only some technical difficulties. 
The formation of the final aggregate, to which all particles belong, is then reached by subsequent events of the type described above. 
Before concluding this part, we introduce the functions 
$\phi_{ij}: \RR^{dN} \to \RR,\ i<j $, for each couple of index $i,j = 1,\ldots, N,\ i<j$, defined by
\begin{equation}\label{eq:phi_chp3}
	\phi_{ij}(\XX) = R_i+R_j - |X_i-X_j|.
\end{equation}
Since the spheres cannot interpenetrate, we impose the following non-overlapping constraints on the system
\begin{equation}\label{eq:continuous_nonoverl}
	\phi_{i j}(\XX)\leq 0, \qquad  i,j=1,\ldots,N,\quad i<j, 
\end{equation}
which will be used in the time stepping method to define our minimization strategy under constraints. In the next section we discuss our new time stepping method.

\section{A new time-stepping method}
\label{sec:TS}  
Before introducing our method we briefly recall the main features of the so-called Event Driven approach which has historically been the first method to be explored in the context of molecular dynamics (MD) \cite{alder1959MDI} and which is nowadays still considered the benchmark for many MD simulations \cite{rapaport2004art}. The {basic version of this approach} will be used, in the second part of the work, to study the computational performances and the accuracy of our new Time Stepping Minimization method which is later discussed in the second part of the section. The details of the ED method are reported in the supplementary materials \ref{sec:ED}. 

The idea behind an event-driven method consists of solving {directly the dynamical system through integration of equations (\ref{eq:continuous_velocity}) up to the instant of the first collision event, to compute successively the solution of a single collision event and then to integrate again equations (\ref{eq:continuous_velocity}) up to the next interaction}. Since, particles follow straight line trajectories and the collision dynamics can be solved analytically, in principle, this approach may seem the best one to follow for simulating growing {aggregates. In fact}, in the simple setting detailed in Section \ref{Sec:Cont_mod}, such method would provide exact solutions. {However, one main difficulty when studying the dynamics of particles is related to the lack of information about the precise state of the system.} It is, in fact, clear that in order to rely on the exact computation furnished by an ED method, one should be able to know the exact initial state of the system as well as the exact geometry of the objects which aggregate and of the domain where they live. Moreover, the exact type of interactions between the agents composing the system should also be known with precision. The direct consequence of these sources of uncertainty is that the results should always be intended in a statistical sense, i.e. instead of relying on a single output produced by the {method}, one should rely on expected values and on high order moments of the probability distribution obtained by producing several different simulations. {This of course remains true even if other simulations method are adopted}. See the book \cite{UQ} for a nice overview on the subject of uncertainty, while a more focused state of the art on uncertainty related to macroscopic hyperbolic or mesoscopic model can be found for instance in \cite{UQ2}. 

The second difficulty is related to the complexity of the simulating algorithm when the number of colliding objects becomes very large. In general, the main challenge is related to look for an efficient technique to compute the exact time at which the next collision event occurs. The standard approach consists in, at a given time $t$, computing the times of collisions between any pair of particles that would collide with each other in a possible future state as if there were no other particles in the system. Then, once the smallest time among all possible events is found, the system is updated to that time. Successively, the single event is resolved and the list of the possible future events is recomputed. Since the system is bounded by hypothesis, all $M(t)$ clusters can collide with any other cluster at a given instant of time. This corresponds to $M(t)(M(t)-1)/2$ possible interaction events. However, in principle, even if the number of clusters $M(t)$ monotonically decreases with time, one has always to compute the possible interaction among all particles belonging to different clusters making the computation of the number of possible collisions highly non linear. {A last important aspect is related to to the fact that, in the simple setting chosen, the dynamical system describing the aggregation dynamics can be solved exactly. However, when the modeling choices make the model more realistic then the ED approach may have additional errors related to the numerical integration of the resulting system which may cause additional computational cost and inaccuracies in the results which have to be taken into account.} The details of such a method are reported in the supplementary material \ref{sec:ED}.

As already stated in the Introduction, instead the principle of a Time Stepping method is different. In this case, the continuous dynamics is replaced by a discrete one in which time is divided into intervals independently of the underlying dynamics. Let $\Delta t = \Delta t^n>0$ be that time increment at the $n$th iteration, which may or may not be kept fixed over the whole dynamics. We summarize in the following the key features of the TS method we propose. 
\begin{itemize}
	\item[(i)] At each time-iteration the particles move in straight trajectories from time $t^n$ to $t^n+ \Delta t^n$ according to their velocities.
	\item[(ii)] If a particle hits the boundary during this time interval, it is specularly reflected back into the domain. If a particle belongs to a cluster, all particles belonging to the cluster are reflected back into the box.
	\item[(iii)] After $\Delta t^n$, overlapping between the spheres is detected and a force is activated between each pair of overlapping (colliding) spheres after the advection step.
	\item[(iv)] A nearby non-overlapping configuration is obtained through a local minimization of an energy associated to attractive forces.
	\item[(v)] The velocities are actualized according to the collision rule~\eqref{eq:new_velocity}.
\end{itemize}
In practice, the main difference with respect to ED methods is that in TS methods multiple collisions are solved simultaneously. Let also notice that, as we will see later in Section~\ref{sec:NR}, the minimization method of step (iv) physically corresponds to the search for a mechanical equilibrium in which non-overlapping constraints and attraction forces balance each other. One important aspect to stress about this point is that the minimization phase may alter the topology of the set of spheres which interact as a consequence of the free ballistic motion. Thus, the new cluster which is formed may change in structure during the flow of the algorithm which searches for an equilibrium position, as it will be made precise later. This is due to the fact that particles move during this search for an admissible configuration and so it may be possible that they overlap with other nearby spheres which were not in contact before the start of the minimization phase. The consequence is that both the particles composing this set as well as their number may change. This increases the complexity of the algorithm. We give now the details of our TS method.

\begin{figure}[h!]
	\centering
	\includegraphics[scale=0.3]{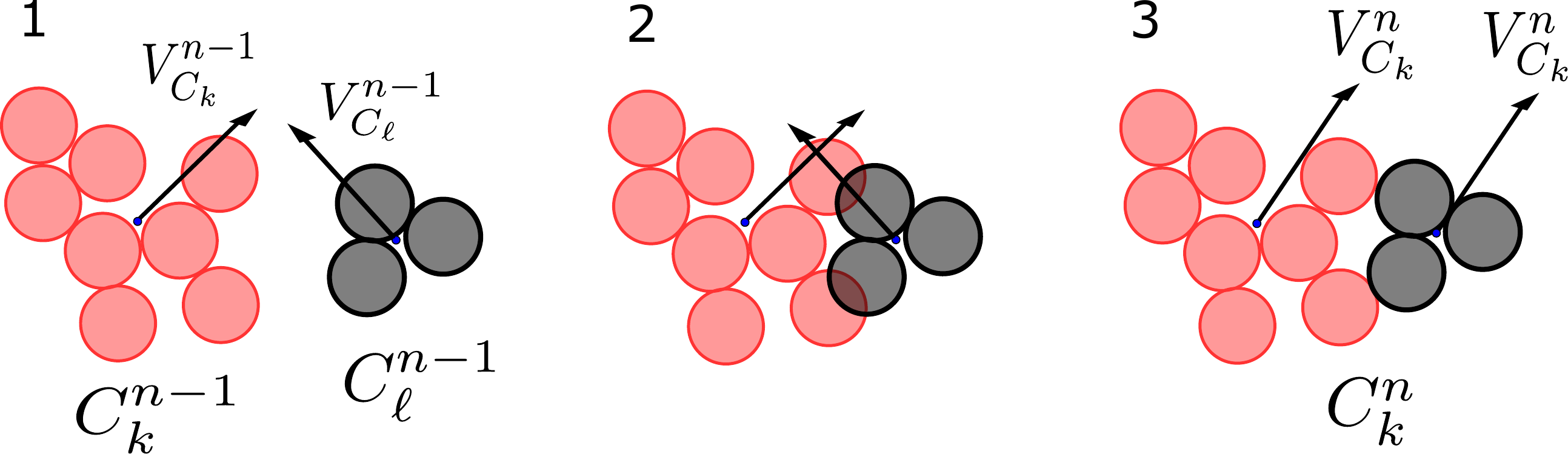}
	\caption{Illustration of the three main steps which constitute one time-iteration of the TS method: 1) free advection through a time-discretization of equations~\eqref{eq:continuous_velocity} 2) overlapping detection, creation of an adhesion force and adjustment of the positions to correct overlapping and 3) actualization of the velocities according to the collision rule~\eqref{eq:new_velocity}.}
	\label{fig:coll3}
\end{figure}

Let now in general $S^n$ be the set of pairs of spheres that are linked through adhesive forces as a result of the ballistic motion at time $t^n$ and let $\# S^n$ be the number of elements in the set $S^n$. This set will correspond to a new cluster. Let also $W_{S^n}: \RR^{2\# S^n} \rightarrow \RR,\ \XX^n \mapsto W_{S}(\XX^n)$ be a given potential associated to those links, the form of which will be specified later. Suppose that $W_{S^n}$ has an attainable minimum in the set defined by the non-overlapping constraints given by the conditions \eqref{eq:continuous_nonoverl}. Finally, suppose that the state of the system $(\XX^{n-1},\VV^{n-1},S^{n-1},M^{n-1})$ at time $t^{n-1}$ is given where $M^0=N$ is the initial number of clusters, $N$ the total number of particles and $M^n$ the number of clusters at time $t^n$. We are now ready to detail one time step of the method from $t^{n-1}$ to $t^n$ with time step $\Delta t^n$. {This is obtained by performing in order the following three steps (see Algorithm \ref{tsm} for a detailed description of each step and  Figure~\ref{fig:coll3} for a visual explanation): 1. Free advection, 2. Interactions, and 3. Actualization of the velocities.}
\begin{algorithm}{Time Stepping method}\label{tsm}
	\begin{enumerate} 
		\item Free advection. The position of each particle evolves independently from the other particles to position $\hat{\XX}^n$: $$\hat{\XX}^n=\XX^{n-1}+\Delta t^n \VV^{n-1};$$
		\item Interactions. The new pair $(\XX^n,S^n)$ of positions and  pairs of spheres belonging to a given cluster after the ballistic displacement is obtained by an iterative procedure. The iterative method flow is the following:
		\subitem{a)} { Initial condition.} Let the positions of the spheres at the first iteration $ \XX^{n,0} = \hat \XX^n$ be fixed by the free motion. This determines $S^{n,0}$, i.e. the set of spheres which are in contact as effect of the free ballistic motion and thus it determines the new attractive potential.
		\subitem{b)} { $(q-1)$th iterate knowledge.} Suppose now that $(\XX^{n,q-1}, S^{n,q-1})$ have been computed, where the apex $q$ identifies the iteration number. We then need to compute $(\XX^{n,q}, S^{n,q})$. 
		\subitem{c)}  { $q$th iterate.} Search for a non-overlapping configuration $\XX^{n,q}$ in the neighborhood of $ \XX^{n,q-1}$ that locally solves the minimization problem:
		{	\begin{equation}\label{eq:minProb}
				\XX^{n,q} \in \underset{\phi_{ij}(\XX)\leq 0,\ (i,j)\in S^{n,q-1},	|X_i^{n,q-1}-X_i|\leq C R_i}{\text{argmin}} \ W_{S^{n,q-1}}(\XX),
		\end{equation}}
		with $C$ a given constant. This furnishes for a given set $S^{n,q-1}$ of interacting spheres a possible equilibrium configuration which we identify with $\bar\XX^{n,q}$.
		\subitem{d)} Actualize the set of spheres belonging to a given cluster by adding to $S^{n,q-1}$ the new links between spheres that are in contact as a consequence of the minimization step $c)$:
		\begin{eqnarray}
			S^{n,q} =  S^{n,q-1}\ \cup\ \{ (i,j) | i<j,\ \phi_{ij}(\bar\XX^{n,q})\geq 0 \}, \ i=1,..,N
			\label{eq:setS_nonRigid}
		\end{eqnarray} 
		where $\phi_{ij}$ is the non overlapping function defined in~\eqref{eq:phi_chp3}. Let notice that the possible change of structure of the set $S^{n,q} \ne  S^{n,q-1}$ is due to the fact that as a result of the minimization procedure (\ref{eq:minProb}) the spheres move and it may happen that in dense aggregates this motion causes an overlapping with other particles which before the minimization were not in contact and thus not belonging to $S^{n,q-1}$.
		\subitem{e)} { Stopping condition.} If $S^{n,q}=S^{n,q-1}$ 
		then an equilibrium state is found. Otherwise, this means that new pairs of overlapping spheres are identified which were not in $S^{n,q-1}$ and that now belong to the set $S^{n,q}$. Thus, one continues to iterate c) and d) up to the moment at which the condition $S^{n,q} = S^{n,q-1}$ is satisfied.
		\item Actualization of the velocities:
		\subitem{a)} Compute the new number of clusters $M^n$ and classify particles, i.e. establish for each particle the cluster $C_1^n,\ldots,C_{M^n}^n$ to which they belong.
		\subitem{b)} Compute the new velocities according to collision rule~\eqref{eq:new_velocity}: $$\forall k=1,\ldots,M^n,\ \forall i\in C_k^n, \ V_i^n=\frac{\sum_{j\in C_k^n} m_j V_j^{n-1}}{\sum_{j\in C_k^n} m_j}.$$
	\end{enumerate}
	%\caption{{ Description of the three main steps of the TS method at the $n$th iteration.}}
\end{algorithm}

This method deserves some remarks. First, in order to avoid the possibility that spheres cross each other without any collision being detected, the time step $\Delta t^n$ has to be chosen small enough. One possibility consists in choosing $\Delta t^n=\Delta t$ satisfying  
\begin{equation}
	\Delta t^n = \min_{k=1,\ldots,M^n} \frac{R}{\max_{k}|V_{C^n_k}|}.
	\label{eq:Deltat}
\end{equation}
where $V_{C^n_k}$ is the velocity of each cluster at time $n$ and $R=\max_i(R_i)$.
This condition is enough to ensure that each particle $i$ during one time-step does not travel more than its own radius, however it may be a non optimal condition since it would require the computation of the maximum value of the particle velocities at each time step. Moreover this choice may be too restrictive for large box sizes. In the next section we will discuss the role of this parameter and explore different values for $\Delta t^n$ showing that it indeed plays a leading role in the results.

Second in order to speed up the algorithm, one may subdivide the domain into boxes of given length  $\ell> 2 \max_i R_i$ and then modify Step 2c) by searching for overlapping between spheres that lie within the same box or in neighboring boxes only. This allows to reduce the number of operations needed to have information about the cluster and transforms the computational cost of {order $N^2$, to order $N\log_2(N)$ \cite{Nature}. Let observe that this represents only a crude estimate of the number of possible collisions at time $t$ since the precise number is certainly lower and depends on the shape of the clusters and thus it cannot be determined in advance. In fact, particles within the same cluster are excluded from potential interactions, meaning that only particles from different clusters are considered for interaction. This approach simplifies the computational process by reducing the number of possible interactions that need to be evaluated, focusing only on inter-cluster interactions.}

During the minimization algorithm in Step 2c), the clusters increase their dimension trying to avoid overlapping of the spheres. This expansion may lead to new overlapping with nearby clusters that were not in contact at the previous iteration. This is the reason for which in Step 2d) we actualize the set of adhesive spheres and solve the minimization problem again with this new set. As expected, the change in the topology of the cluster may arrive during the first steps of the iteration procedure. Then, as a consequence of the convergence of the minimization method towards an equilibrium configuration the displacement of the spheres at each iteration becomes smaller and smaller thus the possibility of new overlappings decreases and consequently clusters tend to reach a fixed size.

Finally, let us observe that the clusters involved in a collision event may either be regarded as rigid objects that move as a whole, or as non-rigid object constituted by several spheres that move individually under the effect of the new adhesion forces and new non-overlapping constraints which are created as a result of the interaction. The second possibility may be interpreted as a modeling choice in which an additional phenomenon, the possibility of separation due to a collision event, is added to the dynamics. This situation may typically arrive when the impact parameter between two colliding objects assumes large values \cite{ashgriz}. However, we make precise that in the present paper the possibility of breaking the links between spheres is not considered. 

Let us close this part by observing that the non-overlapping constraints may be described by inequalities involving non-smooth or conversely smooth functions. While they both lead to equivalent constraints at the continuous level, this different choice could induce some differences at the discrete level. A smooth form of the constraints reads for instance as
\begin{equation}\label{eq:Sconstraint}
	\phi_{ij}(\XX) = (R_i+R_j)^2 - |X_i-X_j|^2.
\end{equation} 
In particular, we observed that the non smooth choice \eqref{eq:phi_chp3} slows down the convergence of our method and thus we prefer to use \eqref{eq:Sconstraint} at the numerical level. 

A final remark deserves to be given. The modifications to the original dynamics we perform with the Time Stepping Minimization approach obtained by substituting the exact collision dynamics with the minimization method give rise naturally to different aggregates of spheres. This may induce to think that such approach is not the correct direction to follow to simulate such aggregation dynamics. However, we stress that clearly the original system lacks robustness to perturbations of the initial conditions as discussed for instance in \cite{maury2004aggregation}. This is mainly due to different sources of uncertainty as discussed in the previous section. The initial condition of such a complex system, to make an example, cannot be known with precision. Thus, one may assume that the modification introduced into the dynamics by the TS approach plays the same role as different sources of uncertainty in which the system is studied. Moreover, one is typically only interested in the average quantities which can be measured using this kind of models and clearly not to one specific realization. Thus, in Section~\ref{sec:NR} we will try to justify this key assumption by showing that the TS and ED methods produce statistically similar results when the shapes of the clusters as well as the time needed to reach a final configurations are the quantities of interest which are measured.

\subsection{{Minimization problem}}\label{sec:min_alg}
In this part, we give the details of the minimization algorithm employed in step 2c) of the TS method described in previous Section~\ref{sec:TS}. We start by describing the so-called non-rigid cluster situation in which all the spheres belonging to the same cluster may move during the search for a non-overlapping configuration in which the potential energy reaches a minimum. Then, in Section \ref{sec:R}, we give details about a possible rigid-cluster variant where the adhesion forces are localized. 

Let now $\hat \XX^{n}$ be the given (non-admissible) overlapping configuration obtained at time $t^n$ of the aggregation dynamics after the displacement of the particles at speed $\VV^n$. Let also $S^{n,q}$ be the set of links (adhesive forces), which defines the cluster configuration during the iterative procedure in Step 2 Section~\ref{sec:TS} for the different iterations identified by the apex $q$. In the minimization algorithm detailed below, we suppose that the set of particles interacting during the iteration $q$ is constant and given by $S^{n,q}$ as stated in Step 2 of Section \ref{sec:TS}. Other choices are possible as discussed in Remark \ref{rem:outer-inner-loop}. In this situation, we look for a sufficiently close configuration $\bar \XX^{n}$ such that the non overlapping constraint is satisfied with the condition that the spheres belonging to the same cluster remain close to each other. This problem can be recast as an iterative procedure indexed by $r$ in the sequel which looks for a minimum of a potential function. This potential force models the adhesive forces between the spheres while the system is subject to a constraint of impenetrability. In practice, our choice is to model the adhesion contact by an attractive potential reading
\begin{equation}\label{min1}
	W_{S^{n,q}}(\XX^{n,q})= \frac{1}{2}\sum_{(i,j)\in S^{n,q}} |X^{n,q}_i-X^{n,q}_j|^2, 
\end{equation}
and to look for a configuration $\bar \XX^{n}$ in the neighborhood of $\hat \XX^{n}$ that locally minimizes this potential under the non-overlapping constraints 
\begin{equation}\label{min2}
	\phi_{ij}(\XX^{n,q}) \leq 0,\ (i,j) \in S^{n,q},
\end{equation}
where $\phi_{ij}$ is defined in \eqref{eq:Sconstraint} and where $\XX^{n,0}=\hat \XX^{n}$.

The resulting minimization problem shares many similarities with the so-called packing problem discussed for instance in~\cite{Hifi2009}. In this problem, one tries to pack as much as possible objects inside a domain identified by a given geometry which typically mimics a container or a box. It is well known that packing problems of this kind may give rise to non deterministic polynomial time hard problems (NP-hard) \cite{Hifi2009}. In fact, even if the potential energy to minimize has a convex structure, the constraints transform the whole dynamics in a non-convex optimization problem. For these problems, optimal solutions are known to be in general not unique. This can be made immediately clear for our specific problem by thinking that permutations and rotations of the spheres to be packed together generate equivalent solutions in terms of the potential energy \eqref{min1}. 

We stress however, that one key difference between standard packing problems and our ballistic coalescence model is that, in the first case, one typically looks for a global minimum. This request characterizes the numerical method used. Here, the situation is different since we clearly do not look for a global minimum configuration since this solution would not be the physically relevant one. Instead, due to the type of dynamics we want to mimic, we more likely search for the closest minimum configuration (with respect to the potential produced as a result of the ballistic motion) which satisfies the non-overlapping constraints. This is due to the fact that we consider fixed ballistic times $\Delta t^n$ such that spheres, in average, do not travel more than the size of their radius $R$ and, consequently, we expect the original dynamics to lead to an admissible configuration very close to the one with overlapping obtained after the transport step. An algorithm searching for a global minimum of such dynamics would produce highly packed configurations which will be very far from the physical phenomenon we want to capture. One can imagine that, in our case and in average, for each sphere the following condition to hold true
\begin{equation}
	|X_i^{n,q_f}-X_i^{n,0}|\leq C R_i \quad \forall \ i=1,..,N
\end{equation}
where $q_f$ is the final iterate of the minimization algorithm, i.e. is such that $\XX^{n,q_f}=\bar\XX^n$ and $C$ a suitable constant of order $\Delta t^n V_i$. Classical methods to approach non-convex optimization problems locally are given, for example, by Uzawa-Arrow-Hurwicz type algorithms \cite{Uzawa58}. In the context of such methods, recently, some of the authors of the present work, developed a new algorithm, belonging to the same family, which well adapts to the packing problem \cite{degond2017damped}. This new method is our starting point for studying the ballistic aggregation by the TS approach. We describe it in the sequel highlighting also the main differences employed in the present study. 

%\subsection{{ Arrow-Hurwicz minimization algorithm}}

Let $\# S^{n,q-1}$ be the number of elements in $S^{n,q-1}$ of step $2c)$ in the algorithm \ref{tsm}. This quantity is not modified during the search of an equilibrium configuration. However, it could be modified in the successive step $2d)$ as discussed previously due to new links formation caused by the displacement of the particles.
The minimization problem \eqref{min1}-\eqref{min2} consequently takes place then with a fixed number of spheres and for this reason in the following, to lighten the notation, we drop the dependence on $n,q$ for $S$. The minimization problem can then be recast through the introduction of a Lagrangian function $\LL_S:\RR^{2N} \times (\RR_0^+)^{\# S(\# S-1)/2}$ defined by
\begin{equation}
	\LL_S(\XX^{n,q},\llambda^{n,q}) = W_S(\XX^{n,q}) + \sum_{(i,j)\in S,i<j} \lambda_{ij}^{n,q}\phi_{ij}(\XX^{n,q})\label{eq:Lag}
\end{equation}
where $\llambda^{n,q} = \{ \lambda_{ij}^{n,q} \}_{(i,j)\in S,i<j}$, with $\lambda_{ij}^{n,q} \geq 0$, is the set of Lagrange multipliers associated to the non-overlapping constraints. Now, if $\bXX$ is a solution of our minimization problem \eqref{min1}-\eqref{min2} at time $t^n$, then, there exists $\bllambda\in (\RR_0^+)^{\# S(\# S-1)/2}$ such that $(\bXX,\bllambda)$ is a critical-point of the Lagrangian:
$$
\begin{cases}
	\nabla_{X_i} \LL_S(\bXX,\bllambda) =\ 0,\ i=1,\ldots,\# S  \\
	\left(\nabla_{\lambda_{ij}} \LL_S(\bXX,\bllambda) =0\ \text{ and }\bar\lambda_{ij}\geq 0\right) \text{ or } \left(\nabla_{\lambda_{ij}} \LL_S(\bXX,\bllambda) <0 \text{ and } \bar\lambda_{ij} = 0\right), \\
	\hspace{8cm} (i,j)\in S, i<j,
\end{cases}
$$
which is equivalent to
\begin{equation}
	\begin{cases}
		\nabla_{X_i} W_S(\bXX) + \sum\limits_{(i,j) \in S,i<j} \bar\lambda_{ij}\nabla_{X_i}\phi_{ij}(\bXX)  =\ 0,\ i=1,\ldots,\# S  \\
		\left(\phi_{ij}(\bXX) =0\text{ and }\bar\lambda_{ij}\geq 0\right) \text{ or } \left(\phi_{ij}(\bXX) < 0 \text{ and } \bar\lambda_{ij} = 0\right),\\
		\hspace{6cm} (i,j)\in S.
	\end{cases}\label{eq:systemSaddlePoint}
\end{equation}
Thus, the nonlinear system~\eqref{eq:systemSaddlePoint} give rise to a suitable solution of our problem. However, due to its non linearity, it has to be solved by an iterative algorithm and its solution in general is not unique. In this depicted context, the classical Arrow-Hurwicz method~\cite{Uzawa58} is obtained by recasting the above problem as the steady state solution of a system of first order ordinary differential equations later referred to as the AH system:
\begin{numcases}{}
	\dot X_i^{n,q}  & $=\ -\tilde\alpha \left( \nabla_{X_i} W_S(\XX^{n,q}) + \sum\limits_{(i,j) \in S,i<j} \lambda_{ij}\nabla_{X_i}\phi_{ij}(\XX^{n,q})\right),$ \nonumber\\
	& $\hspace{6cm} \ i=1,\ldots, \# S$ \label{eq:AHSystem1} \\
	\dot \lambda_{ij}^{n,q}  &$=\ 
	\begin{cases}
		0,\ \text{ if } \lambda_{ij}^{n,q}=0 \text{ and } \phi_{ij}(\XX^{n,q})<0\\
		\tilde\beta \phi_{ij}(\XX^{n,q}),\text{ otherwise}
	\end{cases}$,\nonumber \\
	& $\hspace{5cm} i,j = 1,\ldots,\# S,\ i< j,$ \label{eq:AHSystem2}
\end{numcases}
where $\tilde\alpha$ and $\tilde\beta$ are positive constants to be chosen in order to optimize the convergence towards an equilibrium solution. Thus, considering now an artificial small time-step $\delta t$, a semi-implicit Euler discretization scheme of the previous system leads to the so-called Arrow-Hurwicz algorithm (AHA), which is defined iteratively, using $r$ as the iteration index, by
\begin{eqnarray}\label{AHA}
	\left\{
	\begin{array}{ll}
		X_i^{n,q,r+1}          &=\ X_i^{n,q,r} - \alpha \left[\nabla_{X_i} W_S(\XX^{n,q,r}) + \sum\limits_{(i,j)\in S,\ i<j} \lambda_{ij}^n\nabla_{X_i}\phi_{ij}(\XX^{n,q,r})\right],\\
		&\hspace{8cm} \ i=1,\ldots, \# S \\
		\lambda_{ij}^{n,q,r+1} &=\ \max\{0,\lambda_{ij}^{n,q,r} + \beta\phi_{ij}(\XX^{n,q,r+1})\},\ i,j = 1,\ldots, \# S,\ i< j 
	\end{array} \right. \label{eq:uz}
\end{eqnarray}
where $\alpha$ and $\beta$ now correspond to $\alpha =\tilde\alpha \delta t\ $ and $\beta =\tilde\beta \delta t$ and they have to be intended as parameters of an algorithm which searches for a suitable equilibrium solution under the given constraints. In the next paragraph, we introduce a damped version of the algorithm \eqref{AHA} which is more adapted to efficiently study our aggregation dynamics.

\subsection{A modified Arrow-Hurwicz algorithm}\label{sec:NR}
It is known that the classical Arrow-Hurwicz algorithm presents several weaknesses when applied to solving a packing problem. In fact, typically first order steepest descent methods do not change their behavior close to an equilibrium configuration. Thus, when applied to the solution of a constrained minimization problem which needs to be solved iteratively, it becomes very difficult to converge towards a local minimum and solutions to the system (\ref{eq:AHSystem1})-(\ref{eq:AHSystem2}) may present oscillations around the equilibrium state, see for instance~\cite{degond2017damped} for a discussion about this point.  One possibility to overcome this problem is to pass from a first order to a second order dynamical system sharing the same steady state solutions as the AH system, and to include a friction term to equation (\ref{eq:AHSystem1}) with the scope of relaxing faster to an equilibrium configuration in the situations in which the iterative method draws the system close to such configurations \cite{ALVAREZ}. 
To that aim, a second-order damped Arrow-Hurwicz system~\cite{degond2017damped} reads  
\begin{numcases}{}
	\nonumber\ddot X_i^{n,q} & $= -  \alpha \sum_{m=1}^{\# S} \nabla_{X_m}\nabla_{X_i} \LL_S(\XX^{n,q},\llambda^{n,q})\dot \XX^{n,q}_m $+\\ &
	$- \alpha\beta (\sum_{{ (i,j)\in S }} \phi_{ij}(\XX^{n,q})\lambda^{n,q}_{ij}\nabla_{X_i}\phi_{ij}(\XX^{n,q}))-c\dot X_i^{n,q}, \ i=1,\ldots,\# S$ \label{eq:1stDAHS1} \\
	\dot \lambda_{ij}^{n,q}   &
	$=\ 
	\begin{cases}
		0,\ \text{ if } \lambda_{ij}^{n,q}=0 \text{ and } \phi_{ij}(\XX^{n,q})<0\\
		\beta \phi_{ij}(\XX^{n,q}), \ \text{ otherwise}
	\end{cases},\ \ \ \ (i,j)=1,\ldots,\# S $. 
	\label{eq:1stDAHS2}
\end{numcases}
where $c$ is the new damping term. One possible way to link \eqref{eq:1stDAHS1} with \eqref{eq:AHSystem1} is to take the time derivative in \eqref{eq:AHSystem1} and then to replace $\dot \lambda_{ij}^{n,q}$ by $\beta \phi_{ij}(\XX^{n,q})H(\lambda_{ij}^{n,q})$ with $H(\cdot)$ a suitable function preserving the same steady state of system (\ref{eq:AHSystem1})-(\ref{eq:AHSystem2}). We choose in particular in the following $H(\lambda_{ij}^{n,q}) := \lambda_{ij}^{n,q}$ which is enough to assure that when a steady state is reached then $\phi_{ij}(\XX^{n,q})\lambda^{n,q}_{ij}=0$ and that, therefore, the two systems have the same steady state. 

The system \eqref{eq:1stDAHS1}-\eqref{eq:1stDAHS2} however is still not suitable for our scope since the term $\alpha \sum_{m=1}^{\# S}\\ \nabla_{X_m} \nabla_{X_i} \LL_S(\XX^{n,q},\llambda^{n,q})\dot \XX^{n,q}_m $ may give rise to growing modes in the solution of the dynamical system, its sign not being determined a-priori. Thus, a last step which permits to obtain a dissipative dynamical system sharing the same equilibrium states as our original model  (\ref{eq:AHSystem1})-(\ref{eq:AHSystem2}) consists in simplifying the above dynamics by discarding one space derivative. This gives 
\begin{numcases}{}
	\ddot X_i^{n,q} & $= - \alpha (\nabla_{X_i} W_S(\XX^{n,q}) +\sum\limits_{(i,j)\in S, i<j} \lambda_{ij} \nabla_{X_i}\phi_{ij}(\XX^{n,q}))-c\dot X_i^{n,q}$ \nonumber\\
	&$\hspace{1cm} -\gamma (\sum_{{ (i,j)\in S, i<j}} \phi_{ij}(\XX^{n,q})\lambda_{ij}\nabla_{X_i}\phi_{ij}(\XX^{n,q})), \ i=1,\ldots,\# S$ \label{eq:1stDAHS1_2} \\
	\dot \lambda_{ij}^{n,q}   &
	$=\ 
	\begin{cases}
		0,\ \text{ if } \lambda_{ij}^{n,q}=0 \text{ and } \phi_{ij}(\XX^{n,q})<0\\
		\beta \phi_{ij}(\XX^{n,q}), \ \text{ otherwise}
	\end{cases},\ \ \ \ (i,j)=1,\ldots,\# S$,
	\label{eq:1stDAHS2_2}
\end{numcases}
{ where $\alpha\beta$ was replaced by a new parameter $\gamma>0$}. We observe from \eqref{eq:1stDAHS2_2} that a steady state solution $(\bar \XX, \bar \llambda)$ of system (\ref{eq:1stDAHS1_2})-(\ref{eq:1stDAHS2_2}), implies $\bar\lambda_{ij}\phi_{ij}(\bar\XX)=0$.  Consequently, $(\bar\XX,\bar\llambda)$ is a critical point of the Lagrangian, i.e., we have $\nabla_{X_i} \LL_S(\bar\XX,\bar\llambda) = 0,$ and $\nabla_{\lambda_{ij}} \LL_S(\bar\XX,\bar\llambda) = 0,\ (i,j)\in S$.  This implies that the set of steady state solutions of the new system \eqref{eq:1stDAHS1_2}-\eqref{eq:1stDAHS2_2} coincides with the set of critical-points of the Lagrangian function $\LL_S(\XX^{n,q})$. Let us also observe that this reformulation shares similarities with the so-called dissipative heavy ball method \cite{ALVAREZ} frequently used to accelerate the convergence of unconstrained minimization problems. In the sequel, we refer to \eqref{eq:1stDAHS1_2}-\eqref{eq:1stDAHS2_2} as the dissipative Arrow-Hurwicz system (DAHS) when we consider system \eqref{eq:1stDAHS1_2}-\eqref{eq:1stDAHS2_2}.
\begin{remark}\label{rem:mod_multip}
	Following the same path as above, one can introduce a dissipative dynamics also in the Lagrange multipliers equation. This may give rise to a new system reading
	\begin{numcases}{}
		\ddot X_i^{n,q} & $= - \alpha (\nabla_{X_i} W_S(\XX^{n,q}) +\sum\limits_{(i,j)\in S,i<j} \lambda_{ij} \nabla_{X_i}\phi_{ij}(\XX^{n,q}))-c\dot X_i^{n,q}$ \nonumber\\
		&$\hspace{1cm} -\gamma (\sum_{{ (i,j)\in S,i<j }} \phi_{ij}(\XX^{n,q})\lambda_{ij}\nabla_{X_i}\phi_{ij}(\XX^{n,q})), \ i=1,\ldots, \# S$ \label{eq:1stDAHS1_3} \\
		\ddot \lambda_{ij}^{n,q}   &$=
		\beta \phi_{ij}(\XX^{n,q})-c_{\lambda}\dot \lambda_{ij}^{n,q} ,\  (i,j)=1,\ldots,\# S $,\\
		\dot \lambda_{ij}^{n,q}   &$=	0,\ \text{ if } \lambda_{ij}^{n,q}=0 \text{ and } \phi_{ij}(\XX^{n,q})<0,\  (i,j)=1,\ldots,\# S $
		\label{eq:1stDAHS2_3}
	\end{numcases}
	with $c_{\lambda}$ a new numerical parameter. This interesting modification to the algorithm which may additionally improve the performances of the minimization method with respect to the solution of the classical AH method will not be explored here. We postpone such study to the future.
\end{remark}

Now, we have a non linear system to solve and thus, we need an iterative method to find a steady state solution which also minimize the distance between the particles under the non overlapping constraint. Thus, given now a small step size $\delta>0$, the damped Arrow-Hurwicz algorithm (DAHA) consists on a semi-implicit discretization of the DAHS system (\ref{eq:1stDAHS1_2})-(\ref{eq:1stDAHS2_2}). This reads
\begin{numcases}{}
	&$X_i^{n,q,r+1/2}  = X_i^{n,q,r} - \frac{\delta}{2} V_i^{n,q,r}$\label{eq:1stDAHS1_4} \\
	&$V_i^{n,q,r+1/2}  =  - \widetilde\alpha (\nabla_{X_i} W_S(\XX^{n,q,r+1} ) +\sum\limits_{ (i,j)\in S,i<j } \lambda_{ij}^{n,q,r} \nabla_{X_i}\phi_{ij}(\XX^{n,q,r+1}))+$ \nonumber\\
	& $+V_i^{n,q,r}-\widetilde\gamma (\sum\limits_{(i,j)\in S,i<j} \phi_{ij}(\XX^{n,q,r+1})\lambda_{ij}^{n,q,r}\nabla_{Xi}\phi_{ij}(\XX^{n,q,r+1})), \ i=1,\ldots,\# S$\label{eq:1stDAHS2_4}\\
	& $V_i^{n,q,r+1}  = V_i^{n,q,r+1/2}-\tilde cV_i^{n,q,r+1}, \ i=1,\ldots,\# S$\label{eq:1stDAHS2_4b}\\
	&$X_i^{n,q,r+1}  = X_i^{n,q,r+1/2} - \frac{\delta}{2} V_i^{n,q,r+1}$\label{eq:1stDAHS1_4bis} \\
	&$\lambda_{ij}^{n,q,r+1}  = 
	\max\{ 0, \lambda_{ij}^{n,q,r}+\tilde\beta \phi_{ij}(\XX^{n,q,r+1})\},\ 
	\ (i,j) \in S, \ \ i=1,\ldots,\# S $
	\label{eq:1stDAHS3_4}
\end{numcases}
where $\widetilde \alpha=\delta \alpha,\ \widetilde\beta = \delta\beta$, $\widetilde\gamma=\delta \gamma$ and $\widetilde c=\delta c$ 
are positive numerical parameters (in what follows, for simplicity we will drop the tildes over the parameters).

Let us observe that a steady state of the DAHS \eqref{eq:1stDAHS1_2}-\eqref{eq:1stDAHS2_2} satisfies
\begin{equation}
	\nabla_{X_i} W_S(\bar\XX) = -\sum_{{ (i,j)\in S }} \bar\lambda_{ij} \nabla_{X_i}\phi_{ij}(\bar\XX),\ i=1,\ldots,\# S.
\end{equation}  
These equalities show that a solution to the minimization problem corresponds to a balance between attraction forces associated to the adhesive particles (left hand-side) and repulsion forces associated to the impenetrability condition (right hand-side). The dynamics described by this reformulated system corresponds thus to a search for this balance.

\paragraph{{Initial conditions}} The iterative method  described above needs suitable initial conditions to start. The natural choice, which is the one we employ in this work is to use
\begin{equation}
	\XX^{n,q,0} = \bar \XX^{n,q}  \text{  and  } \lambda_{i,j}^{n,q} = \lambda \max(0,\phi_{ij}(\bar \XX^{n,q})).\label{eq:ini_cond_DAHA}
\end{equation}
where $\bar \XX^{n,q}$ indicates the equilibrium configuration obtained as a result of the previous minimization problem with $S^{n,q-1}$ the set of interacting pairs.
The coefficient $\lambda$ is constant satisfying positivity $\lambda \geq 0$. 

\paragraph{{Stopping conditions}} The method needs also a condition to stop the iterations. We choose two different conditions {imposed simultaneously}. The first one is based on a measure of the steadiness of the iterative scheme. {This reads
	\begin{equation}\label{eq:convCriterion_stst}
		|\XX^{n,q,r+1}-\XX^{n,q,r}| \leq \epsilon_s,
\end{equation}}
which states the fact that each sphere does not move more than a given threshold during one iteration. Let observe that this condition may be generalized asking that particles do not move more than a given threshold for more than one iteration of the optimization scheme. However, in practice, we noticed that the condition \eqref{eq:convCriterion_stst} is sufficient to get reliable results. {The second condition is about the accuracy with which the conditions on the constraints are fulfilled   
	\begin{equation}\label{eq:convCriterion_constr}
		\frac{\sum_{(i,j) \in S^{n,q}} \left( \sqrt{(R_i+R_j)^2 - |X_i^{n,q,r}-X_j^{n,q,r}|^2}\right)}{\# S} \leq \epsilon_c,
	\end{equation}
	for small chosen positive constants $\epsilon_c$ and $\epsilon_s$. Let observe that this second condition is about the average overlapping of the system which has to stay below a given threshold.}

\paragraph{{Improving the speed of convergence: dynamical parameters $c$ and $\beta$}} Before concluding this part we stress that an improvement to the algorithm $\eqref{eq:1stDAHS1_4}-\eqref{eq:1stDAHS3_4}$ can be obtained by taking the numerical parameters $c$ and $\beta$ dependent on the artificial time evolution indexed by $r$. In fact, one way to get to this goal and to improve efficiency consists in allowing larger movement to the particles at the beginning of the minimization and to reduce this possibility as the fulfillment of the stopping conditions are close to be reached.
In practice, the damping coefficient $c$ and the coefficient $\beta$ are chosen such that 
\begin{equation}\label{damping}
	c_i=\bar{c}\frac{|\delta V^{n,q,r}_i-\phi_i^A|}{\phi_A},\quad
	\beta=\bar\beta\frac{ \sqrt{N}}{\phi_A}
\end{equation}
where now $\bar c$ and $\bar \beta$ are suitable constants, $\phi_i^A$ is the average overlapping of the sphere $i$ while $\phi_A$ is the average overlapping of the particles involved in the minimization process. 
This is enough to accelerate the convergence process when a minimum is close to be reached. In fact, as the average overlapping constraint is satisfied then the damping coefficient increases in order to slow down the particle displacement. This displacement is also reduced when the virtual speed of the particles $V^{n,q,r}_i$ multiplied by the time step $\delta$ is too large compared to the average overlapping of a particle in such a way to avoid to move too far from a given configuration. The dependence of $\beta$ upon the number of particles involved in the minimization permits to increase the speed at which the Lagrange multipliers evolve. This is done in order to speed up the procedure when the number of spheres increases which otherwise is observed to be very slow. A last aspect to be discussed is about the non uniqueness of the solutions of the dynamical systems considered both for the case of the original Arrow-Hurwicz system as well as for the case of its dissipative version. To that aim, we recall that we are trying to reproduce a ballistic aggregation phenomenon and, consequently, one knows that the position of the particles before the free motion was such that the condition of non overlapping was satisfied. This condition gives naturally the hint that the new configuration of the system has to be searched in a region which is close to the configuration after the advection step in which some overlapping may be present. Thus, the choice of the damping parameter $c$ in \eqref{damping} permits to reduce the set of possible solutions of the DAHS \eqref{eq:1stDAHS1_3}-\eqref{eq:1stDAHS2_3}, i.e. of the set of minima, keeping the solution close enough to the original configuration.

\subsection{An efficient almost rigid clusters formulation}\label{sec:R}
In the previously described method, the relative position of the spheres in one given cluster may change over time since all particles belonging to a new formed cluster after ballistic motion interact through the minimization potential function $W_S(\XX^{n,q})$ and the Lagrangian multipliers $\lambda_{ij}^{n,q}$. This causes two main drawbacks. The first is related to a modeling choice. In fact, as already claimed, we look for minima of the potential function over the non overlapping constraints which are close to the configuration of the cluster after the advection. However, the potential function being as defined in \eqref{min1}, the attraction force tends to form densely packed aggregates. { This shape is expected to be more compact than the one we would obtain by following the event-driven strategy.} The second reason is important as well and is about complexity. It is in fact clear that { the cost involved in the computation of the set of interacting spheres, $S$, scales quadratically with the number of spheres.} In addition, the Lagrange multipliers cause severe storing problems even for moderate values of the number of particles, i.e. $N\approx 10^5$, because they require to store all the values $\lambda^{n,q}_{ij}, \forall \ i,j=1..,\#S$ which has a { $N^2$ memory consumption} and consequently represents a bottleneck in the simulations.

For the above reasons, in this section, we propose a modification of the previous algorithm that partly prevents the spheres belonging to the same cluster before collision to change their relative position as an effect of the minimization. This means in particular that this modification acts in the direction of considering the cluster as a partial rigid body, where some elasticity is admitted but only locally. Let us observe that the kind of dynamics described by such a method would share more similarities with the ED method (a description of such method is given in Section~\ref{sec:ED} of the supplementary material) compared with the method described in Section \ref{sec:min_alg} since the Event Driven approach is a method to simulate the formation of rigid clusters.

\begin{figure}
	\centering
	\includegraphics[scale=0.7]{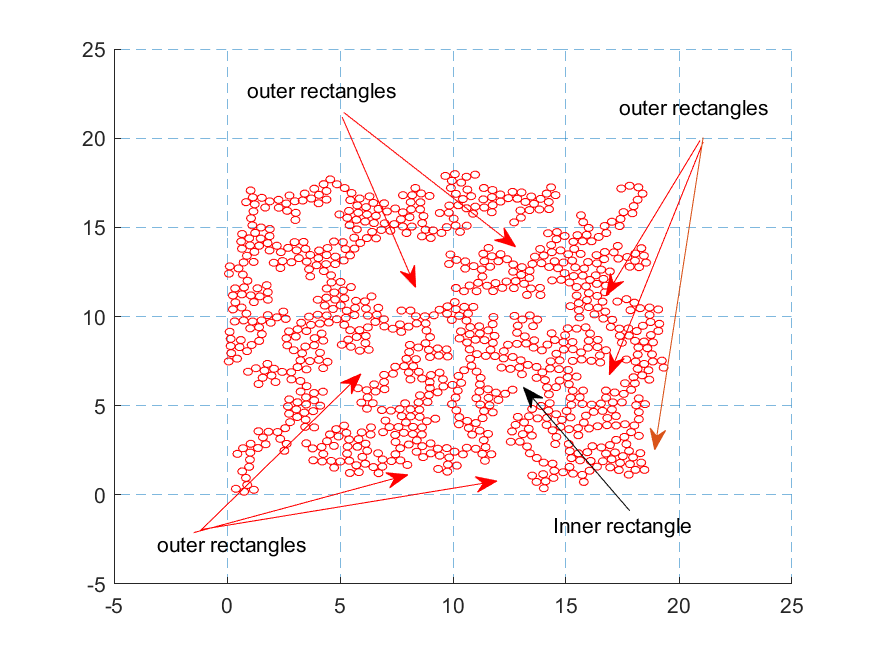}
	\caption{Example of a grid with the inner and the outer rectangles structure. Given a sphere belonging to the inner rectangle, interactions may occur only with spheres belonging to the same inner or with one belonging to the six outer rectangles.}
	\label{fig:grid}
\end{figure}

The modification consists of two steps. First, the potential function $W$ is modified by introducing an indicator function $H_W$ which is responsible for transforming the global attraction to a local one. Then one divides the whole simulation domain through horizontal and vertical lines into rectangles. Each rectangle is then used for registering all particles inside it before performing Step $c)$ of the time stepping minimization method \ref{tsm}. At the same time, as a potential basin for interaction, a wider area is taken into account which consists of all rectangles surrounding the inner rectangle under consideration as shown in Figure \ref{fig:grid}. 
The width of the outer and inner rectangles is determined from the size of the support of the indicator function of the potential $W$ and from the total number of interacting spheres $N$ to improve efficiency. The interaction potential becomes then
\begin{equation}\label{min2new}
	W_S(\XX^{n,q})= \frac{1}{2} \sum_{i\in \tilde S_m}\sum_{j\in \hat S_m} H_W( |X^{n,q}_i-X^{n,q}_j|\leq \Delta_W) |X^{n,q}_i-X^{n,q}_j|^2, 
\end{equation}
with $H_W$ the indicator function and with $\Delta_W$ a given threshold typically of the order of the {maximum displacement of the clusters during the free advection step (Step 1 of the Algorithm \ref{tsm})}.
% radius of the spheres $R$.
In practice, in our simulations, we took $\Delta_W=4R$, but other choices as the one shown in Figure \ref{fig:grid}, are of course possible and may improve the global efficiency of the method. Instead, $\tilde S_m$ represents the reduced cluster composed only of the spheres belonging to an inner rectangle, in the sequel denoted by the index $m$, and $\hat S_m$ the reduced cluster composed by only the spheres belonging to a inner and its relative outer rectangles. Concerning the Lagrange multipliers $\lambda_{ij}^{n,q}$ a similar procedure is adapted. Thus, the possible overlapping considered during the minimization are only those among the particles belonging to a fixed inner rectangle and the ones belonging to that inner and the relative outer rectangles. This means that the equation for the time evolution of the Lagrange multipliers is modified as follows, for $i=1,\ldots,\# \tilde S_m$
\begin{equation}
	\lambda_{ij}^{n,q,r+1}   
	= \max\{ 0, \lambda_{ij}^{n,q,r}+\beta \phi_{ij}(\XX^{n,q,r+1})\},\ 
	i \in \tilde S_m,\ j \in \bar S_m, \ m=1,\ldots,Mr
	\label{eq:1stDAHS3_4mod}
\end{equation}
where $Mr$ represents the total number of inner rectangles, $\tilde S_m$ the subset of the cluster $S$ belonging to a inner rectangles $m$, $\bar S_m$ the subset of the cluster $S$ belonging to the inner, $m$, and its relative outer rectangles. The same modification in the DAHA algorithm described by \eqref{eq:1stDAHS1_4}-\eqref{eq:1stDAHS2_4} holds true for the spheres motion and thus it is not reported. {Let us observe that reducing the number of constraints, which are taken into account in the method, is done only for the purpose of optimizing the computational time. Therefore this step does not change the results of the minimization step as opposite with \eqref{min2new} which instead modifies the shape of the cluster compared with the interacting potential \eqref{min1}.}
Finally, the number $Mr$ of rectangles used depends on the number of interacting spheres $\# S$ and it grows monotonically with this number. For instance, when we have to compute a minimization with $N<10^3$ we take $Mr=1$, while when the number of spheres is of the order $N=10^6$ we consider $Mr=100^2$. A last remark follows.

\begin{remark}\label{rem:outer-inner-loop}
	As detailed before, the methods described in this Section \ref{sec:R} and in Section \ref{sec:min_alg} both freeze the set $S$ of the interacting spheres during the full minimization procedure up to a convergence to a balance between energy and non overlapping constraints. This approach can be revisited by merging step $3 d)$ with step $3 c)$. This would mean to actualize the set $S$ of spheres belonging to a given cluster at each iteration of the minimization procedure. This direction could be pursued to take into account the possibility that new overlapping situations arise: spheres that were not interacting, being not part of the same cluster, overlap as an effect of their displacement during the search of an equilibrium. This will give rise to a class of new methods.
	We decided not explore this interesting perspective in this work and to postpone this possibility to future investigations.
\end{remark}

\section{Analysis of Time Stepping method}
\label{sec:test}
In this section, we perform a detailed analysis of the new Time Stepping Method described previously. 
In the first part, we focus on the performances in terms of computational effort, i.e. the amount of computer work accomplished by the scheme. The second part of the section is instead devoted to the comparison with the Event-Driven approach about the capability of statistically describing the aggregation phenomenon successfully. The Event-Driven method chosen as a benchmark is the classical one \cite{alder1959MDI,rapaport2004art,DEMICHELE} which is described in detail in the supplementary material \ref{sec:ED}.

\subsection{On the efficiency and computational performance of the TS method}\label{sec:modelParam}
We recall that all the numerical simulations are performed in dimension $d = 2$. In this context, the physical model parameters are the length $L$ of the domain, the number of spheres $N$, their mass $m_i, \ i=1,\ldots N$, their velocities $V_i, \ i=1,\ldots N$ and their radii $R_i, \ i=1,\ldots,N$. The above mentioned variables are intrinsic to the physical problem and not to the method employed in the simulations, this means that, in particular, the Event-Driven method shares this specific set of parameters with the TS one. To be more specific, in our study we restrict ourselves to spheres with the same mass and radii, i.e. $ m_i=m = 1$ and $R_i=R,\ i=1,\ldots,N$. These two quantities, in fact, play a role in the collision dynamics only at the level of the post-collisional velocities and for this reason, we expect the Event Driven and the Time Stepping methods to exhibit the same behavior with respect to them, since the post-collisional velocities are computed in the same way for both methods. Instead, we expect the size of the box $L$ compared to the radii $R$ of the spheres as well as the number of particles present in the domain to have a fundamental impact on the behavior of the TS approach both in terms of computational effort as well as in terms of the shape of the aggregates. Finally, also the initial distribution of velocities and the initial configuration of the system may play a role in the aggregation dynamics and may give rise to different behaviors of the TS and ED methods. In addition to the above discussed physical parameters, which characterize the aggregation phenomenon independently of how it is simulated, there exist other quantities related to the sole Time Stepping approach which instead characterize the numerical method developed in this work specifically. In particular, we recall the choice of the damping coefficient {$\bar c$ and of the coefficient $\bar \beta$} appearing in equation \eqref{damping} responsible for the speed at which the Lagrange multipliers evolve, the coefficients $\alpha$ and $\gamma$ measuring the intensity of the adhesion forces and of the overlapping constraints and the artificial time step $\delta$ present in the DAHA algorithm. Finally, the threshold parameters $\varepsilon_c$ and $\varepsilon_s$ measuring the stopping condition in the iterative scheme of Section \ref{sec:NR} also influence the efficiency of the method. 
One last parameter which deserves attention is the time step $\Delta t^n$ chosen for the free flight of the particles in the TS approach. This has an influence on the number of collision events to be handled at each iteration and during the whole phenomenon.

Since the number of variables is very large, we proceed along the following path. The first part of our study consists in finding the set of numerical parameters which are responsible for an optimal behavior of the iterative scheme of Section \ref{sec:NR}, namely {$\bar c$, $\alpha$ $\bar \beta,\gamma,\delta$ and $\varepsilon_c$ and $\varepsilon_s$. }
These parameters are almost independent of the aggregation dynamics phenomenon and needed to be selected for the best efficiency of the Time Stepping procedure. This analysis is not reported here for brevity since it is more related to the minimiziation method and not to the physical problem here studied. In the following, we then employ the values of these parameters which permits to have the best results in terms of computational effort. {The parameter values are:
	\begin{equation}\label{param}
		\bar c=1.3, \quad \delta=0.15, \quad \alpha=\gamma=0.3, \quad \bar\beta=3, \quad \varepsilon_c=10^{-2}, \quad \varepsilon_s=10^{-3}.
\end{equation}}
Instead, in the rest of the section, we focus on the effects of the physical parameters, namely $N$, $L$ and on the initial velocity and position for a fixed radius $R$ of the spheres. The other parameter for which we analyse our results is the time step $\Delta t^n$ chosen for the free flight of the particles. This, in fact, we expect to have an impact on the efficiency of the method. In the following, the parameters $N$, $L$ and $R$ are chosen satisfying the condition
\begin{equation}\label{eq:fitting_condition}
	L^2>4R^2 N,
\end{equation}
which implies that the spheres always fit inside the domain, i.e. there is always an empty subset of the domain not filled by particles. In Table \ref{tab:param} we show the chosen values of the parameters discussed above.
\begin{table}[h!]
	\centering
	\begin{tabular}{c|c|c}
		Description & Symbol &  Value(s) \\ \hline
		Number of particles   & $N$      & $\{10,\ 20,\ 30,\ 50,\  100,\ 200, \ 300, \ 500, \ 1000\}^2$  \\ 
		Volume fraction   & $V_f$      & $\{0.1,\ 0.2,\ 0.3\}$  \\ 
		Size of domain  & $L$    &  $\sqrt{\frac{N \pi}{V_f}} R$\\ 
		Radius of a particle   & $R$   & $0.2$ \\
		Mass of a particle & m  & $1$\\
		Free flight time step & $\Delta t (t=0)$  & $L/(\max_i(|V_i|))\{0.005, 0.01, 0.015\}$
	\end{tabular}
	\caption{Model parameters}
	\label{tab:param}
\end{table} 
The choices for the time steps are such that at the beginning, $(t=0)$, the fastest particles can travel a distance equal to the box length in one hundred time steps. Moreover, as already discussed in the previous section, we choose the time step to increase during the aggregation and in such a way that once a single cluster configuration is reached, this becomes $\Delta t(t=T_f)=2\Delta t(t=0)$ with $T_f$ the final time. This choice of an increasing monotone behavior for the time step is done in order to improve the computational performance of the scheme. In fact, the number of operations needed for the cluster computation, i.e. for detecting  new possible collisions which may arrive after each free travel of the particles, represent a relevant fraction of the full computational cost of the method while, on the other hand, the number of collisions decreases as the physical time increases, up to zero possible collisions, when the final aggregate is reached. 

We discuss now the initial conditions. Particle positions are uniformly assigned in the domain according to a square lattice configuration. This allows us to initialize a dense system where the particles do not initially overlap due to condition~\eqref{eq:fitting_condition}. Randomness in the system is introduced through the choice of the initial velocities $\{ V_i^0\}_{i=1,\ldots,N},$  given by 
\begin{equation}\label{inivel}
	V_i^0 = |V|(\cos(\theta_i^0),\sin(\theta_i^0)), \text{  with  }\theta_i^0\sim\mathcal{U}([0,2\pi]),
\end{equation}
where $|V|$ denotes the velocity unit and $\mathcal{U}(A)$ denotes the uniform distribution with support $A$. The number of realizations is fixed to ten for every value of the model parameters in Table \ref{tab:param} and results are averaged over this number. For the case $N=10^6$ we only perform simulations for the case in which the volume fraction is large, i.e. $V_f=0.3$, which is the case for which we expect to get the best performances of the TS scheme in terms of computational cost.

\begin{figure}[h!]%\label{example1}
	\centering
	\includegraphics[scale=0.12]{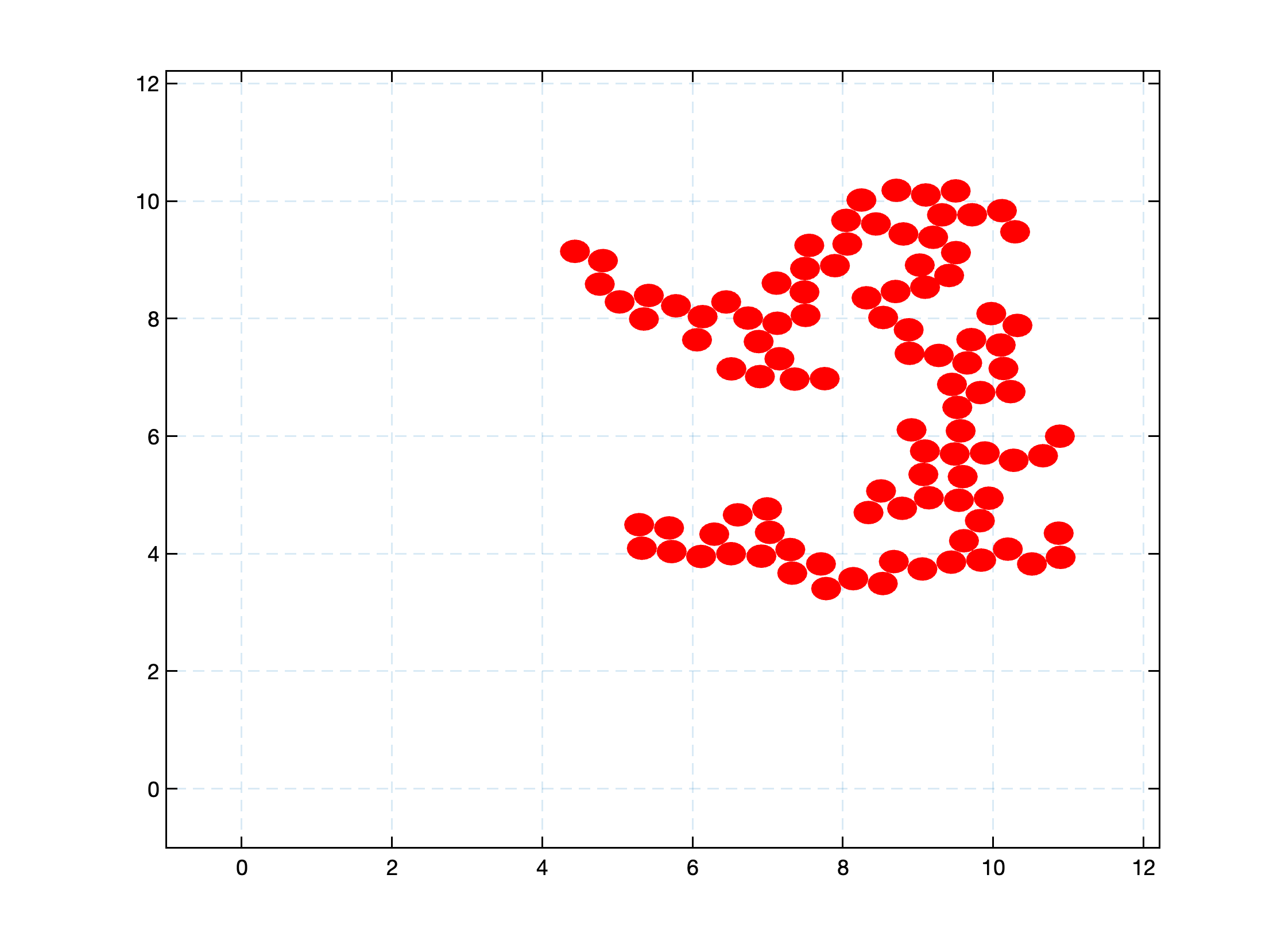}
	\includegraphics[scale=0.12]{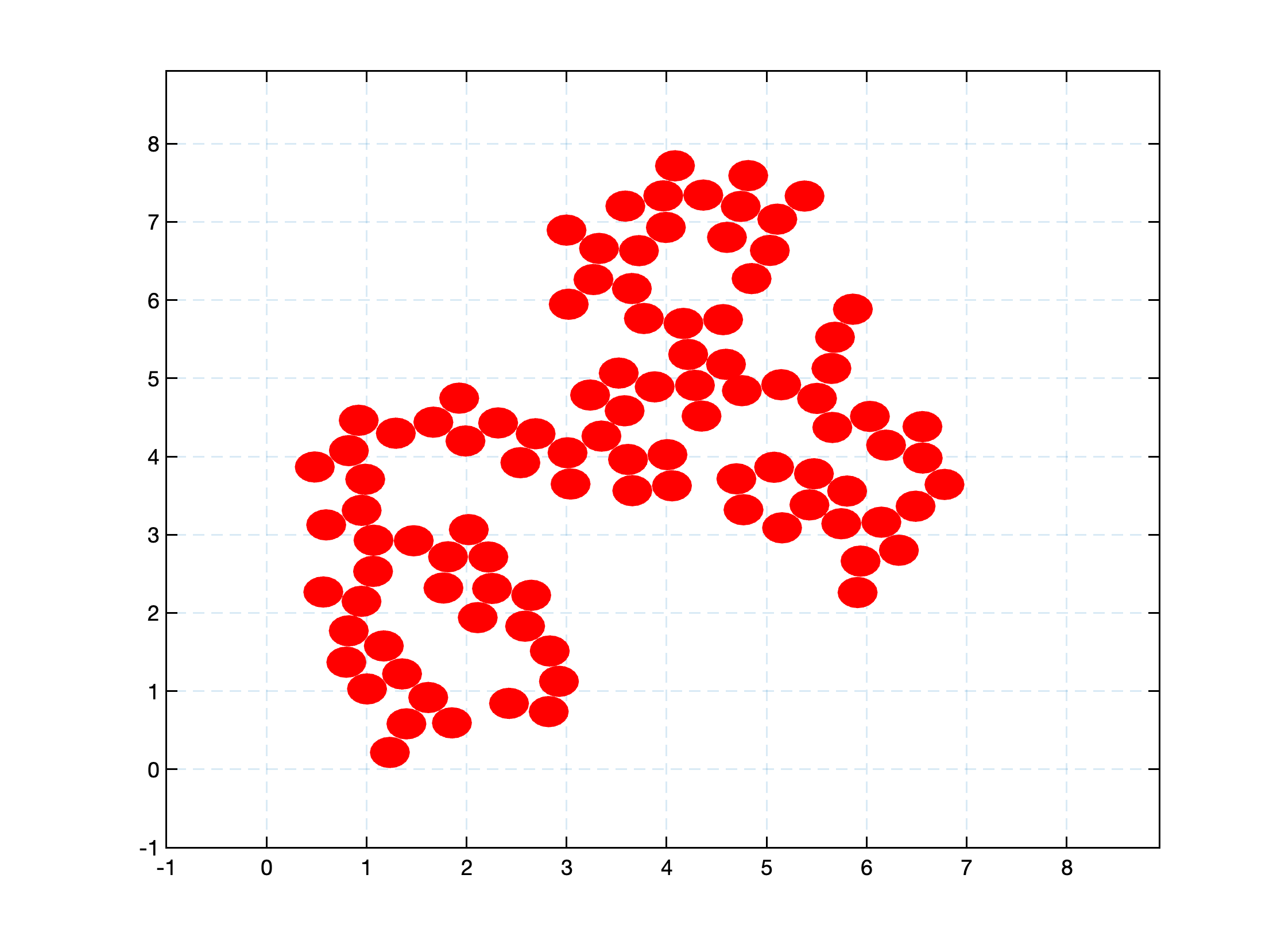}
	\includegraphics[scale=0.12]{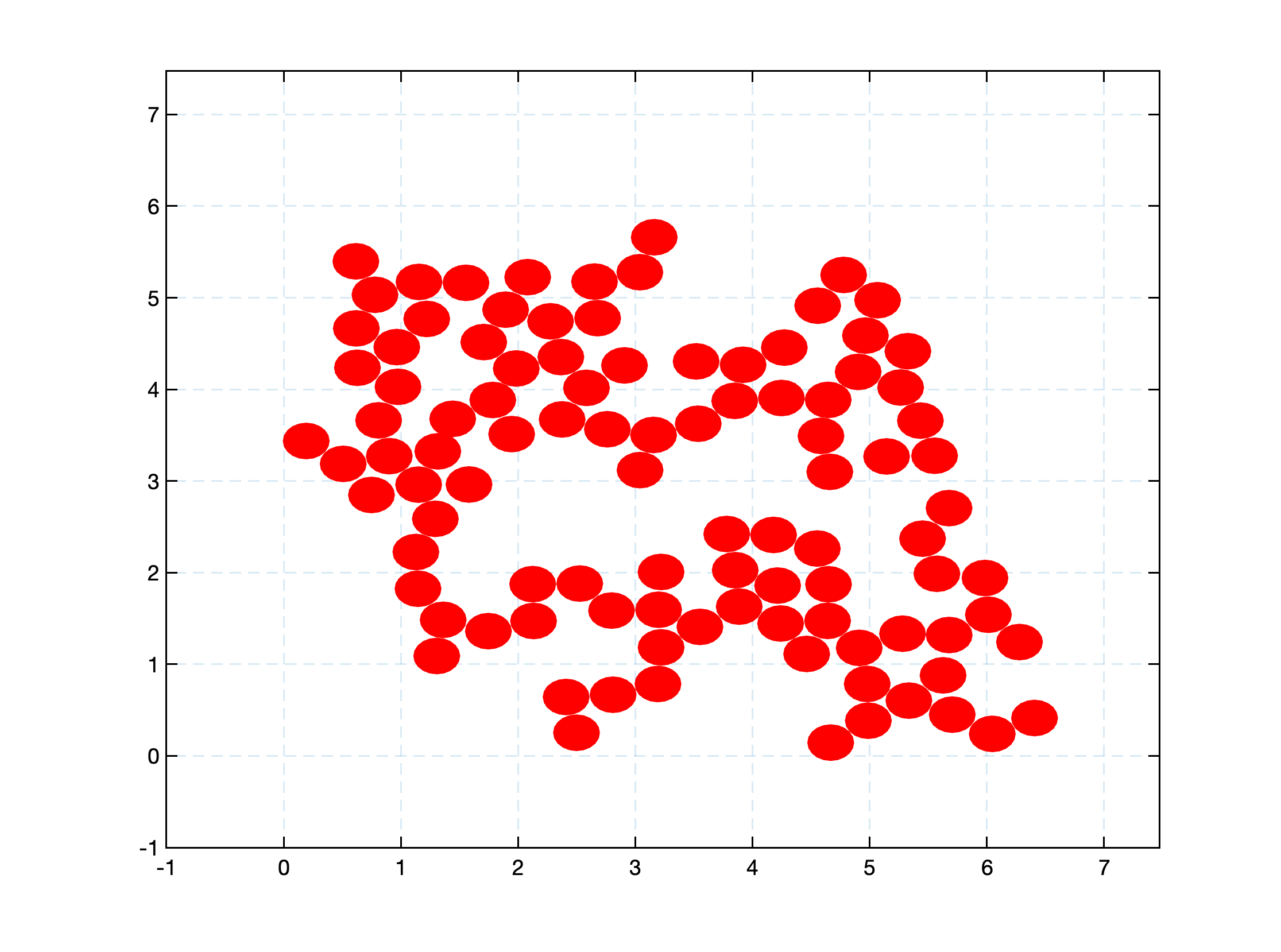}
	\includegraphics[scale=0.12]{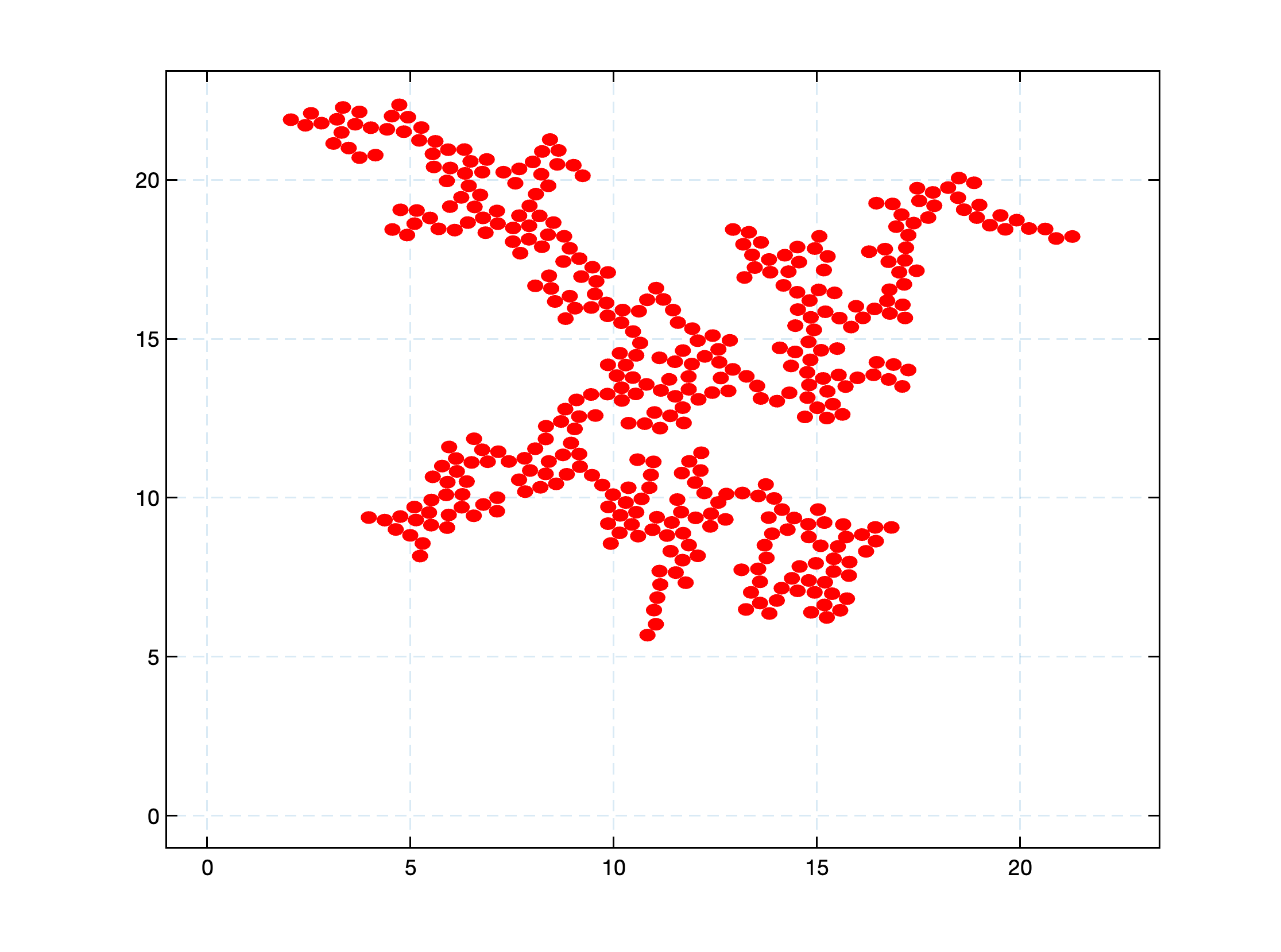}
	\includegraphics[scale=0.12]{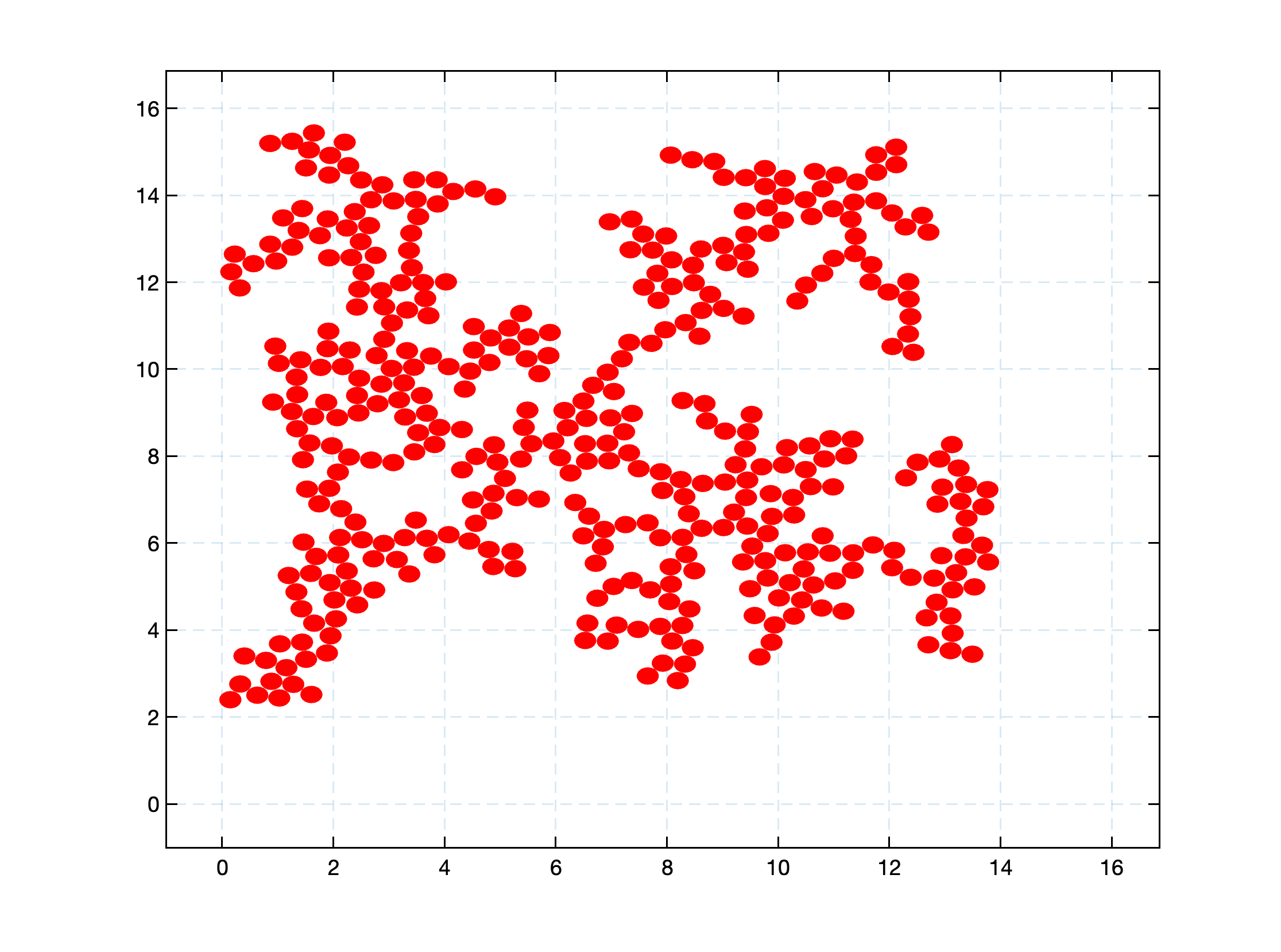}
	\includegraphics[scale=0.12]{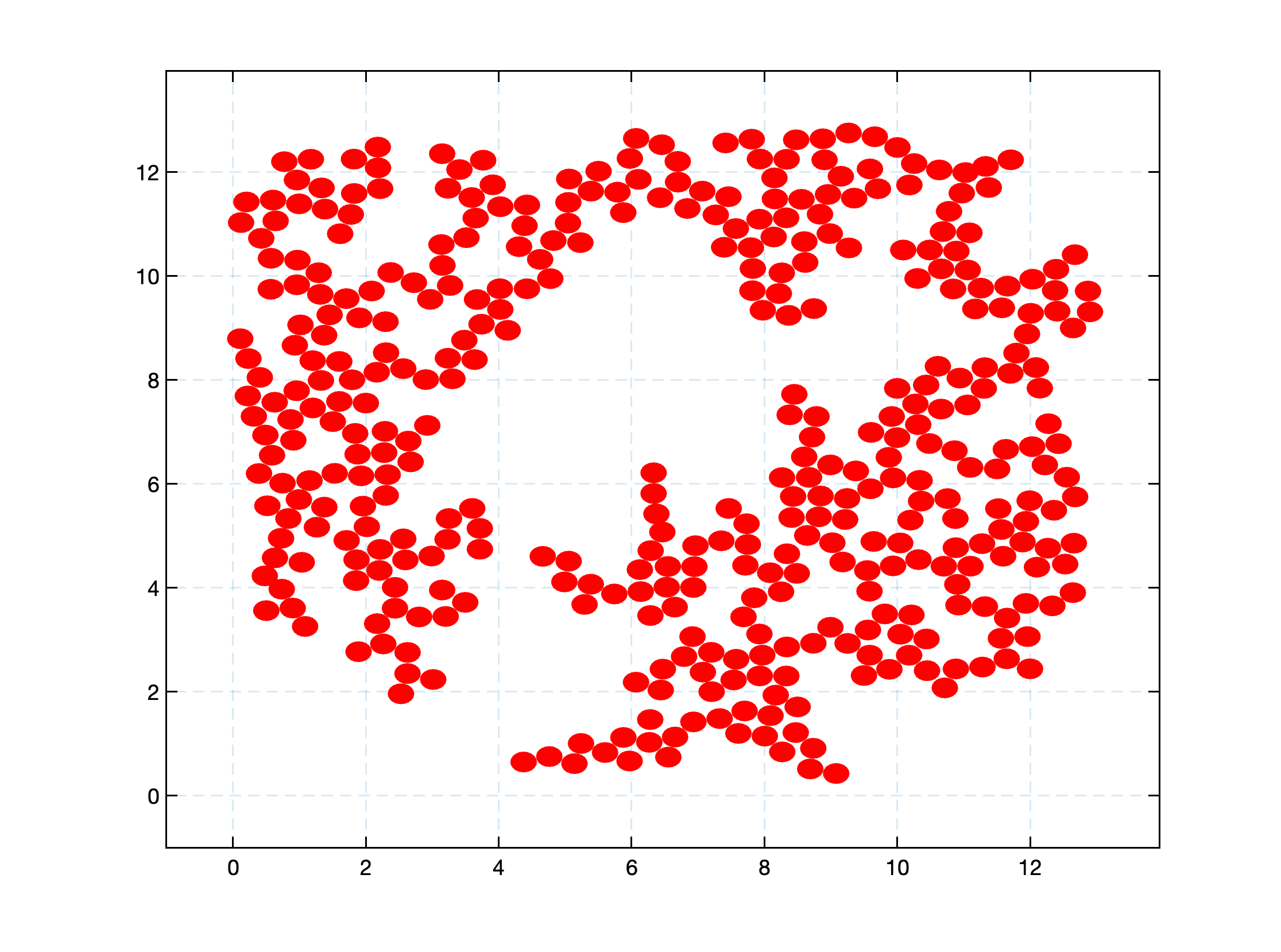}
	\includegraphics[scale=0.12]{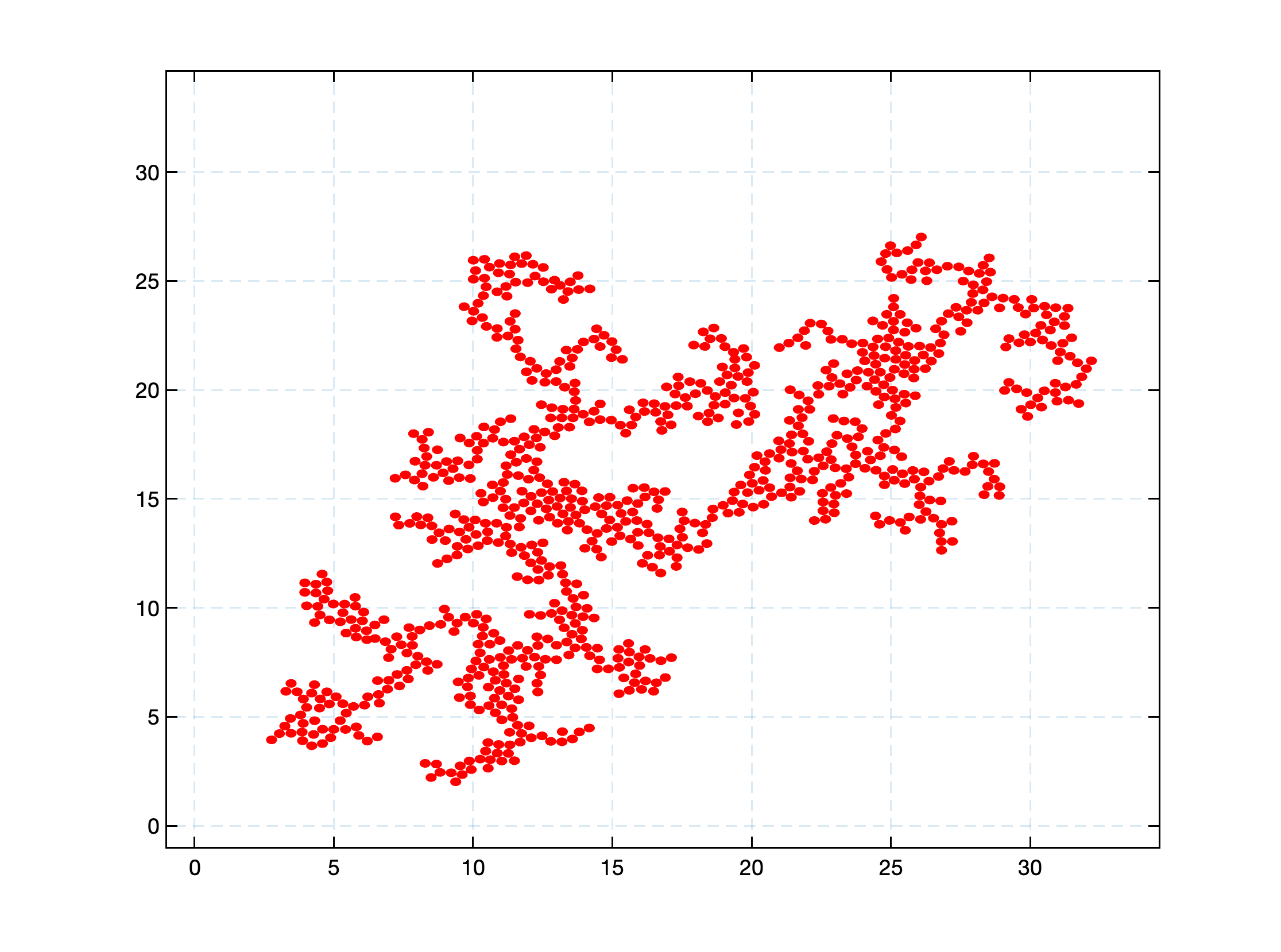}
	\includegraphics[scale=0.12]{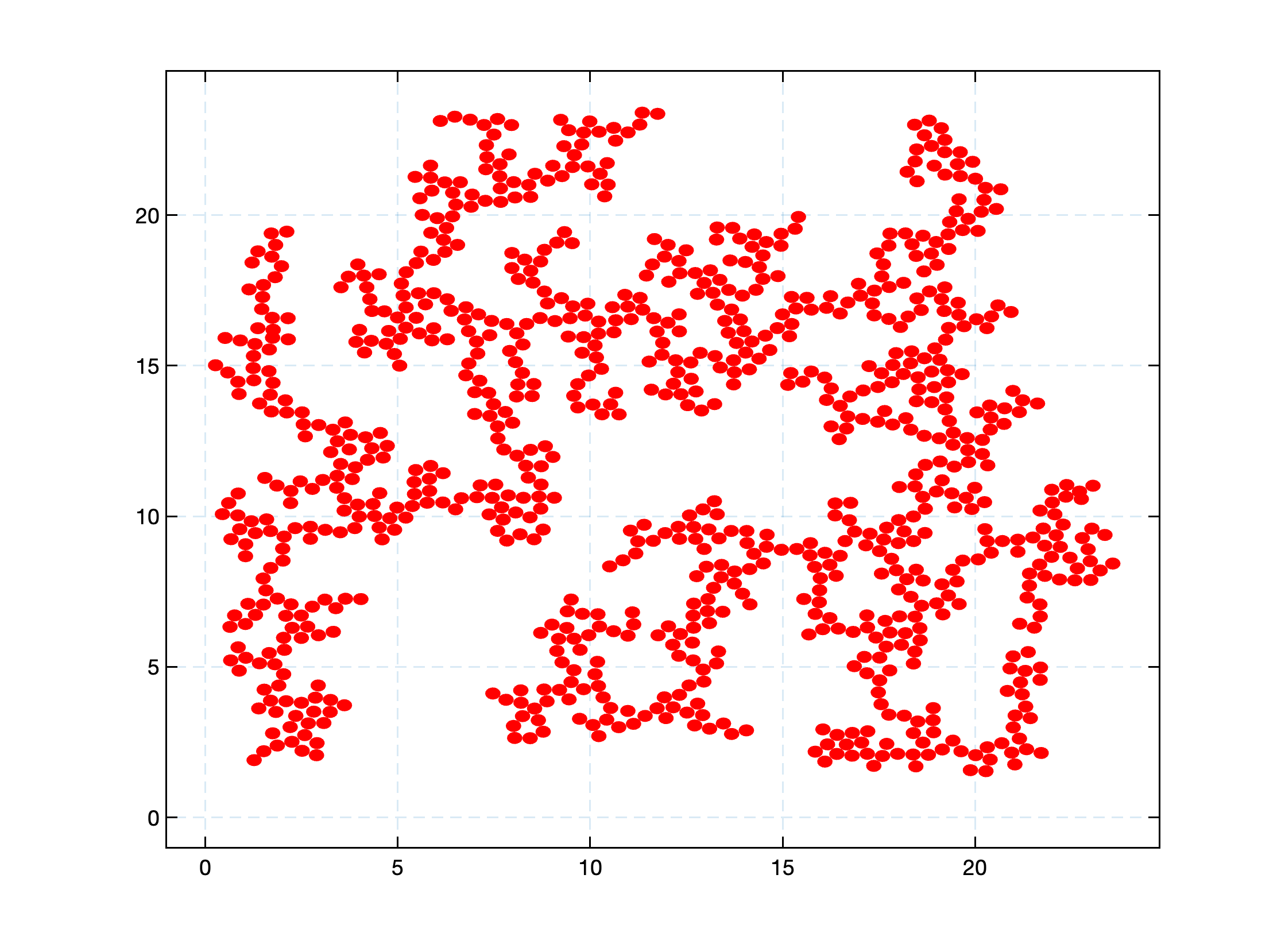}
	\includegraphics[scale=0.12]{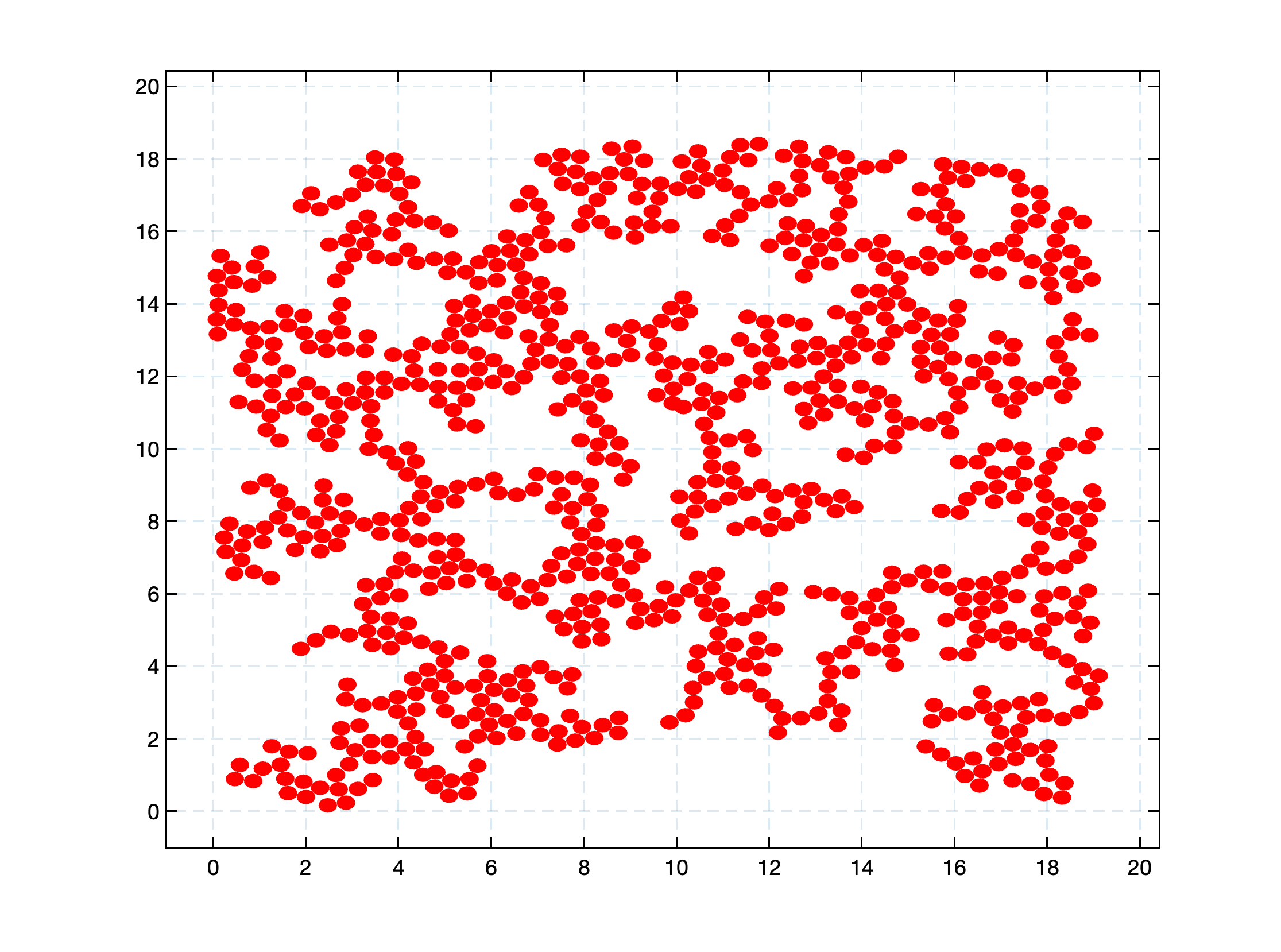}
	\caption{From top to bottom final configurations obtained with the TS scheme for $N=100$, $N=400$ and $N=900$ particles respectively. The volume fraction is equal to $0.1$, $0.2$ and $0.3$ from left to right.}\label{example1}
\end{figure}

\begin{figure}[h!]%\label{example2}
	\centering
	\includegraphics[scale=0.12]{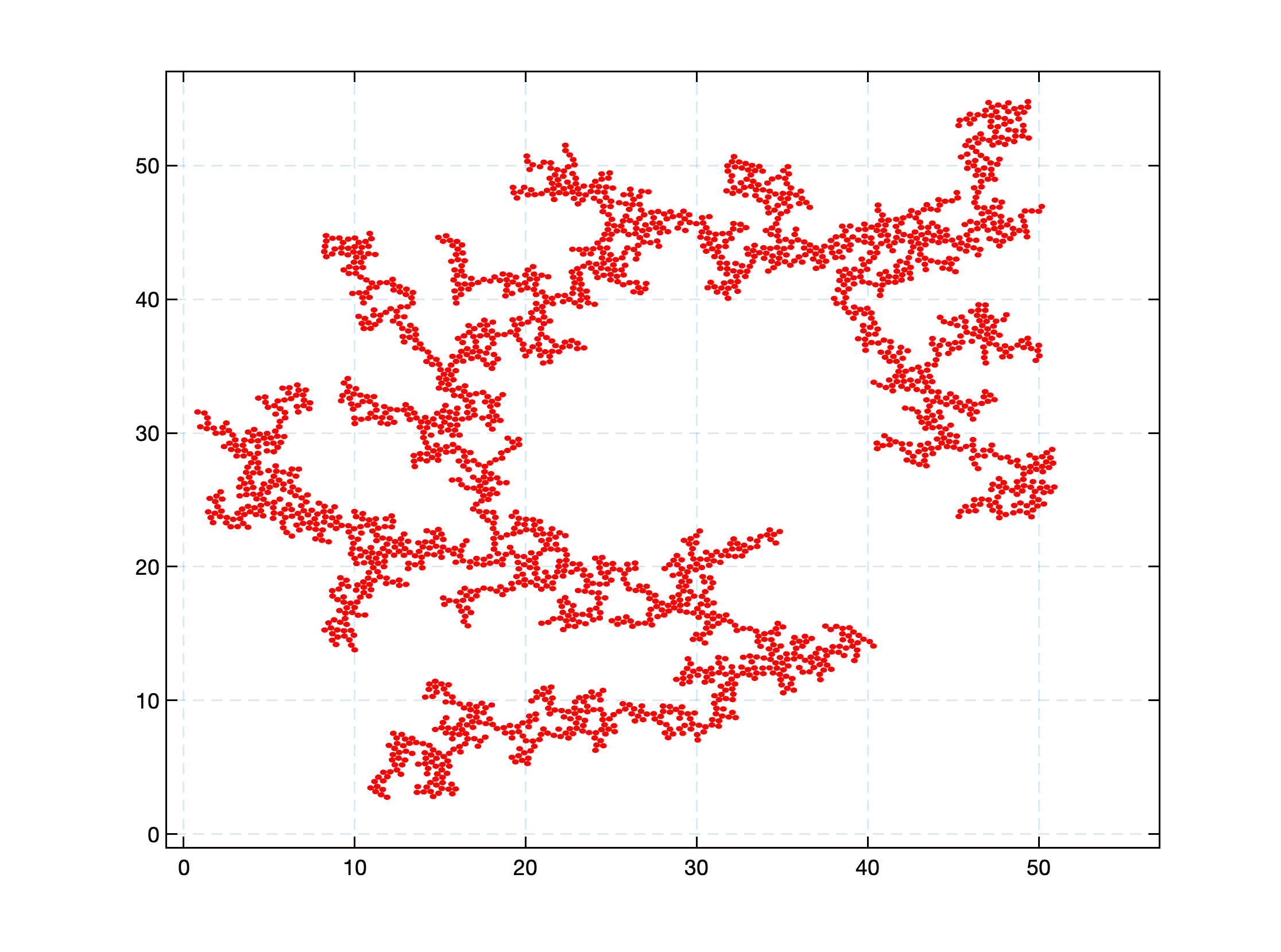}
	\includegraphics[scale=0.12]{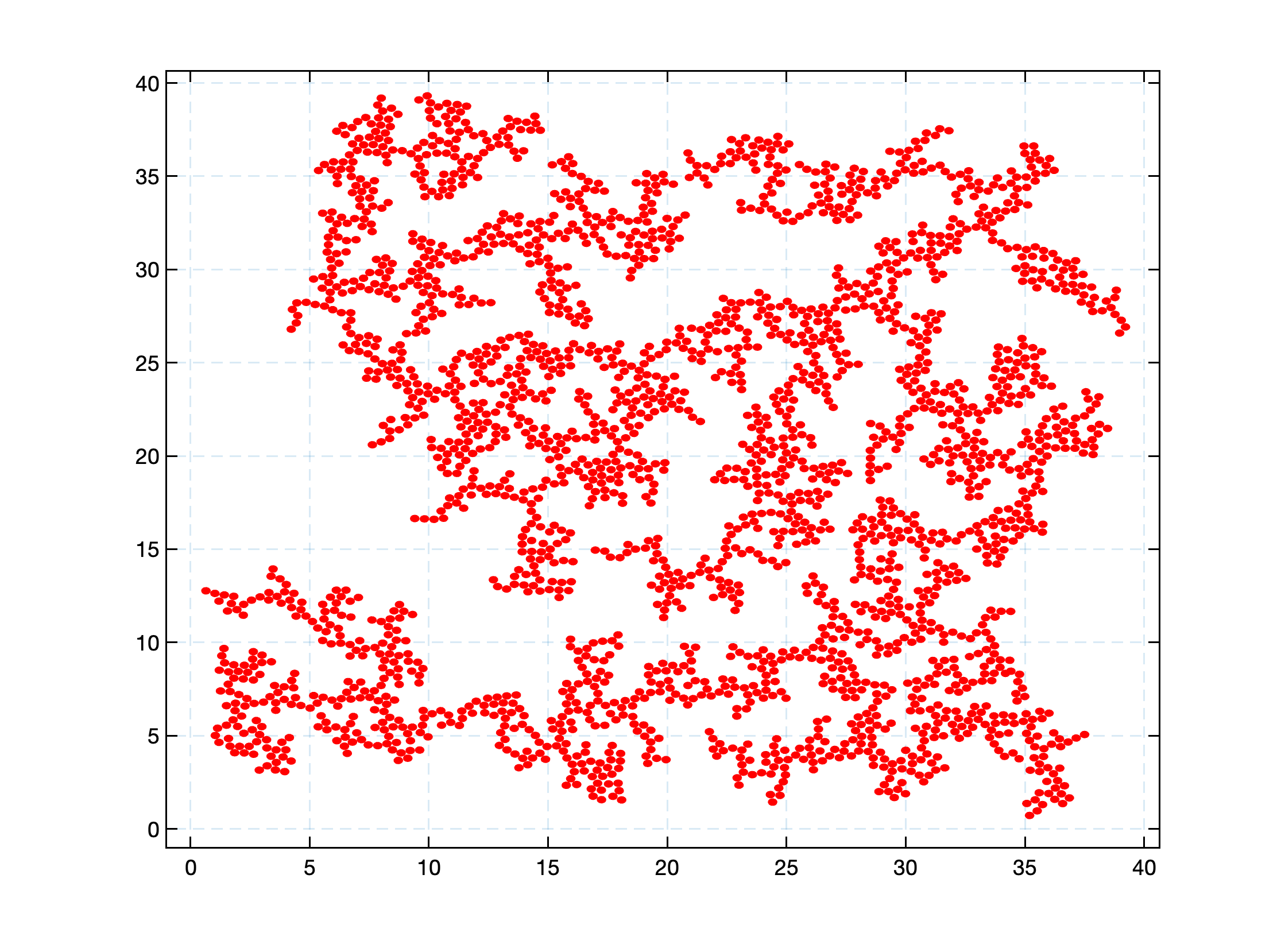}
	\includegraphics[scale=0.12]{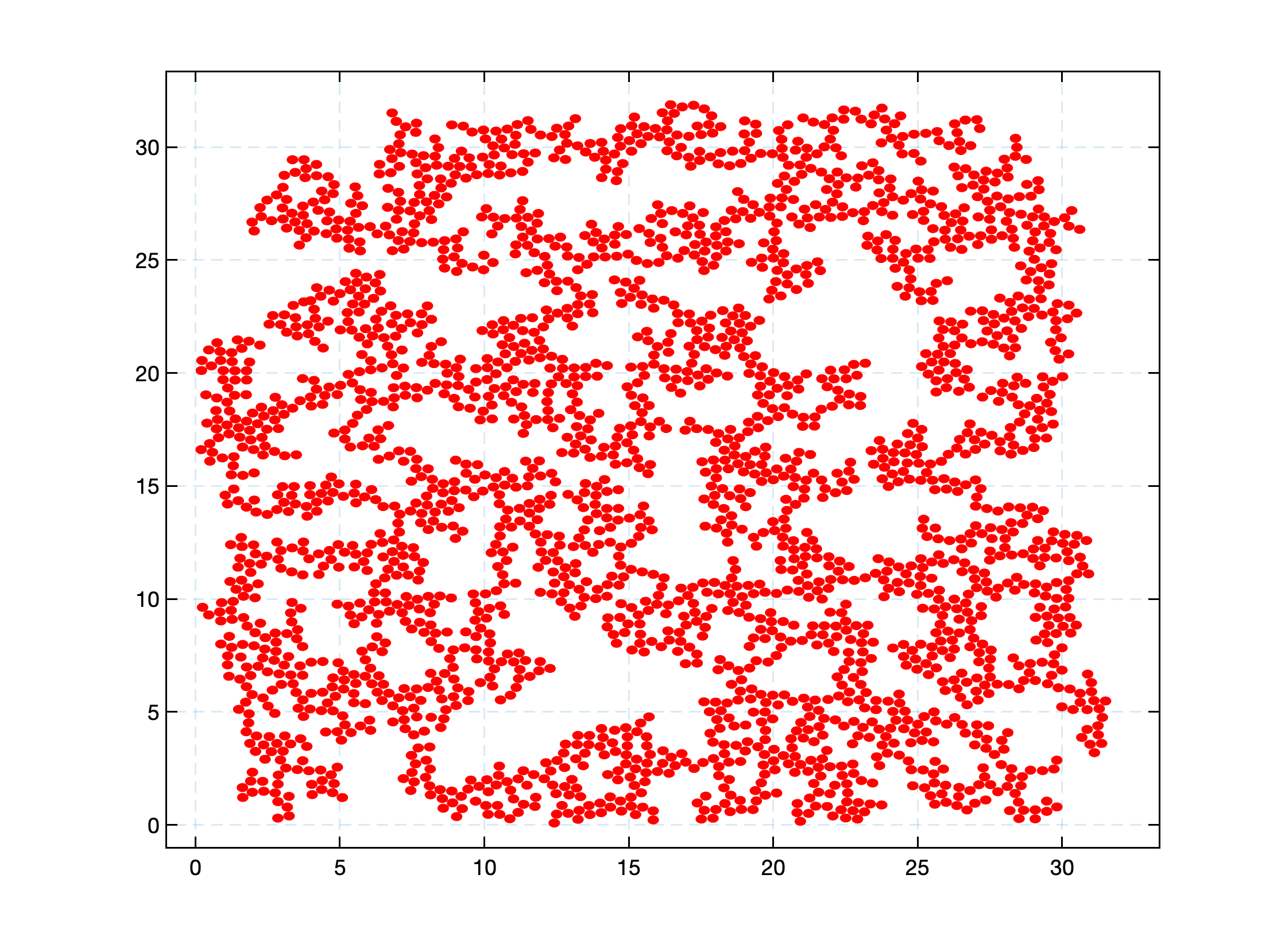}
	\includegraphics[scale=0.12]{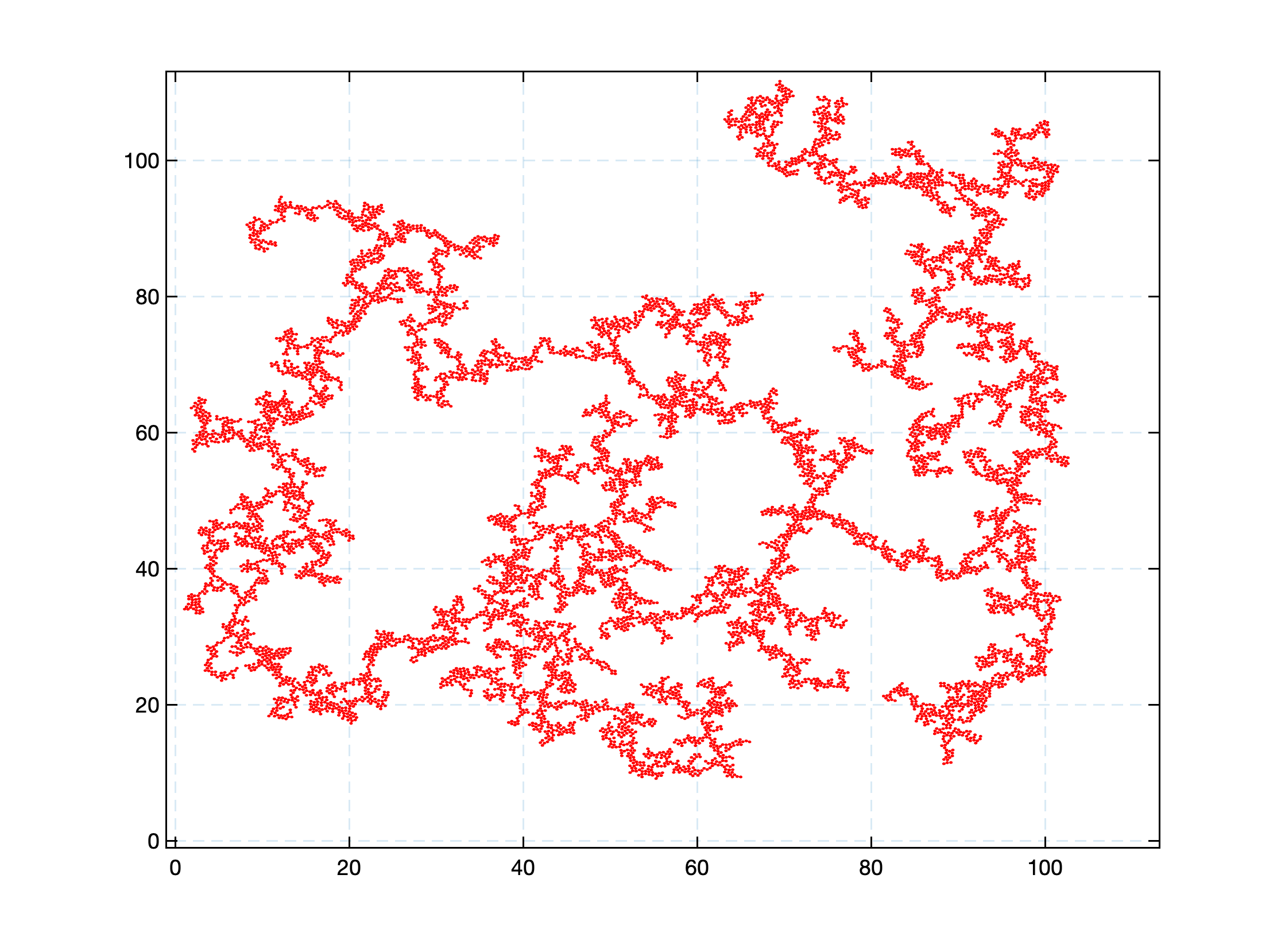}
	\includegraphics[scale=0.12]{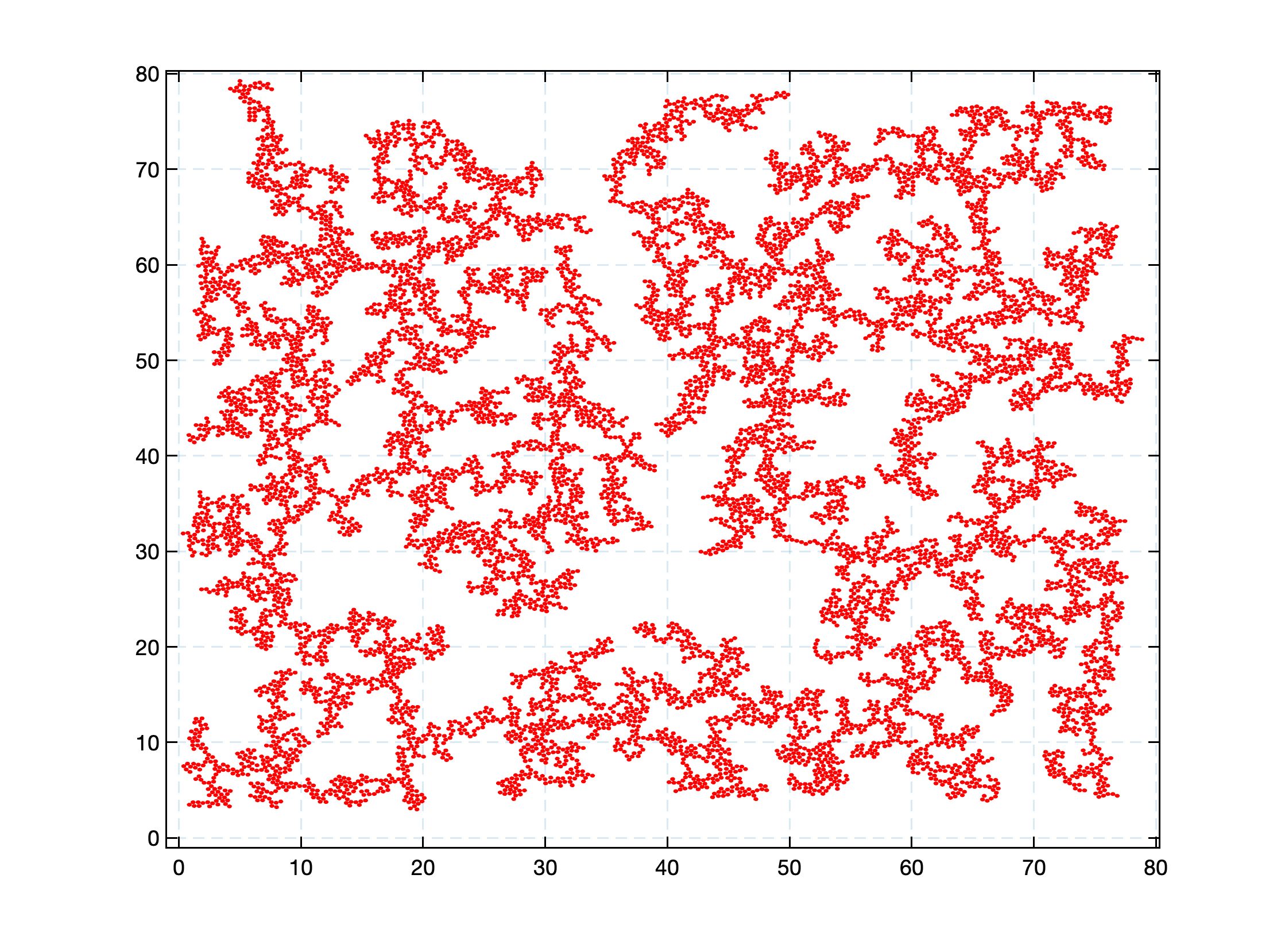}
	\includegraphics[scale=0.12]{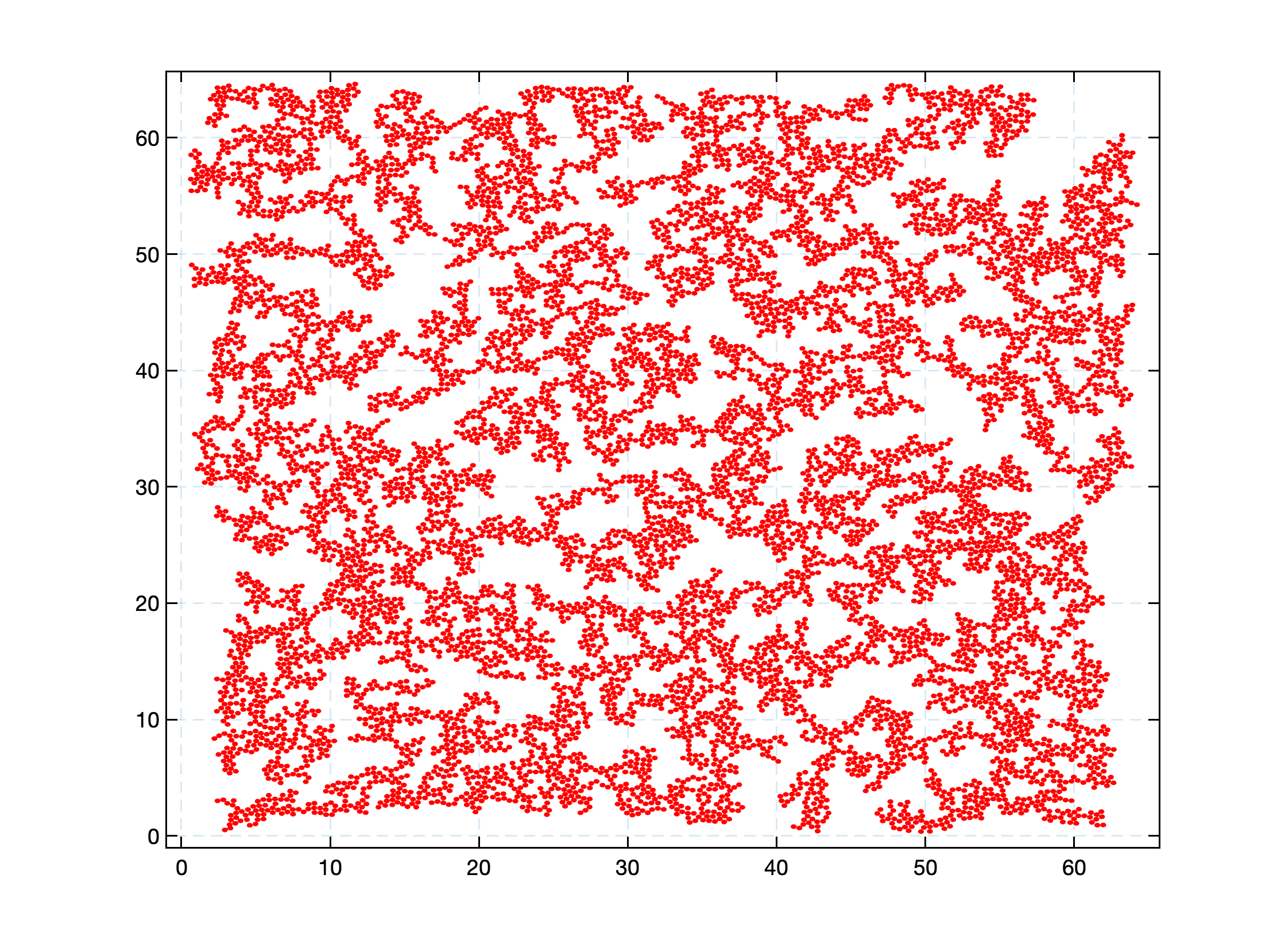}
	\includegraphics[scale=0.12]{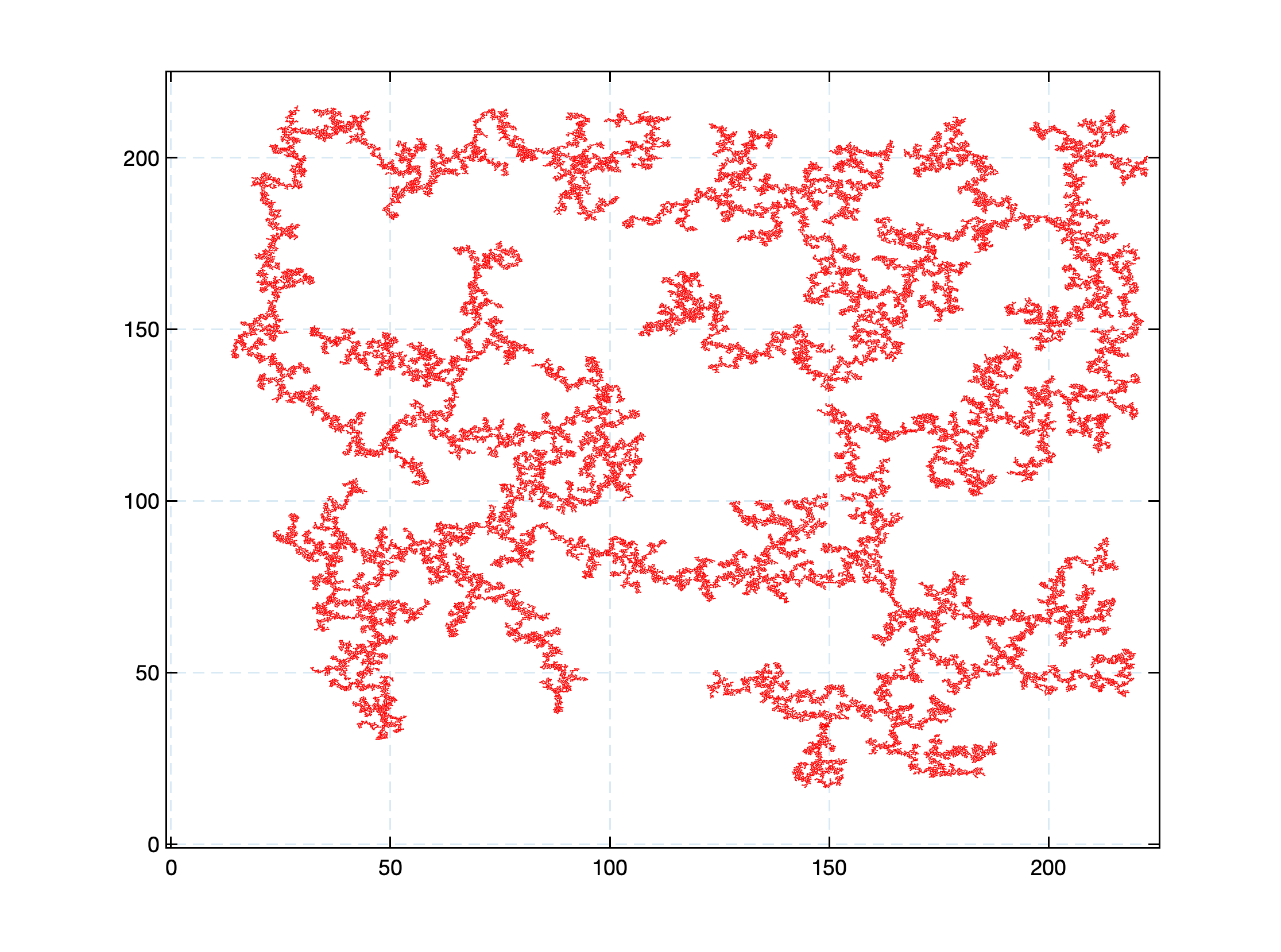}
	\includegraphics[scale=0.12]{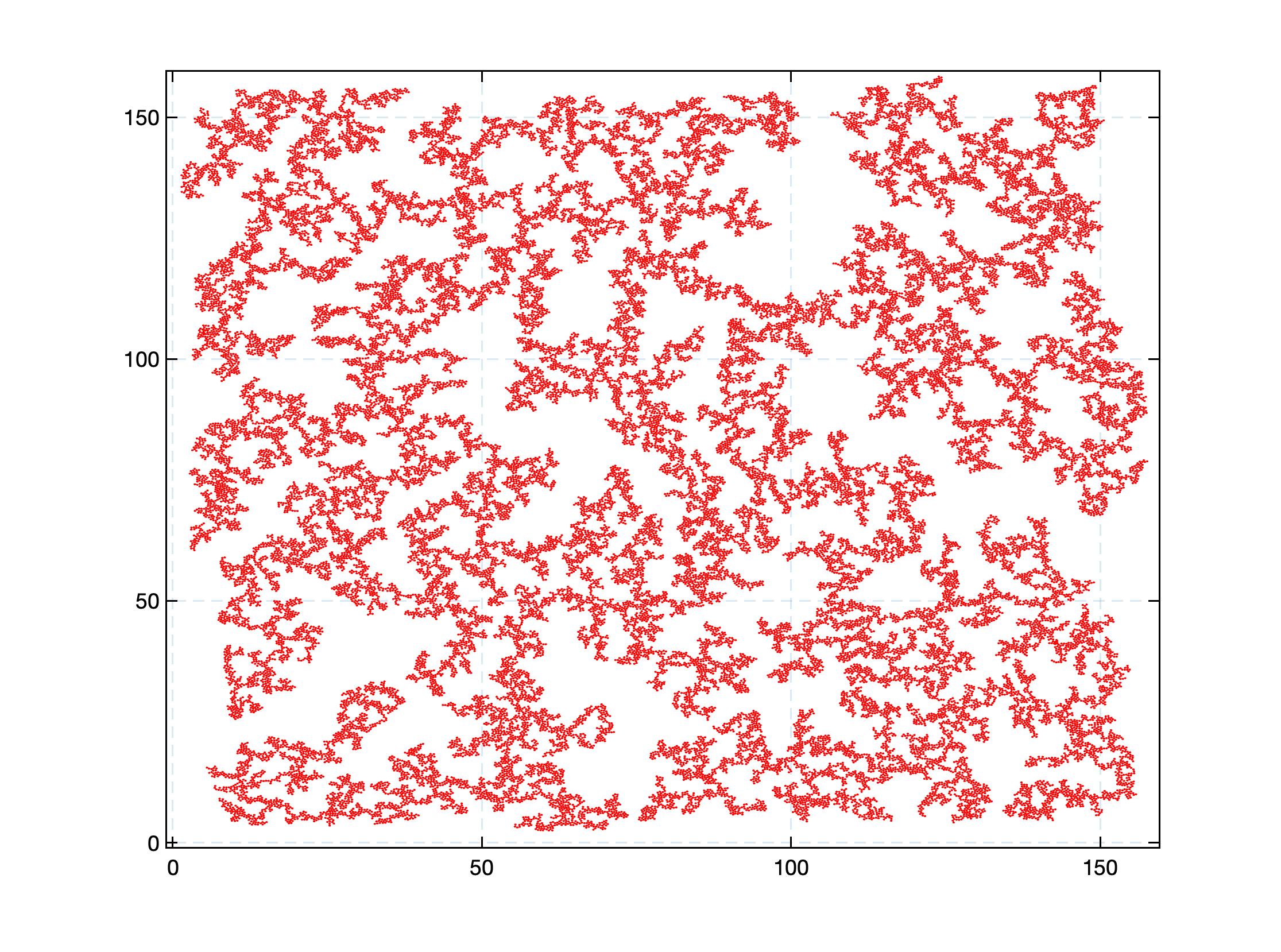}
	\includegraphics[scale=0.12]{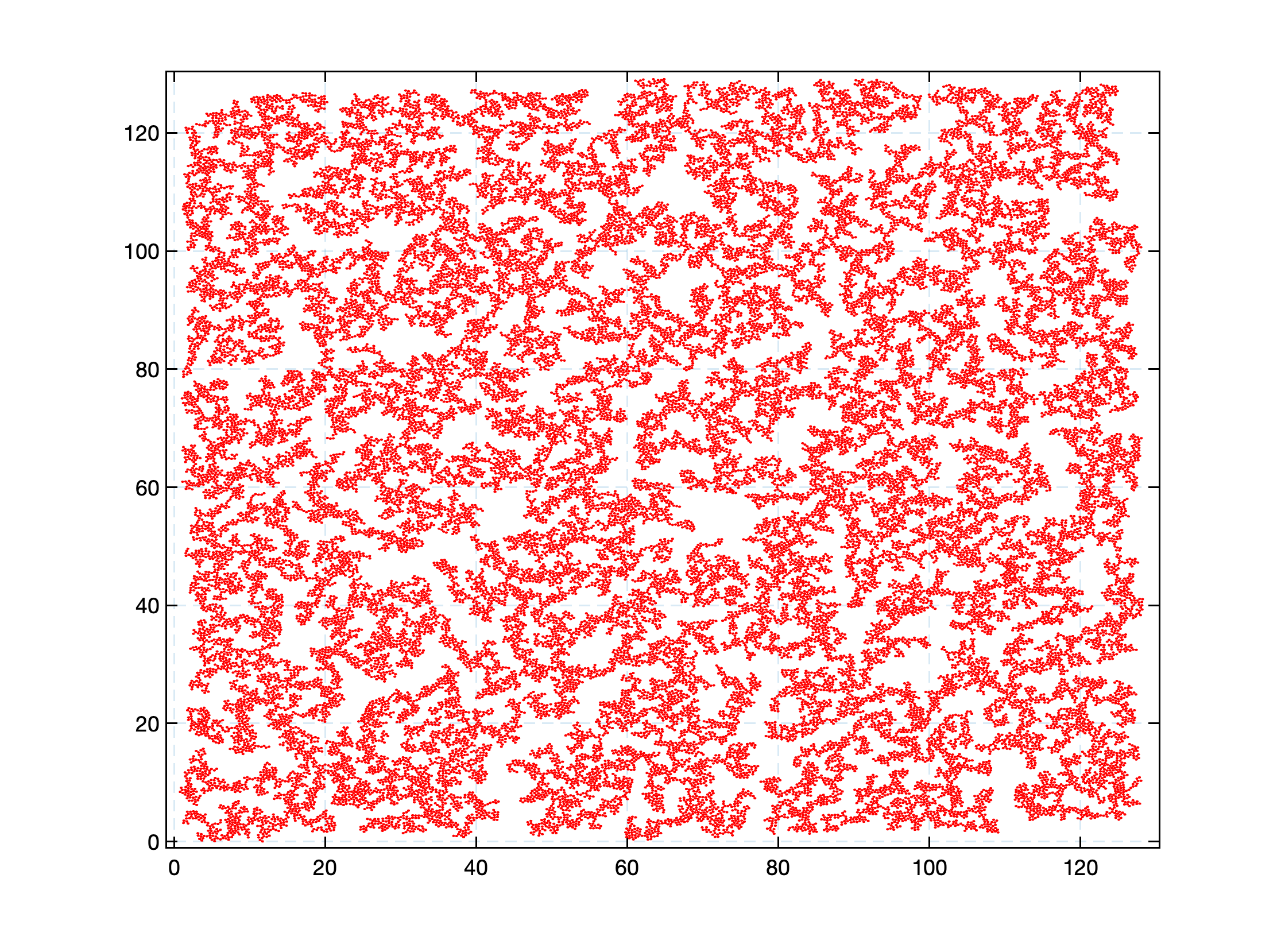}
	\caption{From top to bottom final configurations obtained with the TS scheme for $N=25000$, $N=10000$ and $N=40000$ particles respectively. The volume fraction is equal to $0.1$, $0.2$ and $0.3$ from left to right.}\label{example2}
\end{figure}

In Figures \ref{example1} and \ref{example2} we show some examples of the final configurations, for a given realization, obtained with the Time Stepping method in terms of the different particle numbers, different volume fractions and for a given fixed initial time step, namely $\Delta t (t=0)=0.01\, L/\max_i(|V_i|)$. In the first figure, from top to bottom, the number of particles ranges from $100$ to $900$ while in the second from $2500$ to $40000$ while the volume fractions increases from left to right. The final form of the cluster clearly depends on the box size: spheres are more folded when the volume fraction increases and more elongated when they have more space at their disposal to move.

\begin{figure}[h!]
	\centering
	\includegraphics[scale=0.36]{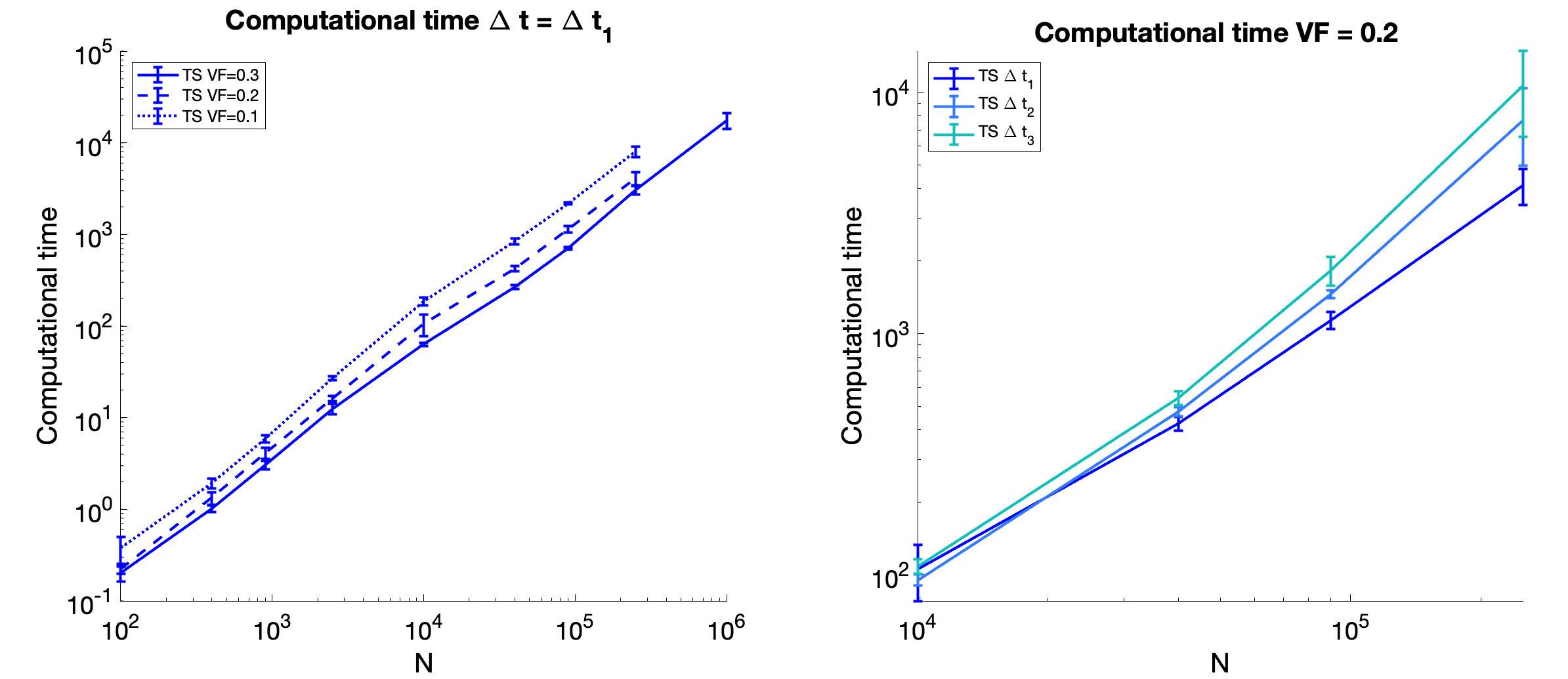}
	\caption{Computational effort for the Time Stepping method. Left: the volume fraction varies from 30\%, 20\% to 10\%. The time step is fixed to $\Delta t_1$. Right: the volume fraction is equal to 20\%. The time step varies from $\Delta t_1$ to $\Delta t_3$. The computational times are averaged over 10 simulations for a number of particles between 100 and $10^6$.  }
	\label{fig:comp_costs}
\end{figure}

In Figure \ref{fig:comp_costs}, we report the computational time over the number of particles $N$ in log-log scale  of the TS method for different values of the volume fraction on the left and for different values of the time step on the right, where the $10^6$ particle case is run only for the largest volume fraction as already stated. We see an approximately linear dependence indicating that the computational time grows polynomially. The degree of the polynomial can be estimated using linear regression giving $1.2241$, $1.2491$ and $1.2862$ for volume fraction of $10\%$, $20\%$ and $30\%$ respectively. This suggests a growth's law of about $N^{1.2}$, so very close to linear. 
We also see from the figure that the larger the volume fraction, the {lower} the computational time. This can be explained by the fact that in denser systems the rate of collisions is higher, and consequently more collisions are solved simultaneously, which improves the total computational time of the TS scheme as fewer minimization steps are required. {One of the bottlenecks of the TS method is determining the new set of linked pairs of spheres, $S^n$, after each free flight step. This determination is crucial for identifying collision events, altering the cluster shape, and initiating the minimization procedure. At low volume fractions, many free flights occur without collisions, increasing the total computational cost. Conversely, at higher volume fractions, the probability of collisions rises, reducing the number of iterations without changes in the cluster shape and leading to faster convergence toward the final solution.
}
Concerning the right image we can infer a different behavior. Namely, for a fixed volume fraction, the larger the time step is, the larger the computational time becomes. The estimated growth is polynomial with degree equal to $1.2491$, $1.3023$ and $1.3401$ for the times steps $\Delta t_1$, $\Delta t_2$ and $\Delta t_3$ respectively estimated again with a linear regression procedure. This is due to the fact that for larger time steps, the number of spheres overlapping as a result of the free transport in average grows and this causes the minimization problems to be solved with a larger set of spheres which in turn need more iterations to converge. These results are shown only for a volume fraction $V_f=0.2$, the results being similar for the other tested situations. % and reported in the Appendix \ref{app_quant}.
{Concluding, one can interpret the results in the following manner. In general the TS proposed works particularly well for large volume fractions in terms of computational effort, however for a {lower} volume fraction is better to use lower time steps. This is due to the computational cost required to determine to which cluster each particle belongs. This operation must be done at the end of each free transport step and it is very expensive.  Consequently, for low volume fractions, this operation must be done frequently and often unnecessarily, as particles do not frequently encounter each other in this setting.}

\subsection{Accuracy analysis of the Time Stepping method.}
\label{sec:numerical}
We introduce in this part some statistical quantifiers which will be used to measure the analogies and the differences between the shape of the clusters obtained with the TS and the ED method. We take the Event Driven as our benchmark since it corresponds to the direct simulation of the physical problem under investigation.

%\subsubsection{Quantifiers}\label{sec:quantifiers}
The shapes of the final aggregates are measured by five indicators defined in the following. Namely these correspond to: the aspect ratio, the orientation of the cluster, the average number of contacts per sphere, the pair distribution function and the fractal dimension. All results of the simulations are averaged over ten different realizations obtained by randomly changing the initial velocity of the particles through the values defined in equation \ref{inivel}.\\
The {\bf aspect ratio $A_r$} is defined as
\begin{equation}\label{aspect}
	A_r = \frac{\lambda_1}{\lambda_2}.
\end{equation}
where $\lambda_i, \ i=1,2$ are the eigenvalues of the gyration tensor of the cluster centred at the origin and $Y = X-\langle X \rangle$ are the coordinates of the particles once the center of gravity of the cluster is shifted to coincide with the origin of the reference plane. The gyration tensor is defined as 
$$G(Y) = \frac{1}{N} \sum_{i=1}^N \left[ \begin{matrix} y^2_i(1) & y_i(1)y_i(2) \\ y_i(1)y_i(2) & y^2_i(2) \end{matrix}\right].$$ 
This indicator gives information about the possible elongation of a cluster in a given direction.\\
The {\bf orientation $D$} of the cluster is defined as
\begin{equation}\label{orientation}
	D = \frac{v_1}{|v_1|},
\end{equation}
where we recall that $| \cdot |$ denotes the Euclidean norm, $v_i, \ i=1,2$ are the eigenvectors corresponding to the eigenvalues defined previously and $\lambda_1,v_1$ are the leading eigenvalue and eigenvector. This indicator gives information about the principal direction of elongation of the final cluster.\\ 
The {\bf relative average number of contacts per sphere $N_c$} is defined as
\begin{equation}\label{cont}
	N_c  = \frac{1}{N}\frac{1}{N_{max}}\sum_{i,j=1}^N \mathbbm{1}_{ \{ | X_i-X_j | \leq \alpha_R R\}}
\end{equation}
where $\mathbbm{1}_A$ is the indicator function of the set $A$ and $\alpha_R$ is a constant chosen equal to $2.1$, i.e. the indicator function is one when the centers of two spheres lie at a distance smaller or slightly larger than the diameter of a particle. We did not choose $\alpha_R$ to be exactly equal to $2$ since the minimization problem is an iterative method always affected by some numerical error. Thus, two particles with center $X_1$ and $X_2$ are considered to be in contact if $| X_1-X_2 | \leq \alpha_R R$. Finally the number $N_{max}$ is equal to the maximum number of particles which may be in contact with a given sphere which in the two dimensional case is equal to six. This indicator measures the relative average number of contacts per particle in the cluster and thus gives an indication about how concentrated or viceversa loose the cluster is. This number we expect to lie between $1/6$ and one.\\ %which corresponds to the maximum number of sphere which can be in contact with a given one.\\
The {\bf pair distribution function $P$} is defined as  
\begin{equation}\label{pair}
	P(r) = \frac{1}{N}\sum_{i,j=1}^N \mathbbm{1}_{ \{ | X_i-X_j | \leq r \}}
\end{equation}
where again $\mathbbm{1}_A$ is the indicator function. This function measures the average number of particles which are at a distance $r$ from a given particle. This describes the distribution of distances between pairs of particles contained in the box, we refer to \cite{JJE19, Jungblut2} for a discussion about this function which is typically defined in a probabilistic sense. In a nutshell, this indicator gives a measure of the local configuration of the cluster with emphasis on the level of the packing. Given a particle $X_i$, this number is computed by finding all the particles $X_j$ at a maximum distance $r$ from it. We choose {{$2 R <r< 20 R$ }} to make a clear distinction between the average number of contacts $N_c$ and this new measure of the shape. Instead, the upper bound is chosen in such a way that at maximum we consider pairs of particles whose distance is not more than {ten times the size of a particle } . \\
The {\bf fractal dimension $D_f$} is defined implicitly by means of the following relation \begin{equation}\label{frac}
	\mathcal{N}(\varepsilon)=a\varepsilon^{-D_f}
	%D_f=\frac{\log\left(\mathcal{N}(\varepsilon_m)/a\right)}{\log(\varepsilon_m)}
\end{equation}
where $\mathcal{N}(\varepsilon)$ is the number of mesh elements of a given Cartesian grid with unit length $\varepsilon$, constructed over the box, containing at least one center of mass of a sphere.
More precisely, we will take discrete values for $\varepsilon$, namely $\varepsilon_m= L/2^m$, $m=1,2,...$, and compute
$$\mathcal{N}(\varepsilon_m) = \sum_{k=1}^{2^{2m}}    \mathbbm{1}_{  \{ X_i:\ i=1,...,N \}\cap g_k  }$$
where $g_k$ is an element of the Cartesian grid $G_m = \{ g_k : k=1,..., 2^{2m}\}$ with side length $\varepsilon_m$. 
The parameter $a$ in \eqref{frac} is the so-called prefactor \cite{fractalcity} corresponding to the measure of the fractal object and it is linked to the base length of the fractal under consideration. 
The relation \eqref{frac} can be equivalently written as
\begin{equation*}
	\log(\mathcal{N}(\varepsilon))=\log(a) - D_f\log(\varepsilon)
\end{equation*}
and the parameters $a$ and $D_f$ can therefore be estimated by applying a linear regression.   %by means of the function {\it polyfit} in {\it matlab}.

The fractal dimension is a quantity which measures the self-similar properties of an object.
%In this case, we say that the cluster is self-similar.
% has a self-similar structure, i.e., the structure of the cluster does not change across different scales.
%This is indeed what we get in our case study, as we will see in the next Section.
% As it is defined,  the fractal dimension is a quantity which measures the self-similarity properties of an object.
To this respect, a given cluster is said to have a self-similar structure if the number of spheres $n(\ell)$ contained within a volume of side length $\ell$ scales according to a power law  $n(\ell)\sim \ell^{D_f}$. In this case, the fractal dimension is defined as the exponent of this law, $D_f$ \cite{MAN83}. 
In practice, the value $D_f$ can be estimated using the box counting method \cite{fractalcity,vic} which is based on the the relation \eqref{frac} described above. In \eqref{frac}, the variable $\ell$ is substituted by the length of the grid $\varepsilon$ which is then varied to identify possible self-similar properties. %Thus the relation which holds in this case is $\mathcal{N}(\varepsilon_m)=a\varepsilon^{-D_f}$ where the introduction of the constant $a$ accounts for the fact that a minimal length in the system, namely, the particle radius is present. 
%In practice, in our analysis the value $D_f$ is estimated from the data using a linear regression by means of the function {\it polyfit} in {\it matlab}.

\subsubsection{Shape analysis}\label{shape}
We present the results for the average number of contacts $N_c$ \eqref{cont}, the pair distribution function $P$ \eqref{pair} and the fractal dimension $D_f$ \eqref{frac}. The results for the main orientation of the cluster and the aspect ratio will be discussed in Section \ref{app_quant} of the supplementary material, since our analysis revealed that they furnish less explicit indications about the possible differences or analogies between TS and ED.

\begin{figure}[h!]
	\centering
	\includegraphics[scale=0.165]{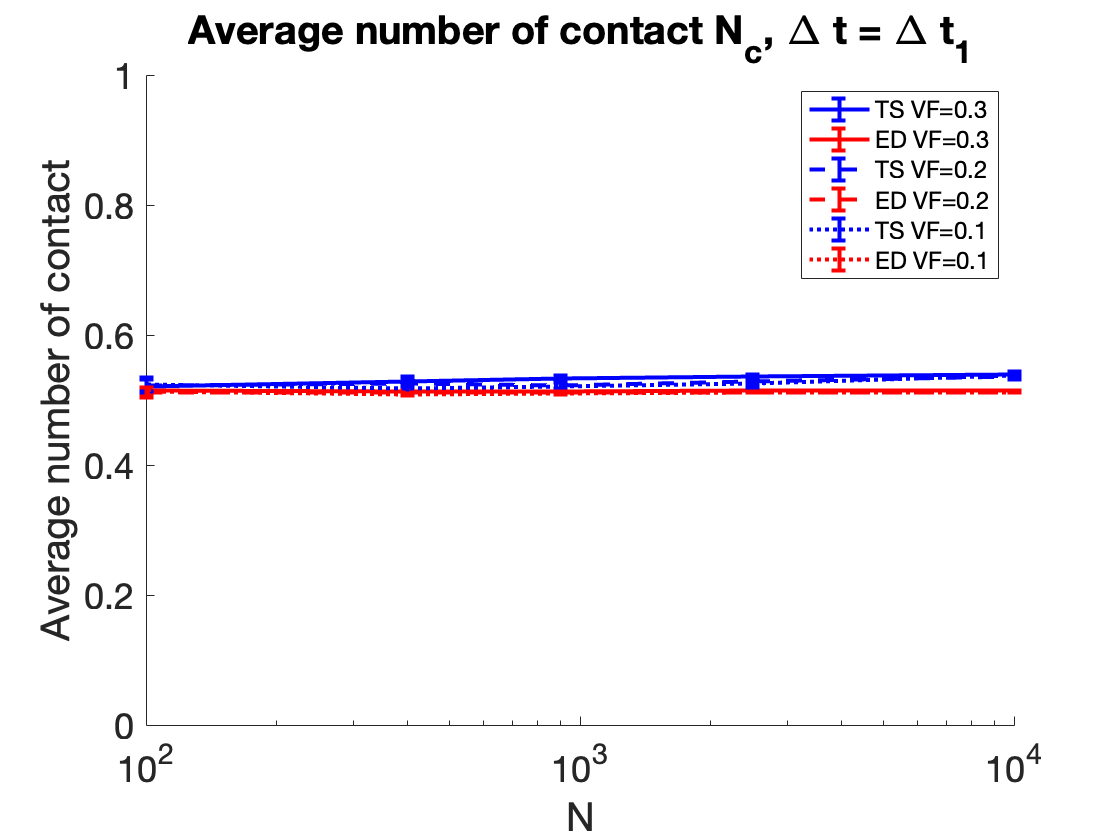}
	\includegraphics[scale=0.165]{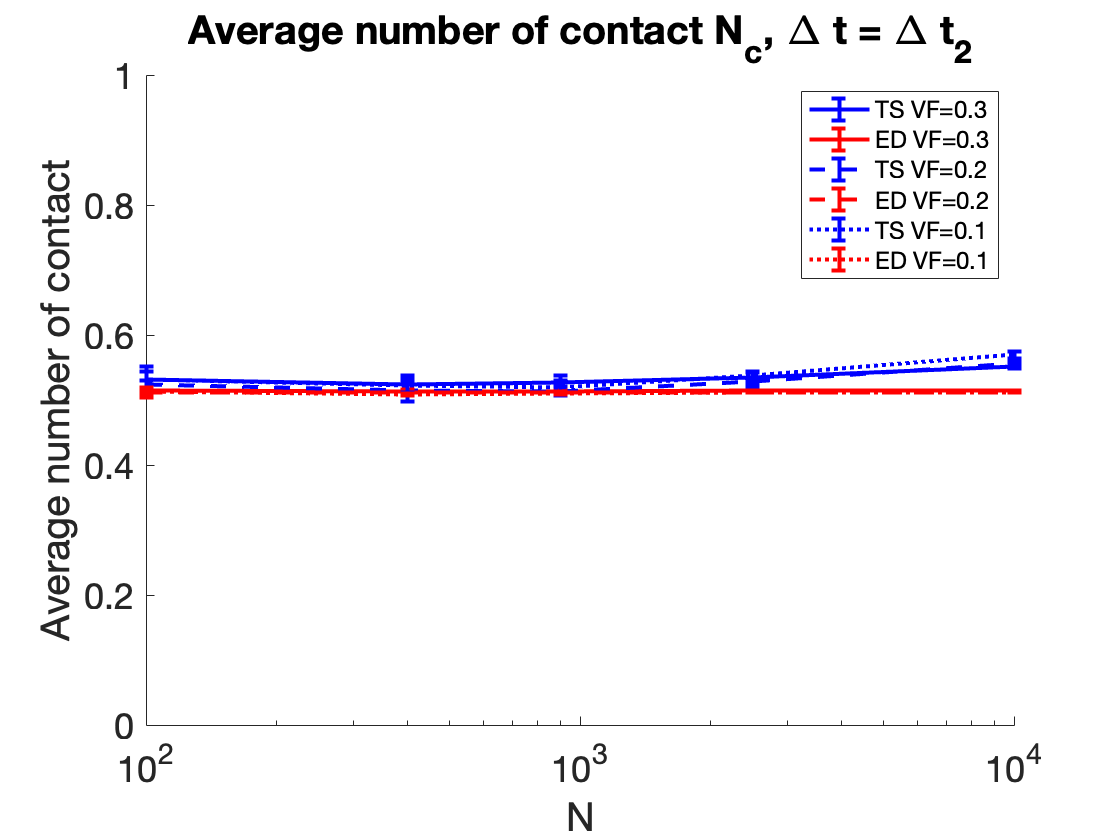}
	\includegraphics[scale=0.165]{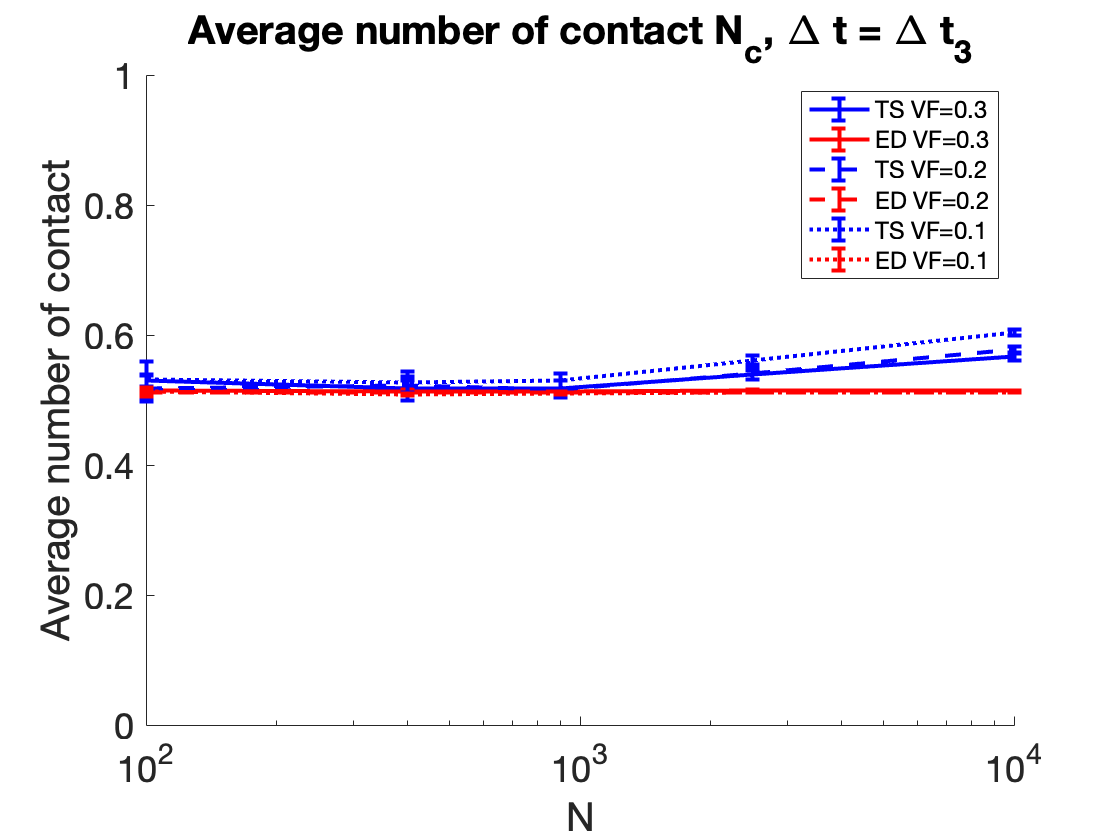}
	\caption{Relative average number of contact for the Time Stepping and the Event Driven method. The time step for the TS method is the first one $\Delta t_1$ from the Table \ref{tab:param}.}
	\label{fig:comp_Cont}
\end{figure}

Figure \ref{fig:comp_Cont} shows the relative average number of contacts per particle respectively for the TS and the ED method. The maximum value for this parameter, corresponding to a full packed situation, is one which corresponds to the case in which one particle is in contact and surrounded by six other spheres. For both methods TS and ED, we see that this number lies around the value $0.5$ for cluster sizes ranging from 100 to 10000, particles meaning that each particle is in contact in average with other three. The agreement between the two methods is very good, showing that the level of packing of the TS is comparable with the one furnished by the ED method. More in detail, the dependence of $N_c$ from the volume fraction is not very significant while for $N$ larger than $2500$ the influence of the time step $\Delta t$ of the TS method is clearly observable. The source of this discrepancy is due to the fact that in average, the larger is $\Delta t$, the larger is the number of spheres which overlap as a consequence of the free transport step. Then, the minimization procedure acts in such a way that between those particles attraction forces are activated and work to keep them in contact, increasing the value of $N_c$. By reducing $\Delta t$,  a reduction of $N_c$ is observed (top left image in \ref{fig:comp_Cont} compared to the bottom image), however the counterpart being the increase in the computational cost of the scheme caused by too small time steps. Thus, we can conclude that with reference to this parameter, it is possible to balance accuracy and efficiency of the TS method by choosing the time step appropriately.

\begin{figure}[h!]
	\centering
	\includegraphics[scale=0.36]{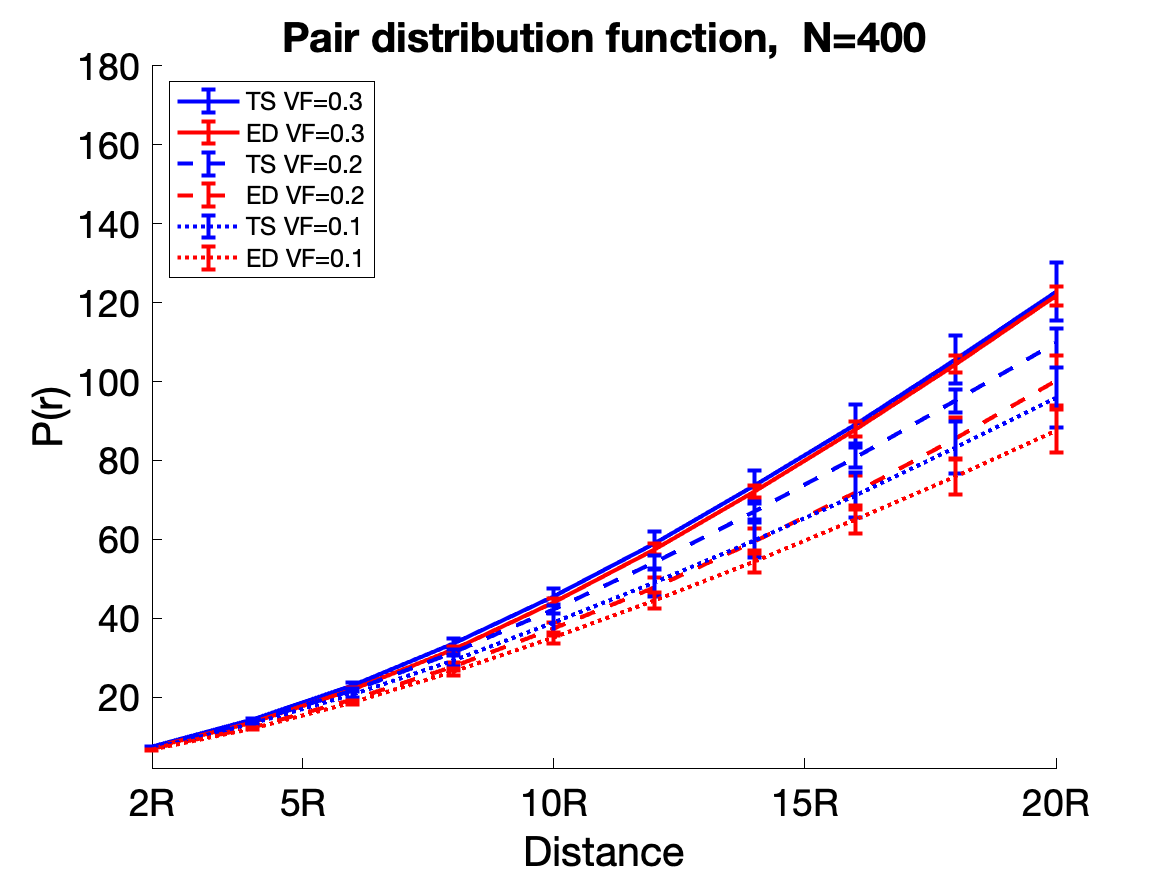}
	\includegraphics[scale=0.36]{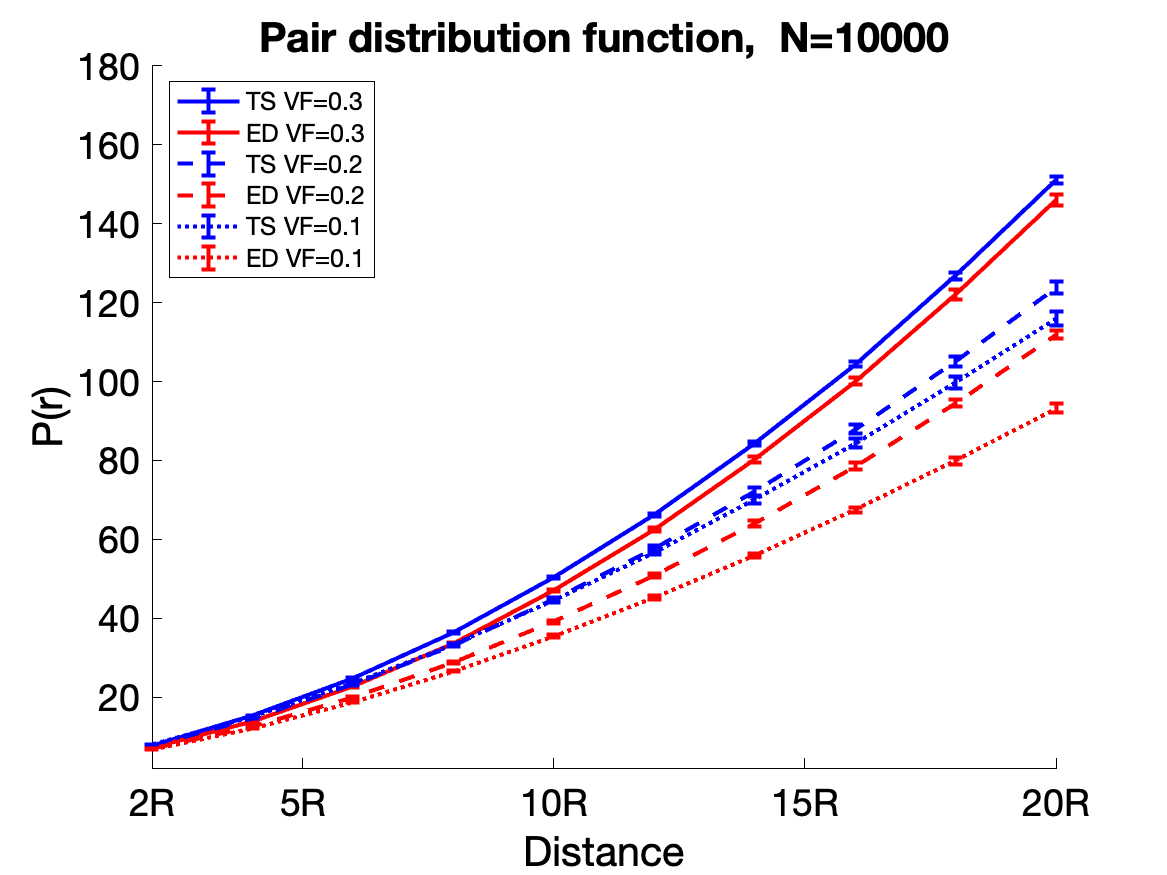}
	\caption{{Pair distribution function} for the Time Stepping and the Event Driven method. The time step for the TS method is the first one $\Delta t_1$ from the Table \ref{tab:param}.}
	\label{fig:comp_Corr}
\end{figure}

The pair distribution function is instead shown in Figure \ref{fig:comp_Corr} as a function of the distance between two pairs measured in terms of the radius of a sphere. On the left, the case with $N=400$ is shown while on the right it is shown the case $N=10000$. The other cases are shown in \ref{app_quant} (supplementary material) and they give similar results. The TS produces slightly more dense clusters than ED. There is however an excellent agreement between the two methods especially for large volume fractions. We deduce that the TS method is more accurate for denser systems. We also observe that the variability regarding the initial conditions decreases with the number of particles. This means that as the number of particles grows the shapes of the cluster tend to a fixed local configuration.

\begin{figure}[h!]
	\centering
	\includegraphics[scale=0.26]{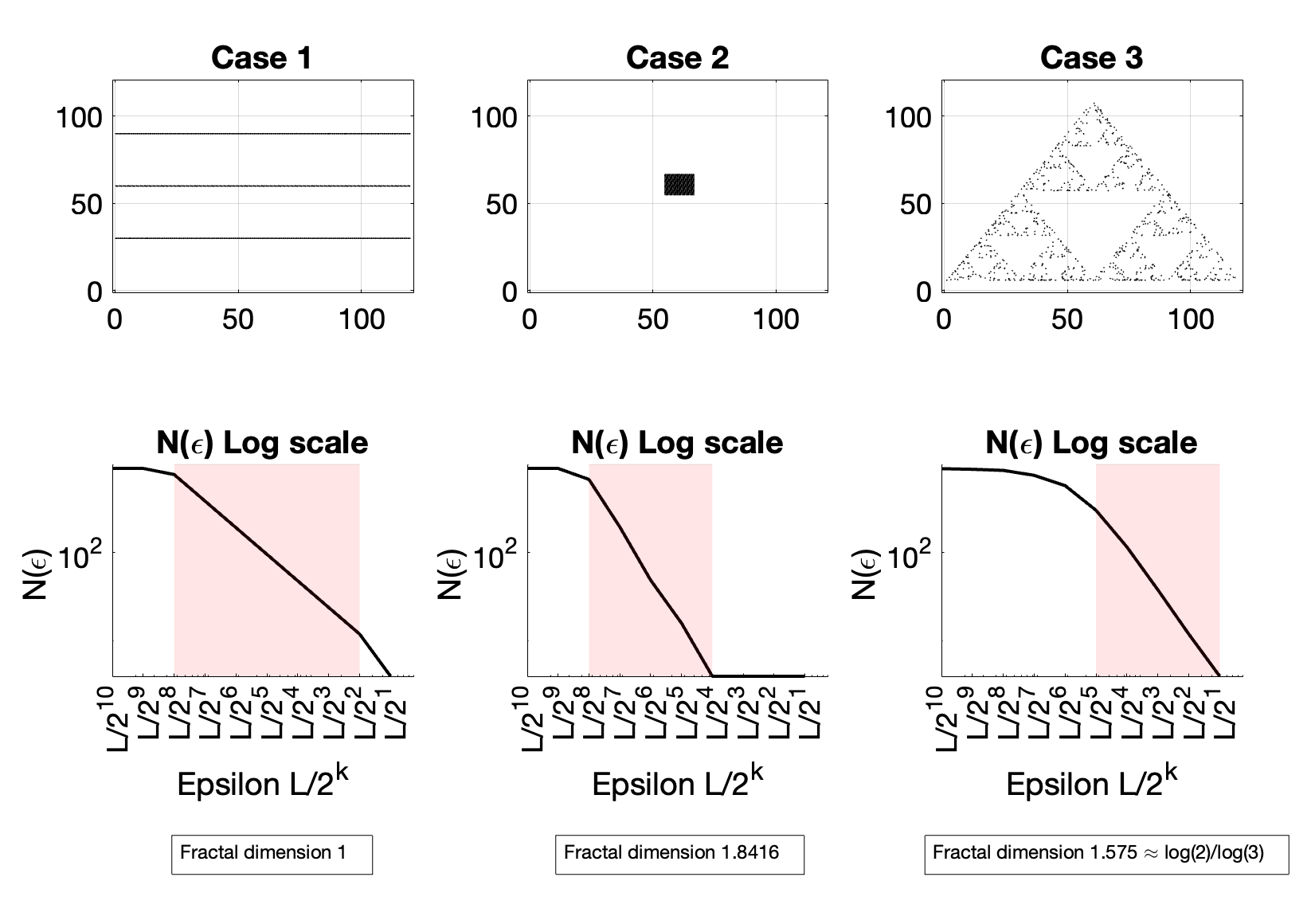}
	\caption{Fractal behavior for different configurations of particles with $N=900$ and box size $L=120$. Top images: from left to right particles are disposed into three horizontal lines, on a square lattice at zero interparticle distance and following the Sierpinski triangle law. Bottom images $\mathcal{N}(\varepsilon)$ as a function of $\varepsilon = L/2^k$ with $k\in [1,10]$ in log-log scale and corresponding estimation of the fractal dimension. The fractal dimension corresponds to the slope of linear fit.}
	\label{fig:comp_FractalDim}
\end{figure}

We finally discuss the results for what concerns the fractal dimension. Before analyzing with respect to this last parameter the behavior of the TS and the ED methods in the simulation of the aggregation phenomenon, we have measured the value of the fractal dimension $D_f$ for given different configurations of particles. About that study, in Figure \ref{fig:comp_FractalDim} on the top, three different configurations of particles are shown: in the first image, on the left, particles are disposed onto three horizontal lines, in the second case the same number of particles is disposed on a square lattice at zero interparticle distance, while the third case corresponds to the Sierpinski triangle, i.e. a fractal of equilateral triangle shape which is then subdivided recursively into smaller equilateral triangles \cite{sier}. The bottom images show the number of mesh elements $\mathcal{N}(\varepsilon)$ containing at least one center of mass as a function of the size of the mesh element in log-log scale and referred to the above described situations. The corresponding extrapolation of the fractal dimension for these configurations is shown as well. We observe for the two first cases two plateaus respectively for small and large mesh sizes. This can be explained by the fact that when the mesh size becomes sufficiently large all particles lie in the same cell giving back a constant and maximum value for $\mathcal{N}(\varepsilon)$. On the other hand, when the mesh size is too small the box counting method finds only a single particle in each mesh element giving back again a constant $\mathcal{N}(\varepsilon)=1$ value. For the Sierpinski triangle the second plateau on the right hand side is not present since the only case in which all particles lie in the same mesh element is when the mesh size equals the length of the domain and in our graph the last mesh size value corresponds to $L/2$. The central parts of the bottom figures correspond instead to the parts where the fractal behavior can be observed and measured. Thus, depending on the slopes of the lines connecting the two plateaus for each configuration, we can infer to have similar or different fractal structures for the different aggregates. This corresponds indeed to the quantity $D_f$ in the definition \eqref{frac}. 

\begin{figure}[h!]
	\centering
	\includegraphics[scale=0.33]{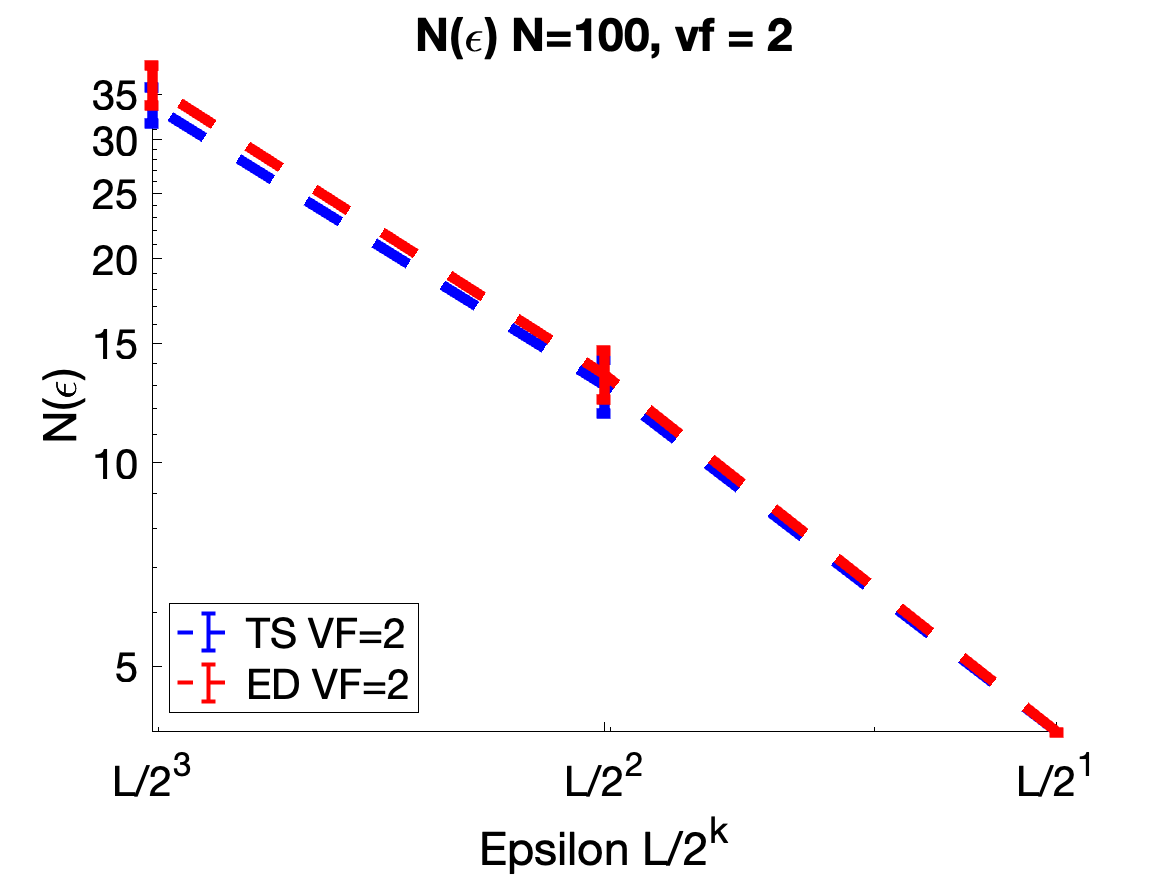}
	\includegraphics[scale=0.33]{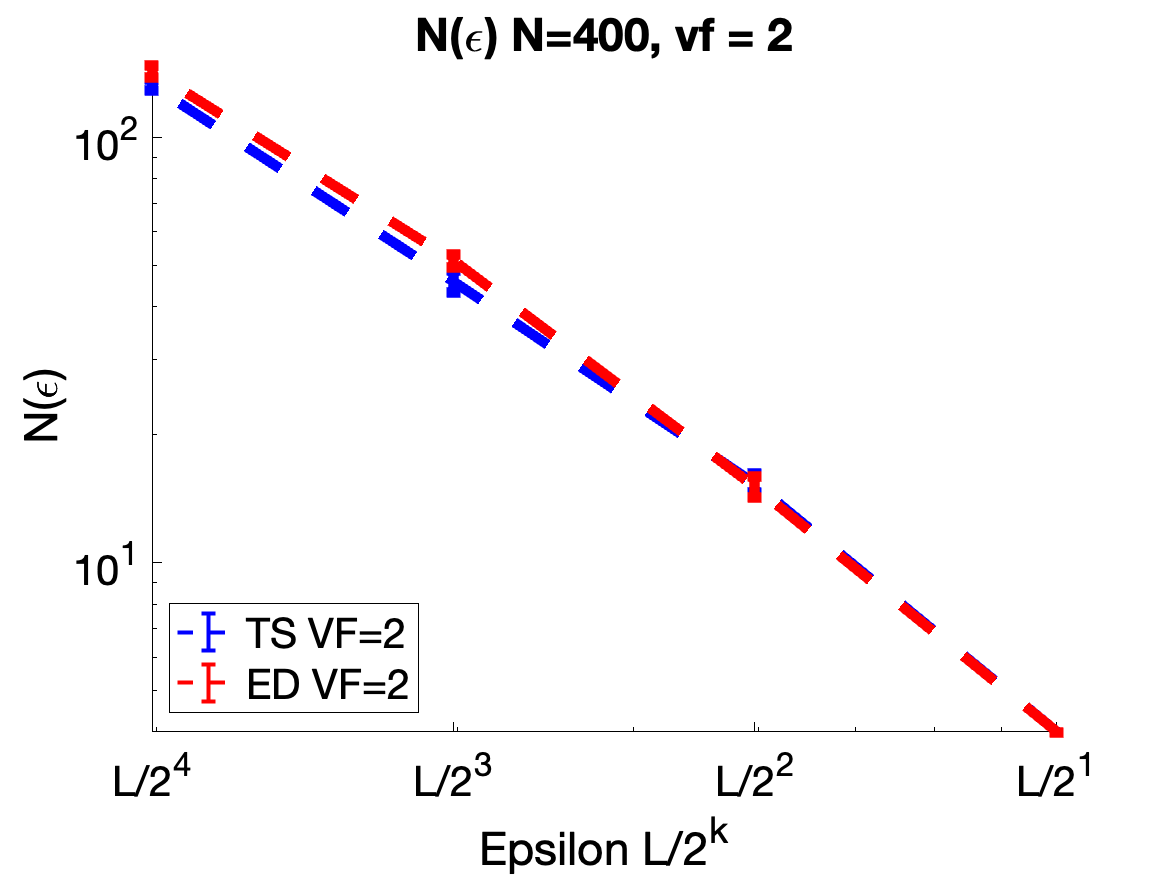}
	\includegraphics[scale=0.33]{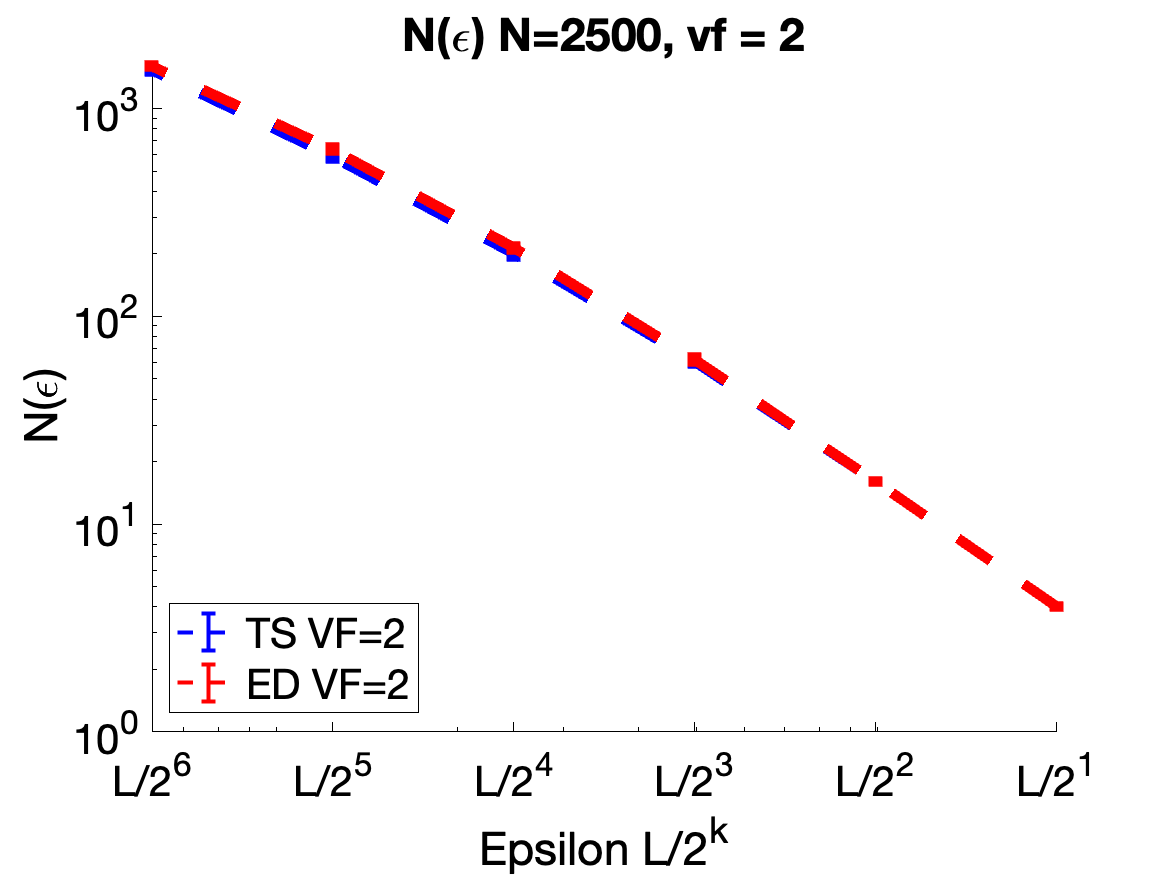}
	\includegraphics[scale=0.33]{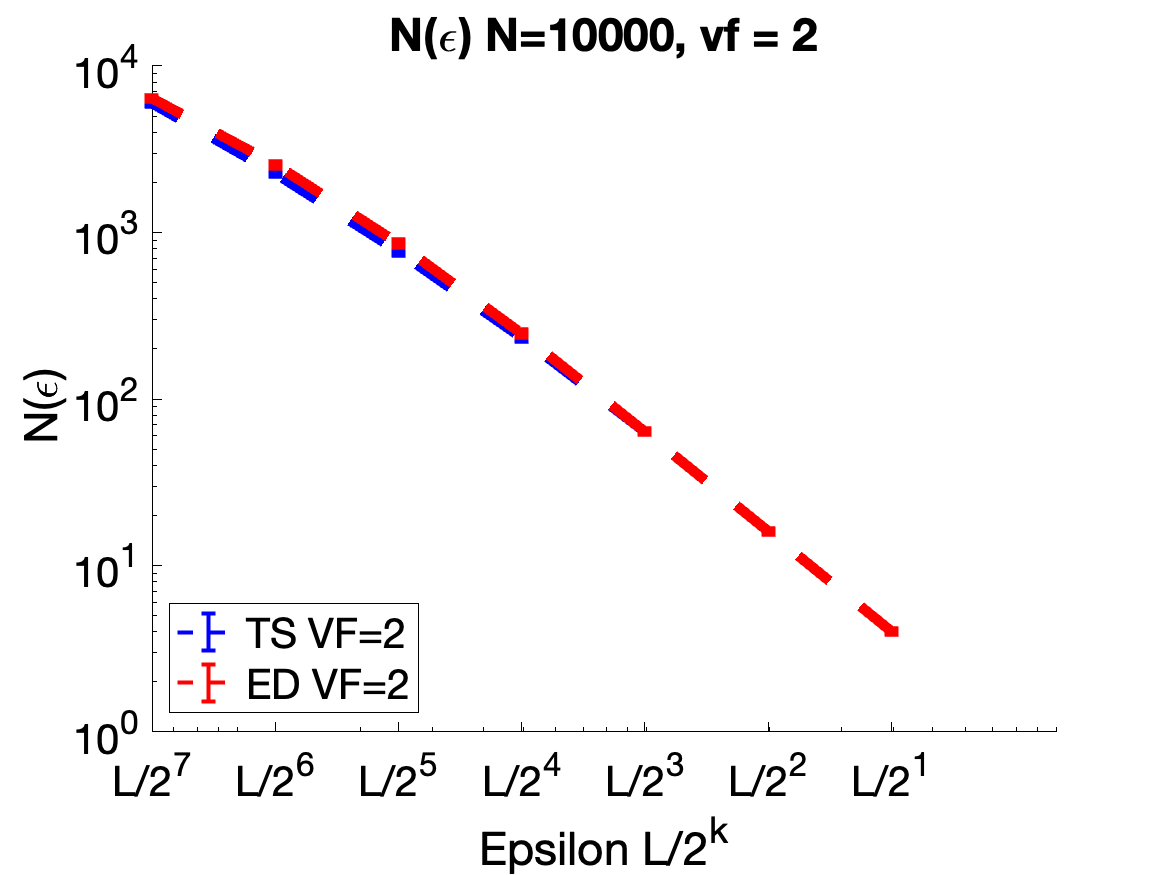}	
	\caption{$\mathcal{N}(\varepsilon)$ over $\varepsilon$ in log log scale for the Time Stepping and the Event Driven method. Volume fraction $V_f$ is fixed to $0.2$, from top left to bottom right the number of particles grows from $N=100$ to $N=10000$.}
	\label{fig:comp_Neps}
\end{figure}

\begin{figure}[h!]
	\centering
	\includegraphics[scale=0.38]{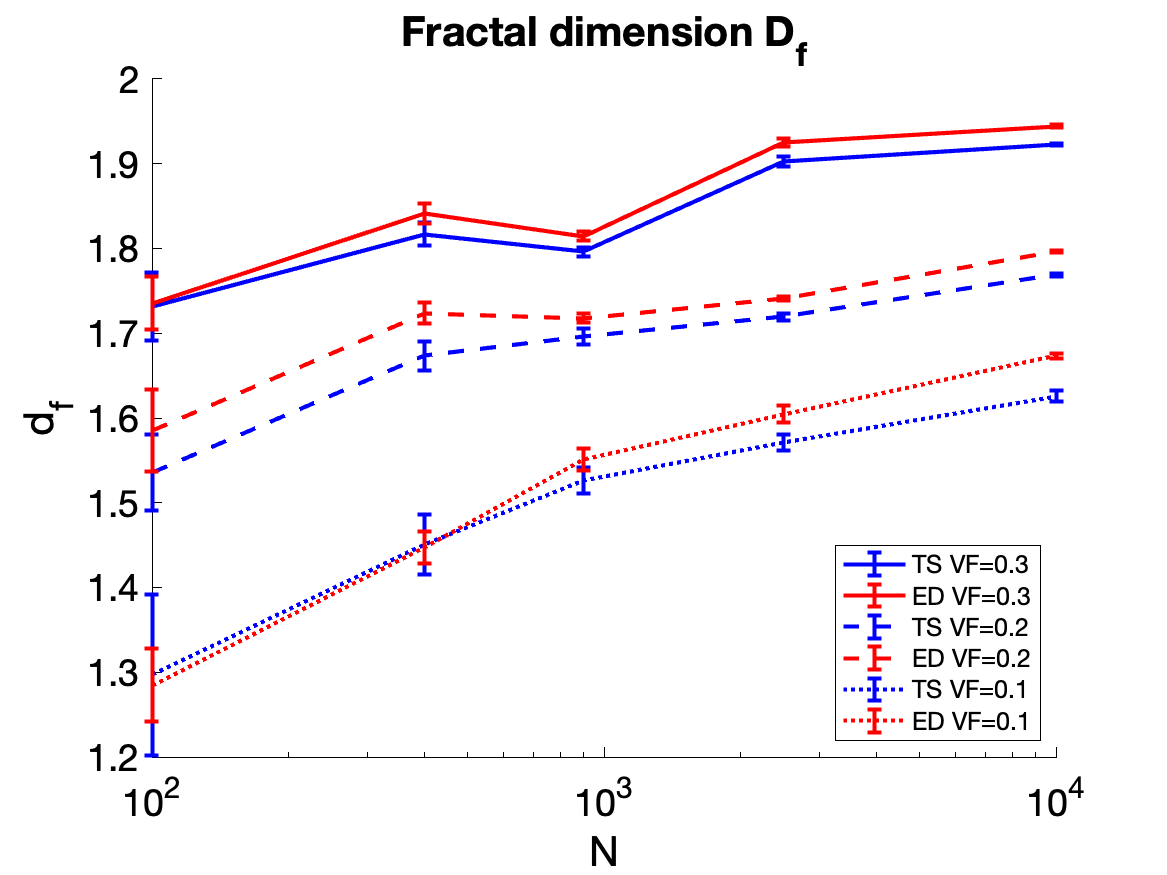}
	\includegraphics[scale=0.38]{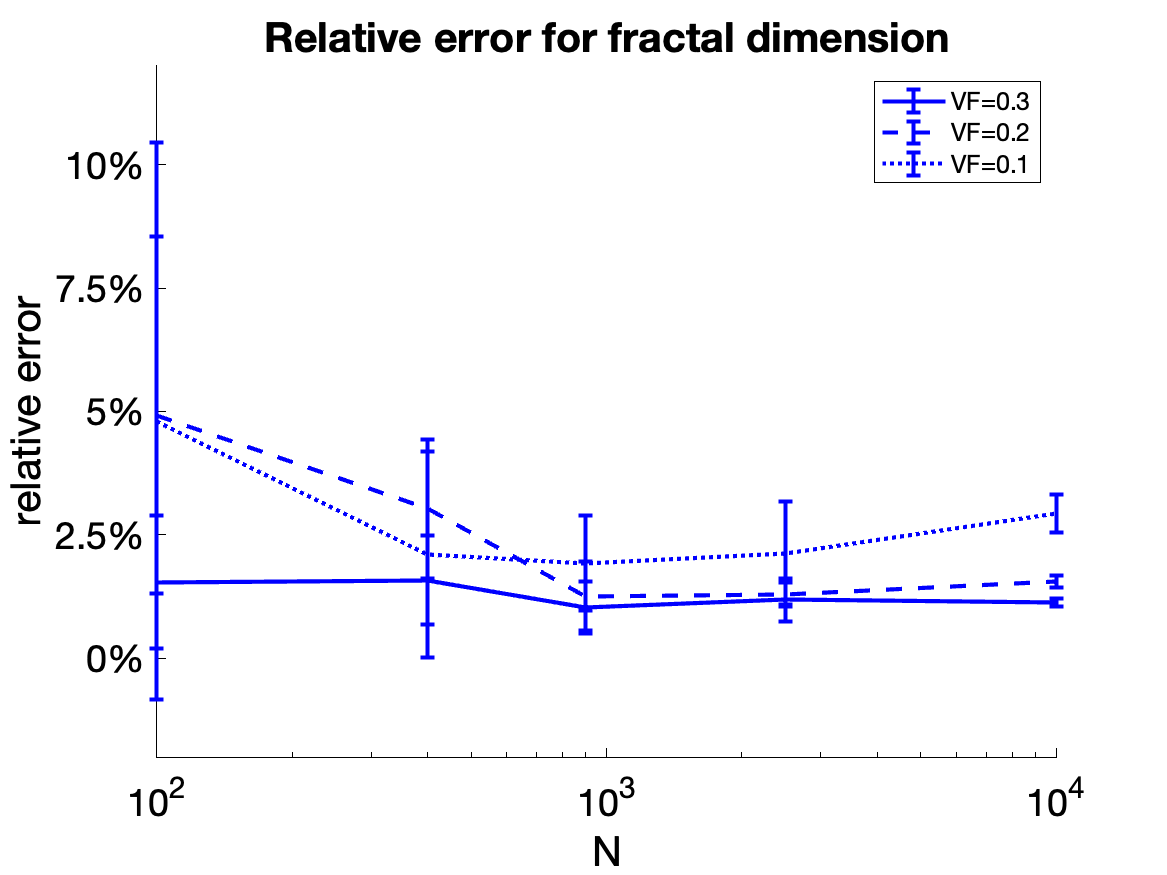}
	\caption{Left image: fractal dimension for the Time Stepping and the Event Driven method. Right image: relative error between TS and ED.}
	\label{fig:comp_FractalDim_err}
\end{figure}

In Figure \ref{fig:comp_Neps}, the number of mesh elements $\mathcal{N}(\varepsilon)$ is shown for the aggregation phenomenon for respectively the TS and the ED methods for different values of $N$ {and $\Delta t_1$}. For all tested cases, we observe the same qualitative behavior of the previous situation shown in Figure \ref{fig:comp_FractalDim}. We then have two plateaus one for very small mesh sizes, in this case all sphere centers lie in different mesh elements, and a second plateau when the mesh element is of size larger enough such that all particles are contained in one single element. %In this case, we have that in average all particles fit in one single cell.
We only show in these images the values of the mesh size for which $D_f$ is monotone decreasing, i.e. we only consider average values for $L/2^k$ cutting the two plateaus corresponding to the extreme situations, { namely, $3R \leq \varepsilon \leq \frac{L}{2}$.} 
The number of particles constituting the cluster is $N=100$ at the top left, $N=400$ top right, $N=2500$ bottom left and $N=10000$ bottom right. The volume fraction is fixed to $V_f=0.2$. The same results for different volume fractions are shown in the supplementary material \ref{app_quant}.  This figure shows a very good agreement for what concerns the two methods as the two corresponding graphs are almost superimposed. In order to get more insight on this fractal measure, in Figure \ref{fig:comp_FractalDim_err} the slopes $D_f$ are extrapolated and depicted in function of the number of particles $N$ on the left, while on the right the relative error $e=\frac{|D_{f,TS}-D_{f,ED}|}{D_{f,ED}}$ is shown. The largest error is obtained for $N=100$ and is about $5\%$ in the case of rarefied aggregates, i.e. $V_f=0.1$, the average error is less than $3\%$.

\section{Conclusions}
In this work, we have modeled large particle systems undergoing ballistic aggregation through a two step approach. Our method belongs to the class of the so-called time stepping algorithms in which the particles and clusters move freely for a fixed and sufficiently small time steps independently of the movement of the rest of the system as if they were alone in space. Then, the method proceeds handling multiple collisions, which result from  the regime of free motion, by means of a minimization approach. The idea is to replace the exact but too expensive dynamics of single pair of collisions with the search of an equilibrium configuration in which cohesion forces under non overlapping constraints coexist. 

The first goal was to show that this alternative way to model aggregation dynamics is competitive and may outperform the standard methods based on the exact simulation of the ballistic phenomenon. The second goal was to show that this method holds the capacity to describe the problem under consideration at least in a statistical framework. Our results have shown through a detailed analysis of the computational performances and of the shapes of the final aggregates that both goals can be achieved by the proposed method. { In particular, the computational time of the approach here developed grows polynomially with respect to the number of particles with polynomial degree between $1.2$ and $1.3$ depending on the volume fraction size while very good agreements between the clusters produced by a direct simulation technique, namely the Event Driven method, and our approach have been found concerning different statistical indicators.}
%In particular, the computational cost and the memory consumption of standard Event Driven methods are overcame by our method as soon as the number of particles is larger than $100$. In a nutshell, results show that, while a single simulation of the standard Event Driven method for $10000$ particles needs around one week on a sequential machine to be ran, our Time Stepping method needs only around one day to simulate the aggregation of one million particles { in the case of dense systems}.

We finally stress that the results here obtained are preliminary, for spherical particles and in a two dimensional setting. In the future, we would like to extend our approach to three dimensional cases considering more comprehensive physical models. We would also like to explore different minimization strategies based on stochastic paradigms which will permit to improve the computational performances of the method proposed. 

\appendix
\section{The event-driven method for particle cluster formation}
\label{sec:ED}
We discuss the details of a standard Event Driven method, the ones used in the simulations, and we derive the exact formulas used to predict the time of the next collision, see for instance \cite{Donev} for more detailed explanations. Let $X_i'$ and $X_j'$, be the positions of two particles belonging to two different clusters $C_k$ and $C_{\ell}$ at the time  $t+\Delta t_{ij}$ of contact. At this time the distance between these two spheres is exactly
\begin{equation}
	\label{eq:dist}
	|X_i'-X_j'|=R_i+R_j.
\end{equation}
On the other hand, at time $t$, the positions, $X_i$ and $X_j$, and the velocities, $V_i$ and $V_j$ , of the two spheres satisfy 
\begin{eqnarray}
	X'_{i}=X_{i}+\Delta t_{ij} V_{i} \qquad \text{ and } \qquad  X'_{j}=X_{j}+\Delta t_{ij} V_{j}.\label{eq:position_colliding_spheres}
\end{eqnarray}
Substituting now the $2d$ equations~\eqref{eq:position_colliding_spheres} into equation \eqref{eq:dist} gives the following expression for the collision time
\begin{equation}\label{eq:time_col1}
	\Delta t_{ij}=\left \{
	\begin{tabular}{cc}
		$\infty$ & \textit{if} \ $\Delta V\Delta X\geq 0$ \textit{ or } $\sigma<0$ \\
		$-\frac{\Delta V\Delta X+\sqrt{\sigma}}{\Delta V \Delta V}$ &   \textit{otherwise},
	\end{tabular}
	\right.
\end{equation}
with $\sigma=(\Delta V \Delta X)^2-(\Delta V \Delta V)(\Delta X \Delta X-(R_i+R_j)^2)$, $\Delta X=X_i-X_j$ and $\Delta V=V_i-V_j$.  Thus, using formula (\ref{eq:time_col1}), one computes the time interval $\Delta t_{ij}$ for each pair $(i,j)$ of spheres that do not belong to the same cluster, i.e. $i,j=1,..N, \ i\in C_{k}, \ j\in C_{\ell}, \ k\ne \ell$. Then, in order to predict the time of the next collision event, it is sufficient to find the minimum between these time increments at time $t$:
\begin{equation}
	\Delta t=\min\limits_{i,j=1,..N, \ i\in C_{k}, \ j\in C_{\ell}, \ k\ne \ell}\ \Delta t_{ij}.\nonumber
\end{equation}
The cost of such approach is clearly proportional to the number of possible events that can take place in the system at a given instant of time. Thus, at $t=0$, when all clusters are of size one we have possibly $(N(N-1)/2)$ collision pairs. The computational cost is then of order $\mathcal{O}(N^2)$. Successively, this cost decreases monotonically with time due to the formation of the clusters and thus to the reduction of the possible pairs of spheres that can interact. 
%Let us observe that the exact evaluation of the cost as a function of the evolution of the system is not possible since this depends on the size of merging clusters and thus it strongly depends on the initial data. It is however possible to estimate at least the number of possible collision pairs between object regardless of the shape of the clusters. In fact, after every collision the number of pairs diminishes by one and we have $$(N-N_c(t))(N-N_c(t)-1)/2$$ 
%collision pairs with $N_c(t+\Delta t)=N_c(t)+1$, $N_c(0)=0$ and $N-1$ total collisions before one single cluster is created. 
However, the computational time of the exact detection of the possibility of one event depends on the shape of the aggregate and thus it is a function of the type of dynamics originated from the initial data. 

In order to speed up the Event Driven approach, a standard improvement consists of subdividing the domain into squared boxes and computing the collision times between the pairs that lie in the same box or in neighboring boxes only. This allows to reduce the number of operations from order $ \mathcal{O}(N^2)$ to order $\mathcal{O}(N\log(N))$ in standard collision algorithms. Let us observe that, however, in order to maintain efficiency in grid methods, one is obliged to fix the time step $\Delta t$ in such a way that one sphere does not travel more than one cell during its ballistic shift. Thus, in practice, the time step is fixed by
\begin{equation}
	\Delta t=\min_{(i,j) \in 1,..,N, \ i\in \mC_k \ j\in \mC_\ell,\ k\ne\ell} (\tilde{\Delta t}_i, \Delta t_{ij}),\label{eq:time_col3}
\end{equation} 
where $\Delta t_{ij} $ is given in~\eqref{eq:time_col1} and $\tilde{\Delta t}$ is the time needed for a particle to reach one boundary
\begin{equation}
	\tilde{\Delta t}_i =   \frac{d_{ib}}{|V_i|}, 
	\label{eq:time_col2}
\end{equation}
with $ |V_i|=\sqrt{V_{i,x}^2+V_{i,y}^2}$ and $d_{ib}$ the distance from the box.
Here, we do not employ the above strategy for the ED method. Thus, we can state that the number of possible events at a certain time remains a quantity { proportional to the square of the number of clusters existing in the system at that time.}  %However, as before, the precise measure of the cost of the interaction depends on the shape of the aggregates, since the admissible collisions are function of the size of the clusters. 

\begin{figure}[h!]
	\centering
	\includegraphics[scale=0.4]{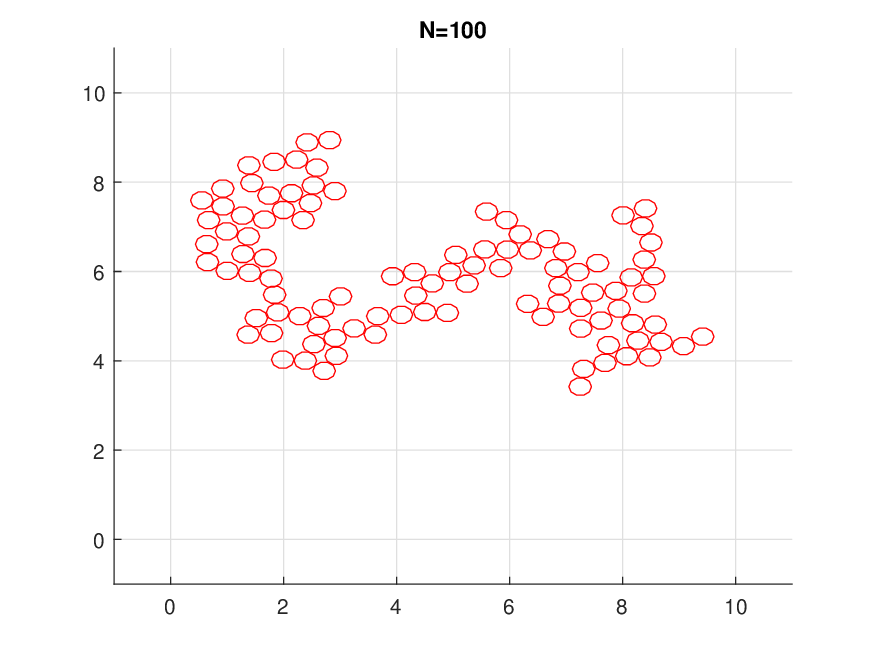}
	\includegraphics[scale=0.4]{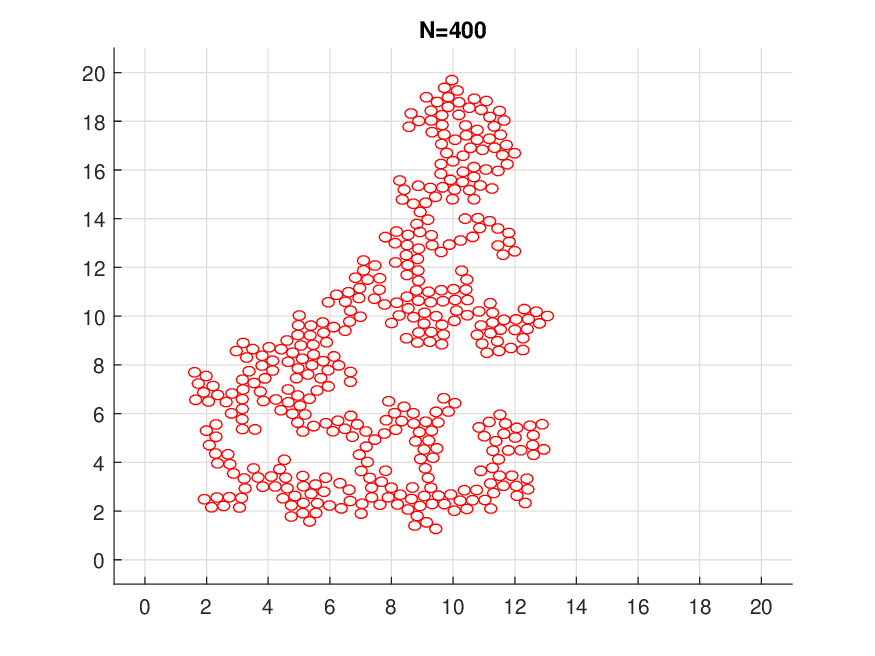}\\ 
	\includegraphics[scale=0.4]{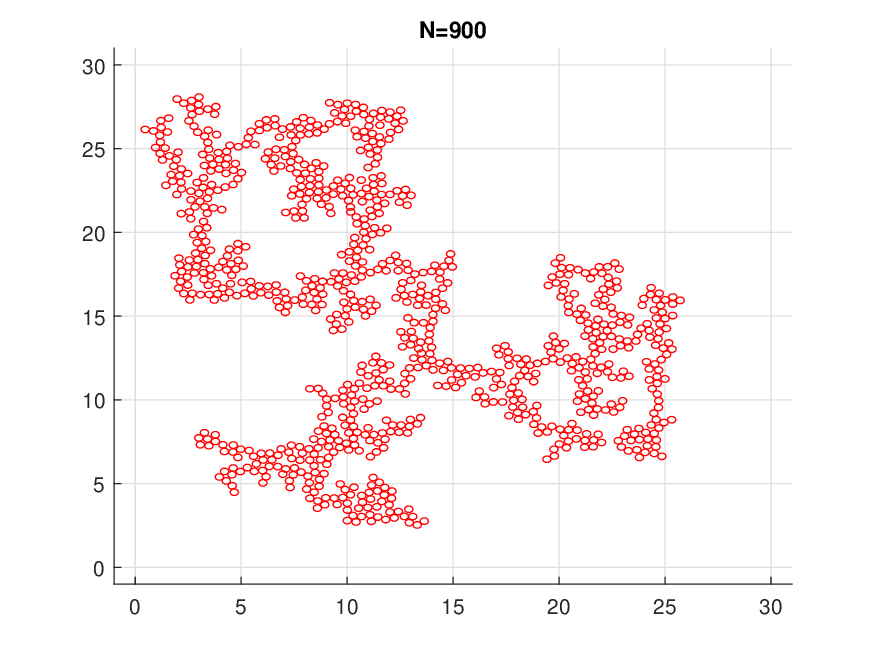}
	\includegraphics[scale=0.4]{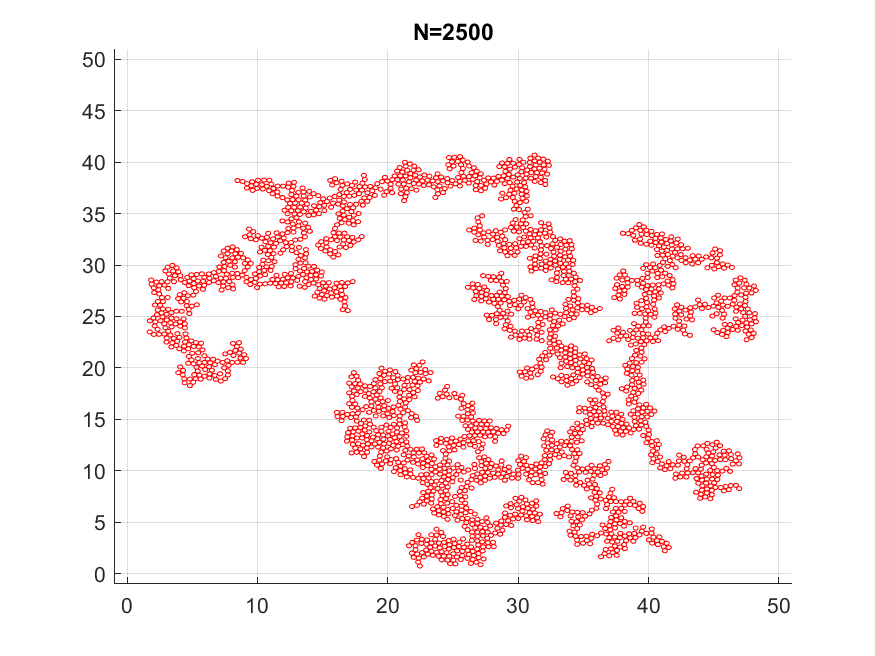}
	\caption{Four examples of cluster aggregation using the Event-Driven method described in Section \ref{sec:ED}. From top left to bottom right the number of spheres is $N=100$, $N=400$, $N=900$ and $N=2500$. }
	\label{fig:ghost}
\end{figure}
Thus, summing up, the system evolves from one event to a successive one as follows. Let the positions $\XX(t )=\{X_i(t )\}_{i=1,\ldots,N}$ and velocities $\VV(t )=\{V_i(t)\}_{i=1,\ldots,N}$ of the particles at time $t$ be given. 
Additionally, let the number of clusters $M(t)$ and the sets of particles that belong to the same cluster  $C_1(t),\ldots,C_{M(t)}(t)$ be given. While $M(t)>1$, the state of the system at the time of the next collision is  obtained by performing three steps: 
\begin{enumerate}
	\item   Computation of the time  $t+\Delta t $ of the next collision through~\eqref{eq:time_col1},~\eqref{eq:time_col2} and~\eqref{eq:time_col3}.  
	\item  Evolution of the positions of particles and the clusters from $t$ to $t+\Delta t$ along the straight line trajectory $$\XX(t+\Delta t) = \XX(t) + \Delta t \VV(t);$$ 
	\item Computation of the new clusters. 
	\subitem{a)} Determination of the new number of clusters $M(t+\Delta t)$. 
	\subitem{b)} Determination of the set of particles belonging to each new cluster $C_1(t+\Delta t),$ $\ldots$, $C_{M(t+\Delta t)}(t+\Delta t)$.
	\item  Computation of the velocities $\VV(t+\Delta t)$ according to the collision law \eqref{eq:new_velocity} or \eqref{boundcoll}.
\end{enumerate}
In Figure \ref{fig:ghost}, we show some examples of cluster formation using the ED method described above. The initial condition which gave rise to the results shown consists in positioning the particles with given radius $R=0.2$ on a lattice with distance in each direction of one unit and in a box of size $[0,L]^2$ with $L=N$,  the number of particles. The initial velocity field is uniformly randomly distributed between $[-0.5,0.5]$: $V(z)=z-\frac{1}{2}$ with $z\sim \mathcal U([0,1]^2)$.

\begin{figure}[h!]
	\centering
	\includegraphics[scale=0.35]{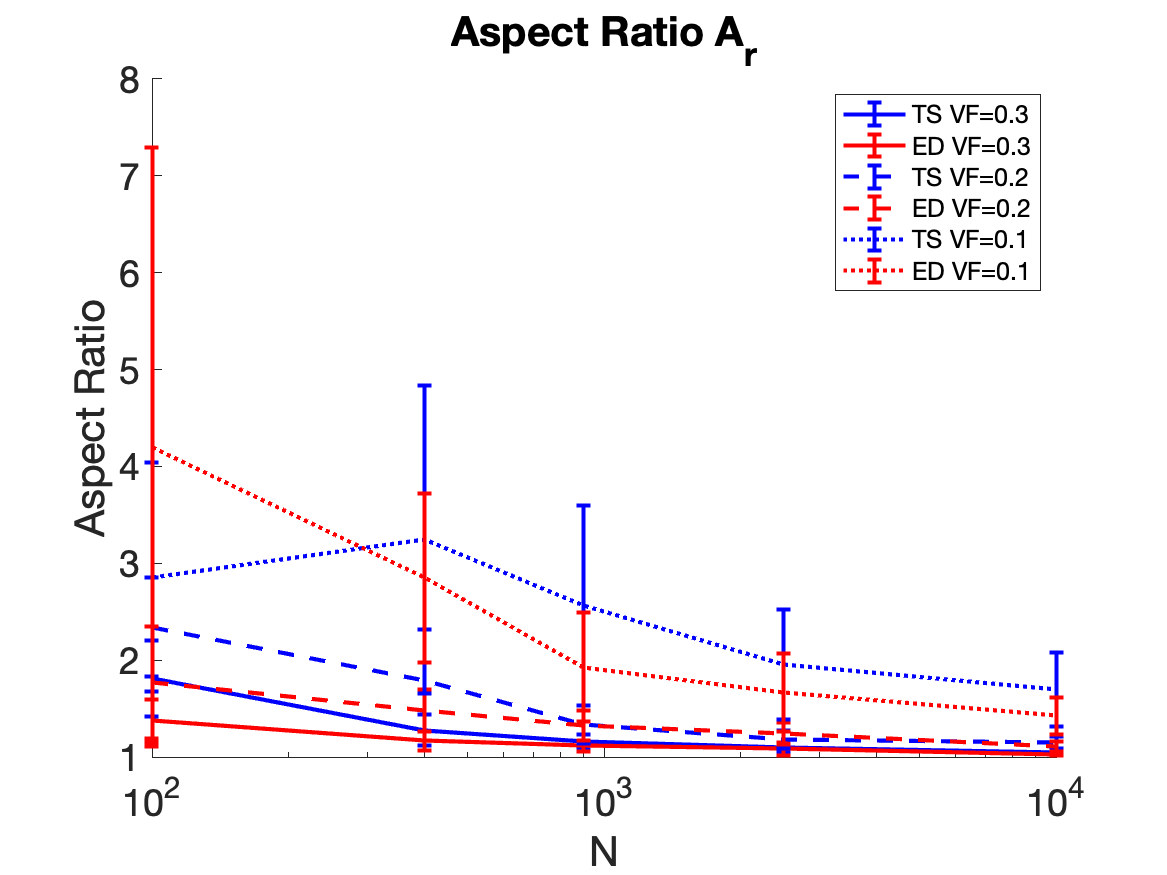}
	\caption{Aspect ratio for the Time Stepping and the Event Driven method.}\label{figAR}
\end{figure}

\begin{figure}[h!]
	\centering
	\includegraphics[scale=0.33]{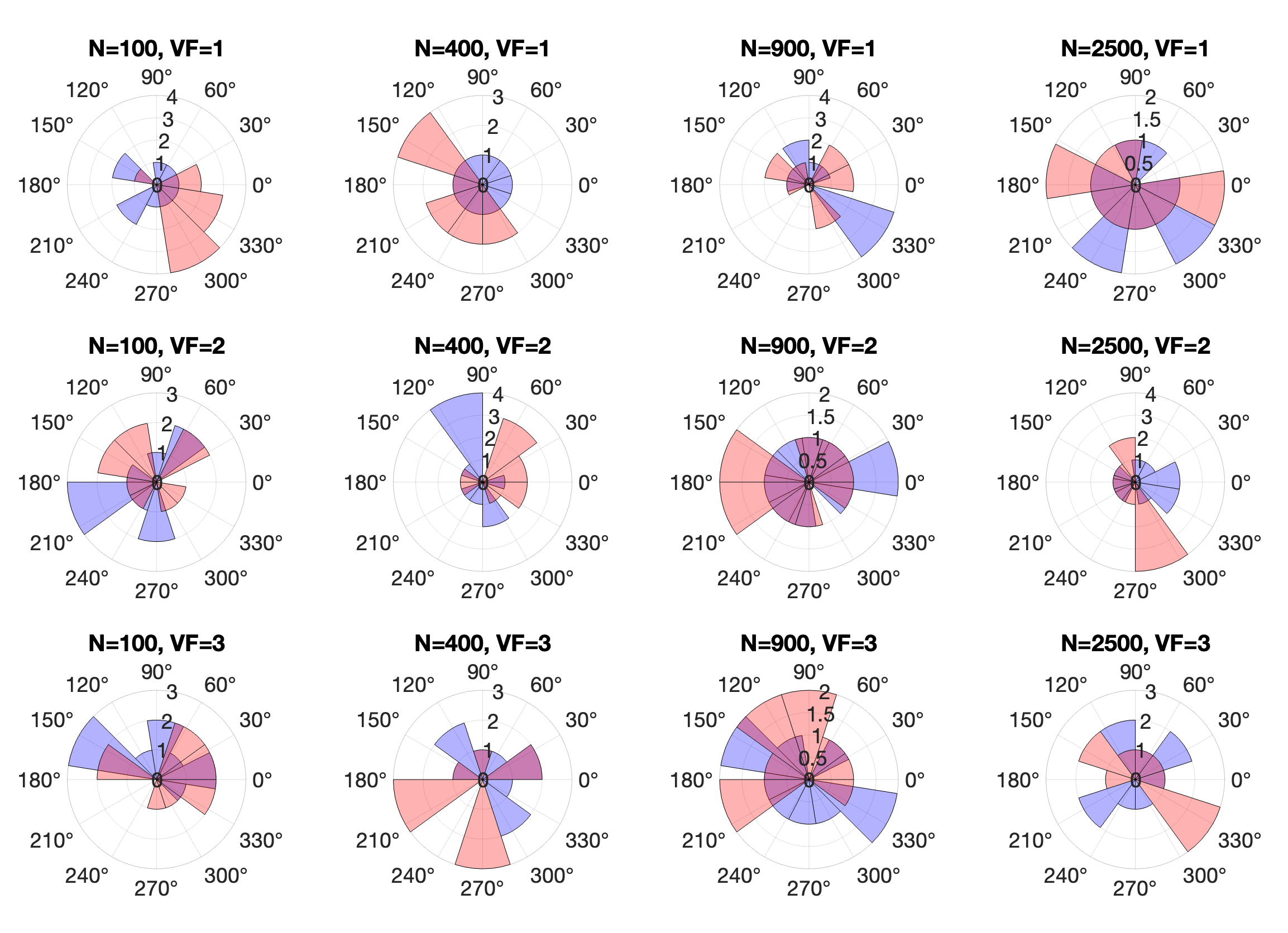}
	\caption{Direction of the leading eigenvector for the Time Stepping and the Event Driven method.}
	\label{fig:comp_LeadingEigenvalue}
\end{figure}
\begin{figure}[h!]
	\centering
	\includegraphics[scale=0.3]{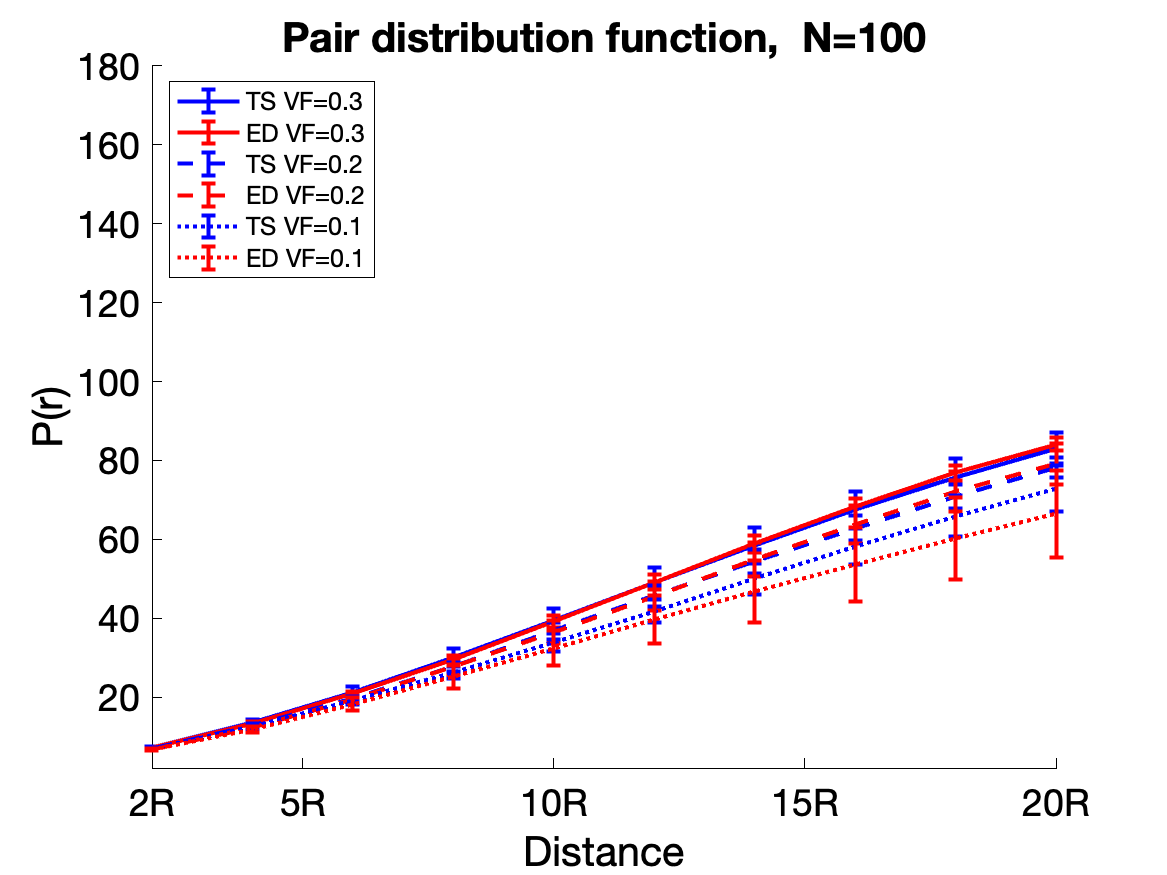}
	\includegraphics[scale=0.3]{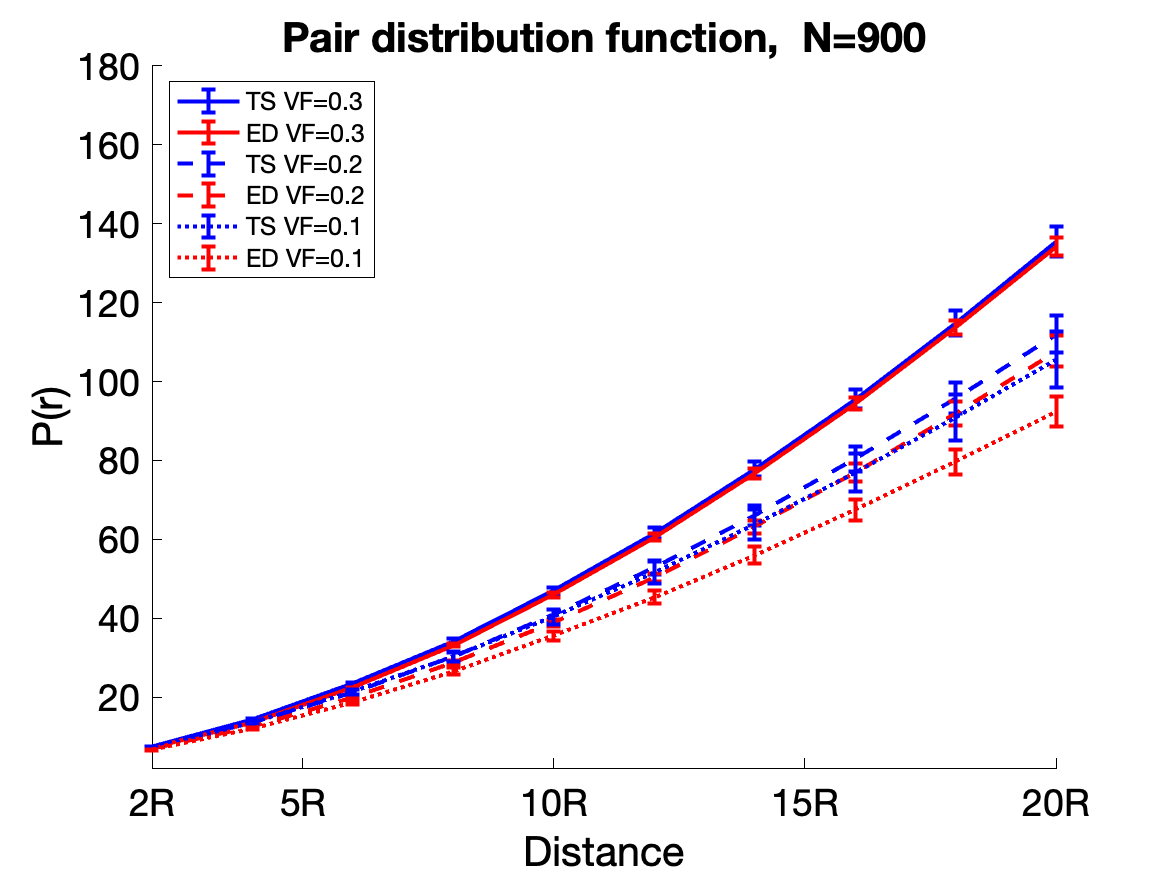}
	\includegraphics[scale=0.3]{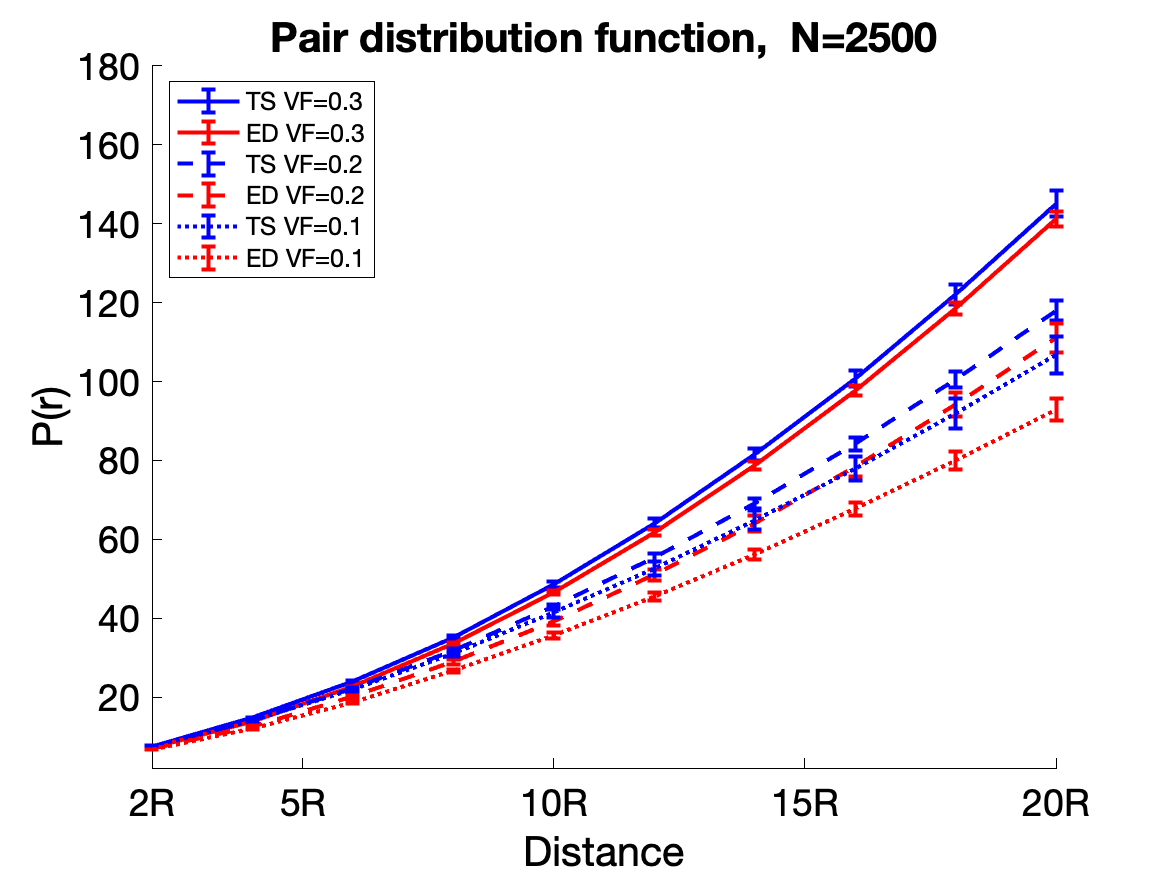}
	\caption{{ Pair distribution function} for the Time Stepping and the Event Driven method.}
	\label{fig:comp_Corr_appendix}
\end{figure}
\begin{figure}[h!]
	\centering
	\includegraphics[scale=0.28]{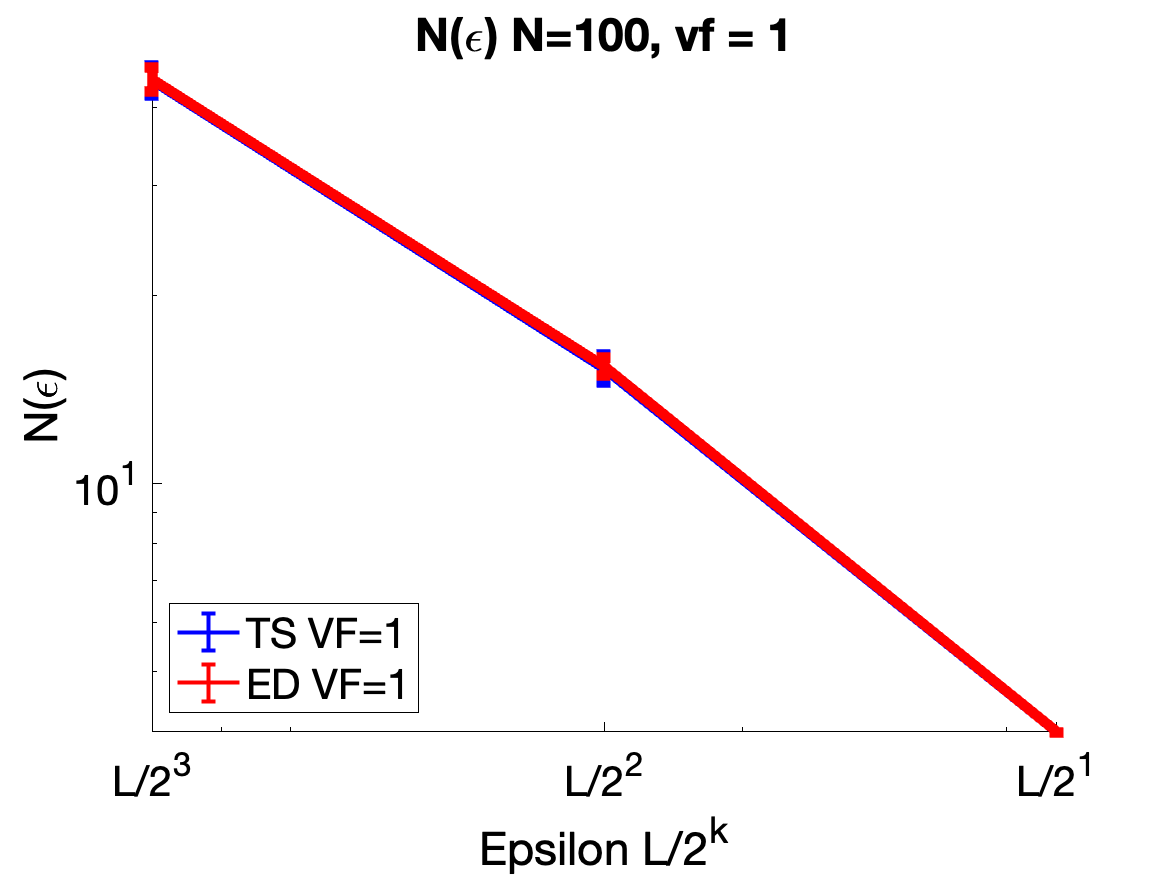}
	\includegraphics[scale=0.28]{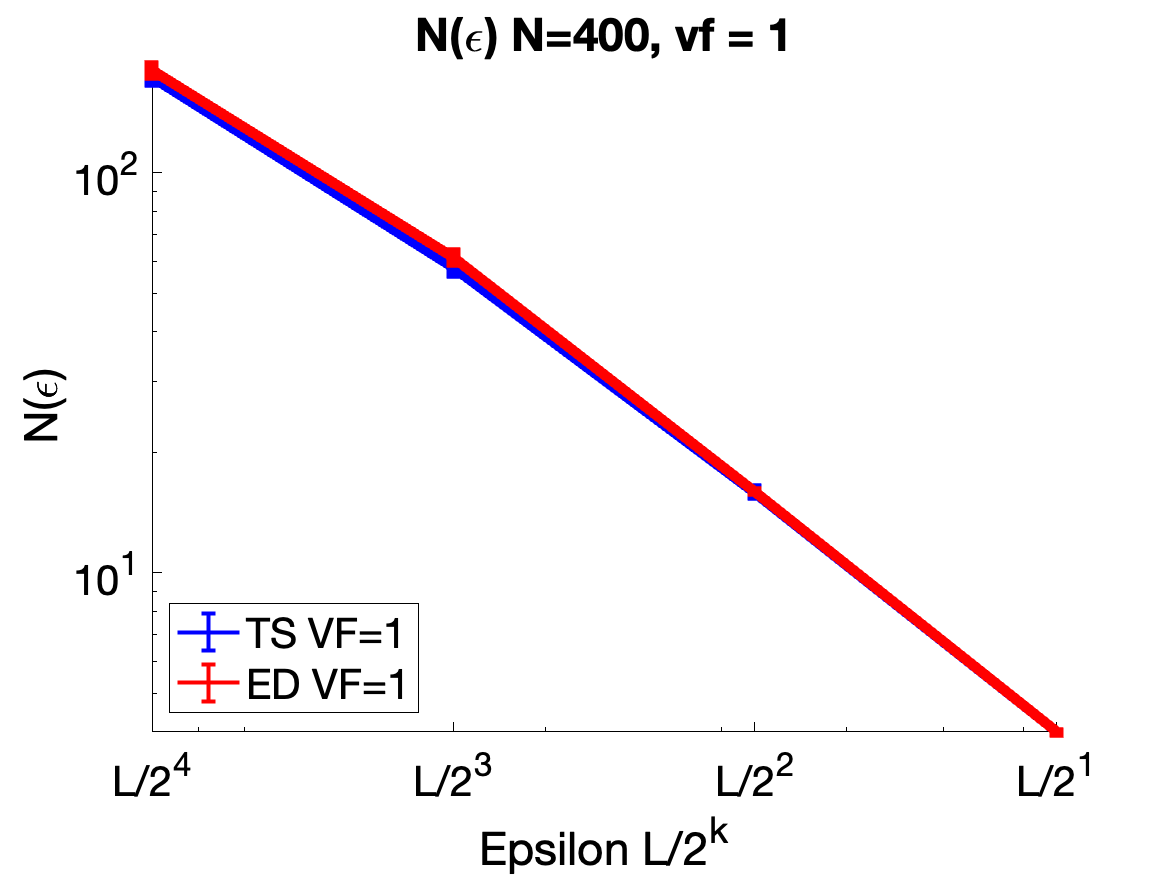}
	\includegraphics[scale=0.28]{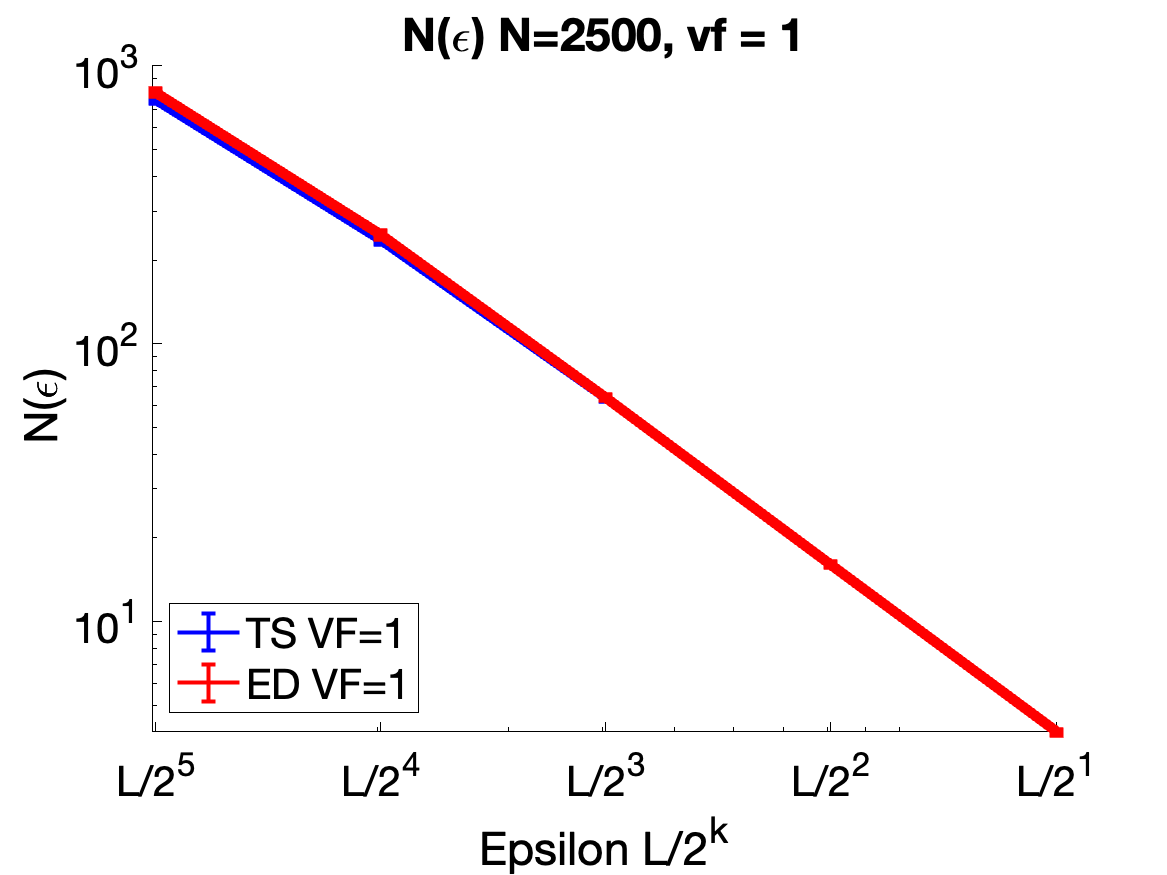}
	\includegraphics[scale=0.28]{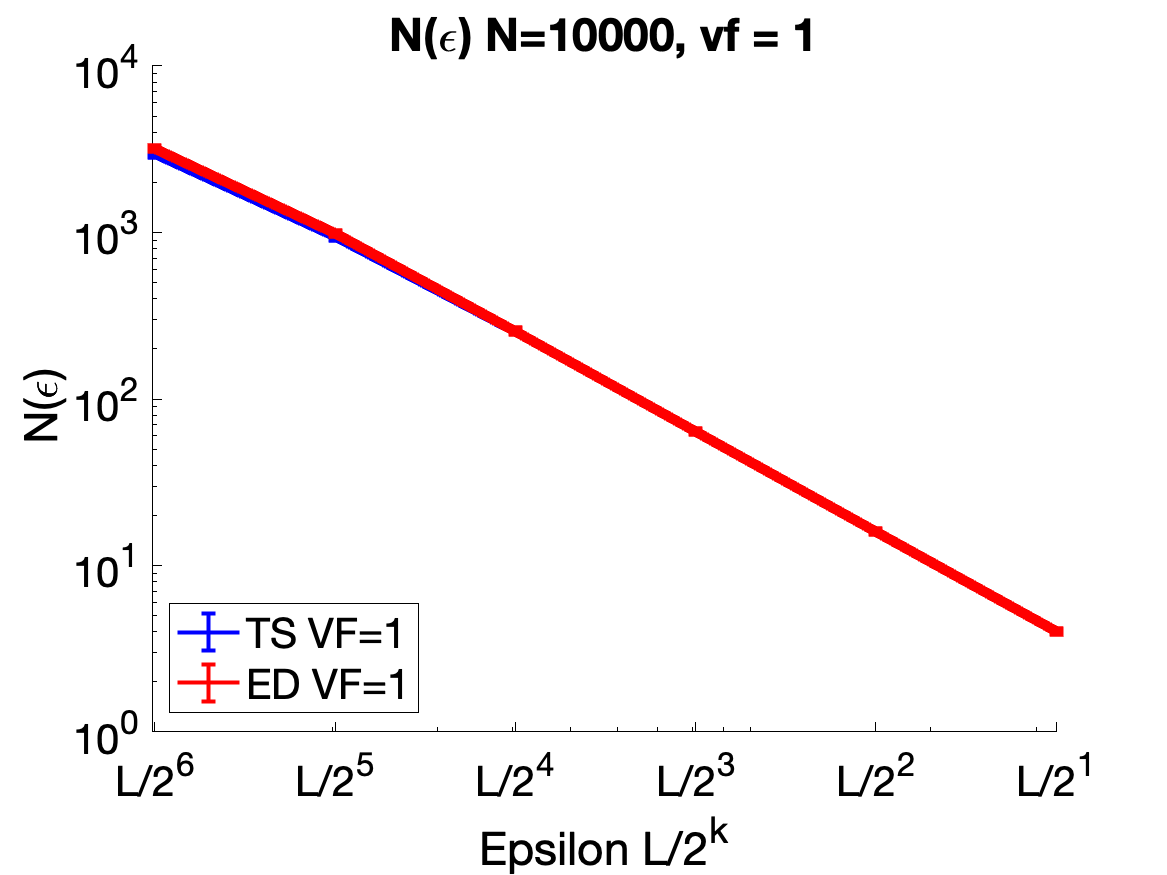}	
	\caption{$\mathcal{N}(\varepsilon)$ over $\varepsilon$ in log log scale for the Time Stepping and the Event Driven method. Volume fraction $V_f$ is fixed to $0.1$, from top left to bottom right the number of particles grows from $N=100$ to $N=10000$.}
	\label{fig:comp_Neps2}
\end{figure}
\begin{figure}[h!]
	\centering
	\includegraphics[scale=0.28]{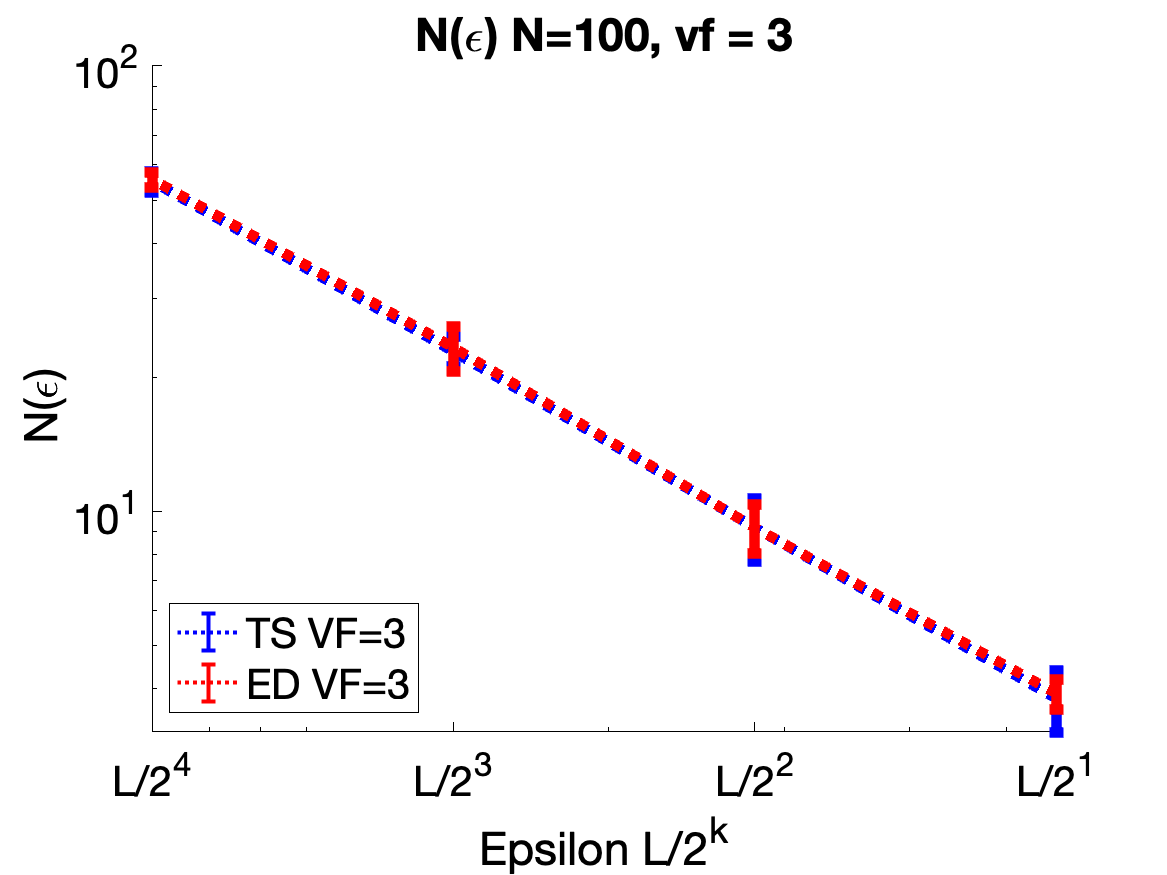}
	\includegraphics[scale=0.28]{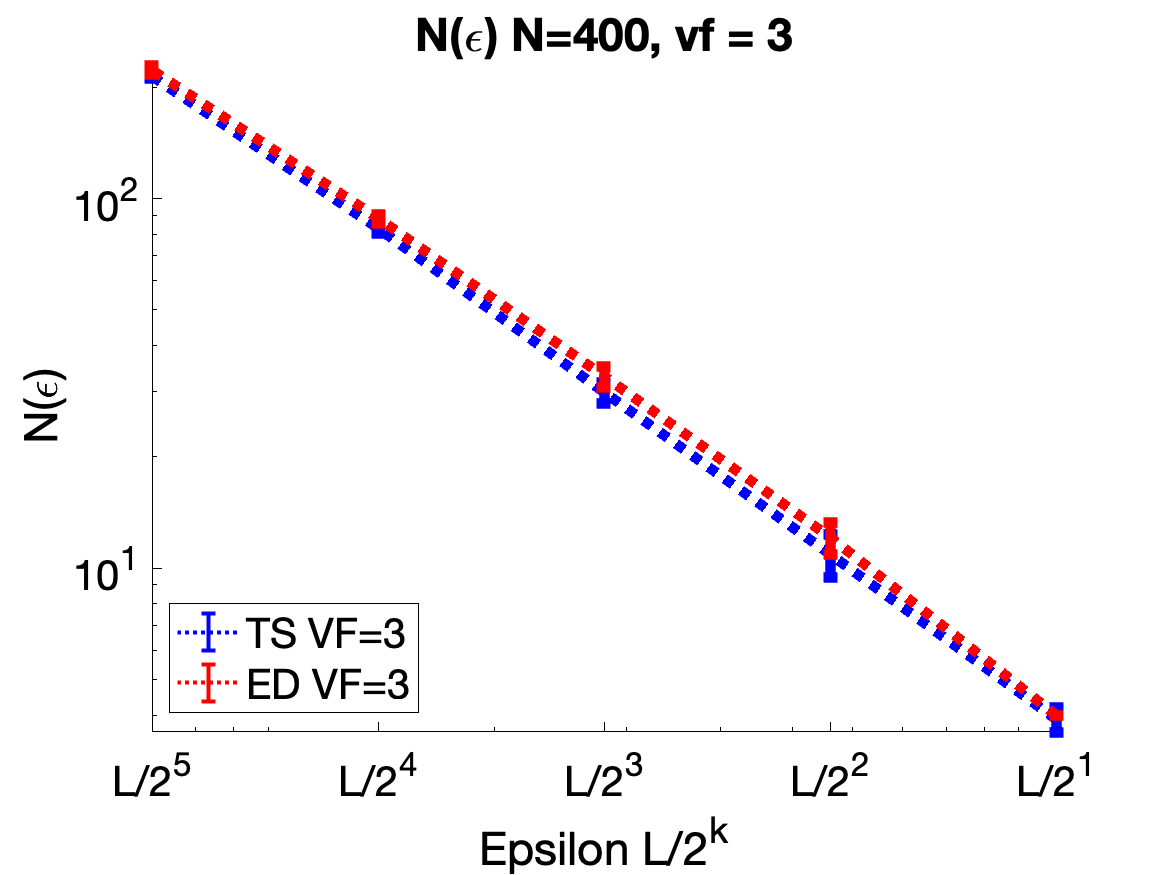}
	\includegraphics[scale=0.28]{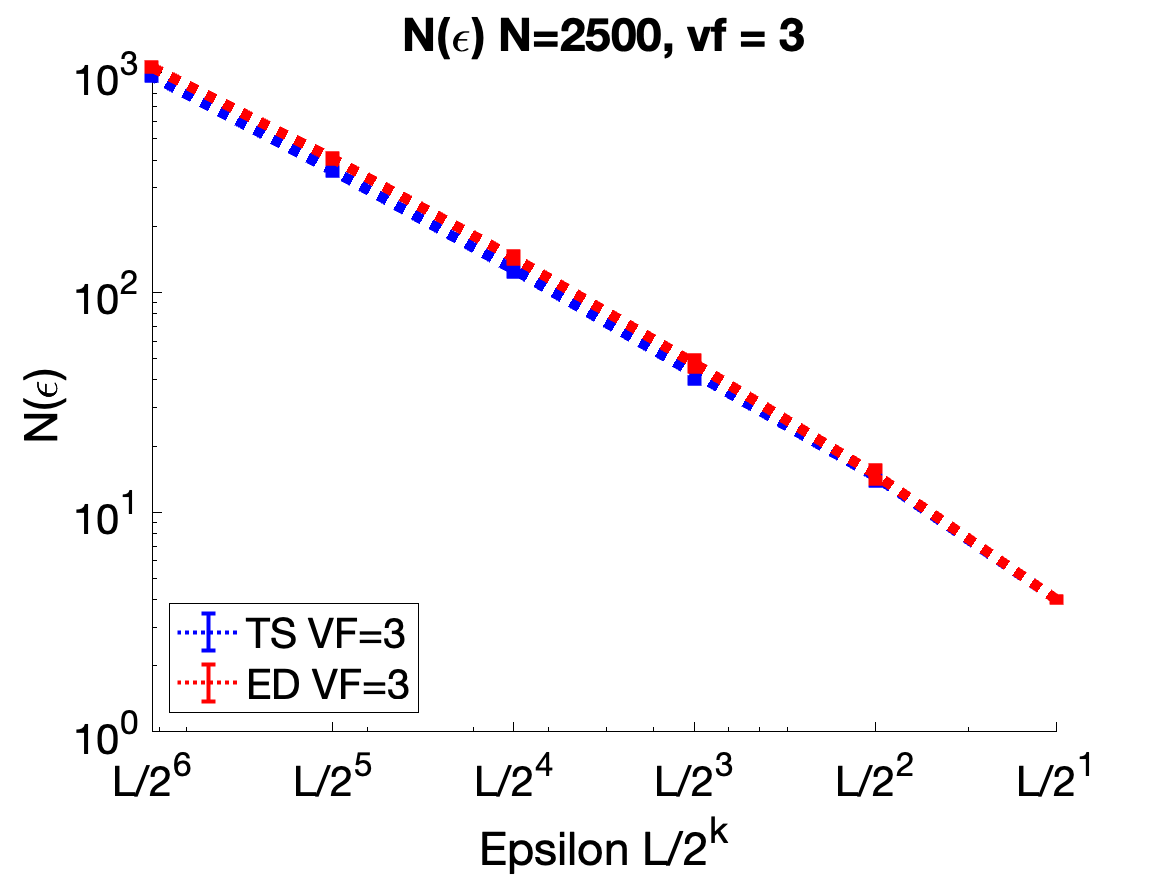}
	\includegraphics[scale=0.28]{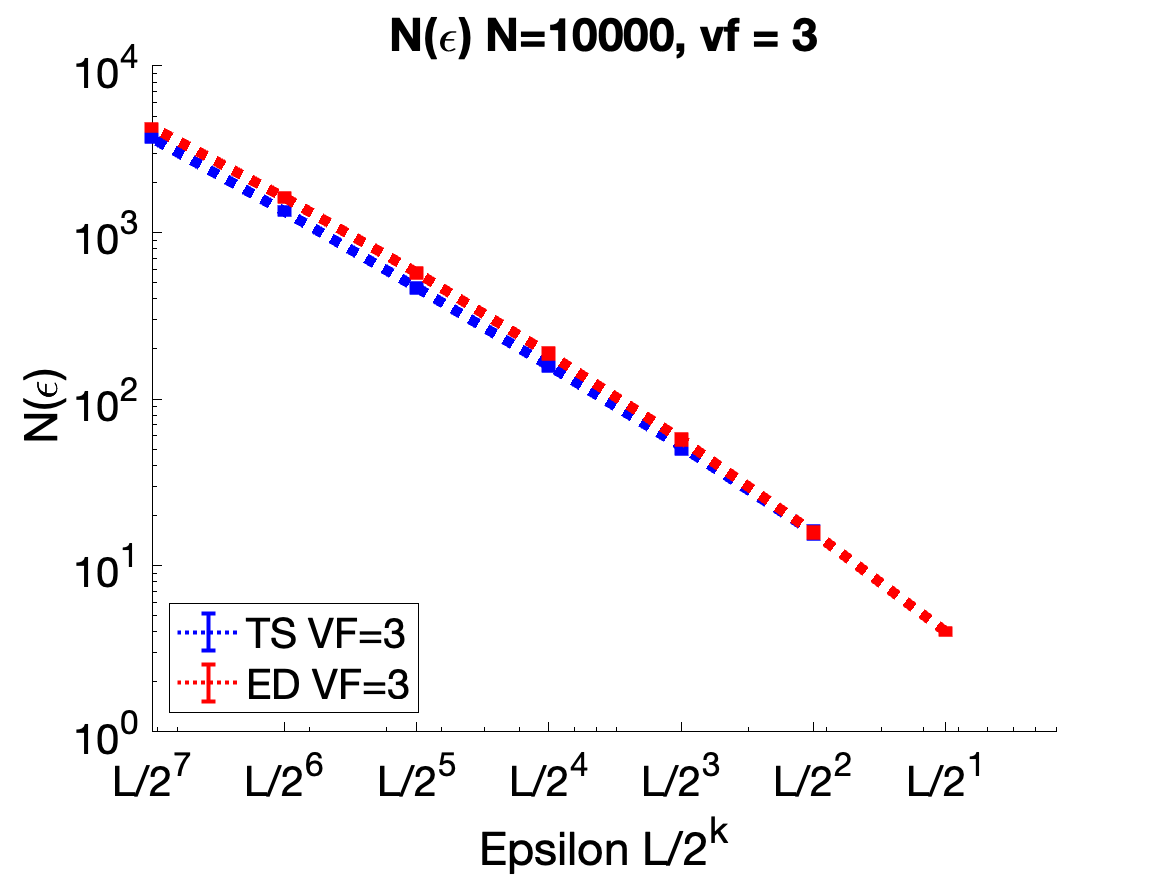}	
	\caption{$\mathcal{N}(\varepsilon)$ over $\varepsilon$ in log log scale for the Time Stepping and the Event Driven method. Volume fraction $V_f$ is fixed to $0.3$, from top left to bottom right the number of particles grows from $N=100$ to $N=10000$.}
	\label{fig:comp_Neps3}
\end{figure}
\section{Additional shape analysis results}\label{app_quant}
We discuss in this part the results of our shape analysis for what concerns the aspect ratio $A_r$ \eqref{aspect} of the final cluster and its main orientation $D$ \eqref{orientation}. We also show the additional results about the pair distribution and the fractal dimension discussed in Section \ref{shape}. 

Figure \ref{figAR} shows the results for the aspect ratio for respectively the ED and the TS methods for different volume fractions as a function of the number of particles $N$. We observe that the shapes of the clusters are more varied when they are composed of a few number of spheres: the variance is much larger for $N=100$ and decays monotonically when $N$ grows. At the same time we observe that $A_r$ tends to slightly decrease with $N$. One last observation is about the role of the volume fraction, clearly as the volume fraction increases the aspect ratio diminishes. These behaviors are similar for the ED and the TS results. They can be explained by the fact that as the volume fraction grows there is less free space to move and naturally the clusters tend to organize themselves in a square like shape as shown in Figures \ref{example1}-\ref{example2}. The opposite situation occurs when the volume fraction is small and clusters have more flexibility to take an elongated shape. The higher variability observed for low numbers of particles can be explained by the role played by the ratio between the radius and therefore the size of the spheres and the size of the box. In fact, for small values of $N$ the ratio between the area occupied by a single particle and the box size plays a major role and as a consequence the shape taken by the final cluster is strongly influenced by the initial conditions giving very different final shapes depending on the initial velocities possessed by the particles. This leads to results affected by larger variance with respect to the cases with larger $N$. For what concerns the TS and the ED methods it is not possible to appreciate statistical differences. We then conclude that for the aspect ratio, the two methods give similar results. The same holds true for the main orientation $D$ identified by the leading eigenvector $v_1$ shown in Figure \ref{fig:comp_LeadingEigenvalue}. The results show that there is not a specific pattern in the leading direction as expected. Lastly for completeness, we show the remaining results for the {pair distribution function} $P(r)$ in Figure \ref{fig:comp_Corr_appendix} and for the fractal dimension in Figures \ref{fig:comp_Neps2}-\ref{fig:comp_Neps3}. The results are on the same path of the ones discussed in Section \ref{shape} suggesting a very good agreement between the TS and the ED methods for all tested situations. 

\section{List of supplementary videos}\label{sec:videos}
This article is supplemented by several videos which can be accessed by following this link: \href{https://figshare.com/articles/media/Modeling_ballistic_aggregation_by_time_stepping_approaches/24081027}{figshare.com/articles/media/Modeling\_ballistic\_aggreg-\\ation\_by\_time\_stepping\_approaches/24081027}. They are listed and described below. In all shown simulations, particles are positioned on a lattice square at time $t=0$ while initial velocity is randomly uniformed distributed over the circle.
\paragraph{Video100\_{ts}} It shows an example of ballistic aggregation with $100$ particles using the Time Stepping method with a volume fraction $V_f=0.2$ and the smallest free flight time step of Table \ref{tab:param}  of Section \ref{sec:numerical}. 
\paragraph{Video100\_ed} It shows an example of ballistic aggregation with $100$ particles using the Event Driven method with a volume fraction $V_f=0.2$ and the smallest free flight time step of Table \ref{tab:param}  of Section \ref{sec:numerical}.
\paragraph{Video400\_{ts}} It shows an example of ballistic aggregation with $400$ particles using the Time Stepping method with a volume fraction $V_f=0.2$ and the smallest free flight time step of Table \ref{tab:param}  of Section \ref{sec:numerical}. 
\paragraph{Video400\_ed} It shows an example of ballistic aggregation with $400$ particles using the Event Driven method with a volume fraction $V_f=0.2$ and the smallest free flight time step of Table \ref{tab:param}  of Section \ref{sec:numerical}.
\paragraph{Video900\_{ts}} It shows an example of ballistic aggregation with $900$ particles using the Time Stepping method with a volume fraction $V_f=0.2$ and the smallest free flight time step of Table \ref{tab:param}  of Section \ref{sec:numerical}. 
\paragraph{Video900\_{ts}} It shows an example of ballistic aggregation with $900$ particles using the Event Driven method with a volume fraction $V_f=0.2$ and the smallest free flight time step of Table \ref{tab:param}  of Section \ref{sec:numerical}.
\paragraph{Video2500\_{ts}} It shows an example of ballistic aggregation with $2500$ particles using the Time Stepping method with a volume fraction $V_f=0.2$ and the smallest free flight time step of Table \ref{tab:param}  of Section \ref{sec:numerical}. 
\paragraph{Video2500\_ed} It shows an example of ballistic aggregation with $2500$ particles using the Event Driven method with a volume fraction $V_f=0.2$ and the smallest free flight time step of Table \ref{tab:param}  of Section \ref{sec:numerical}.
\paragraph{Video10000\_{ts}} It shows an example of ballistic aggregation with $10000$ particles using the Time Stepping method with a volume fraction $V_f=0.2$ and the smallest free flight time step of Table \ref{tab:param}  of Section \ref{sec:numerical}. 
\paragraph{Video40000\_{ts}} It shows an example of ballistic aggregation with $40000$ particles using the Time Stepping method with a volume fraction $V_f=0.2$ and the smallest free flight time step of Table \ref{tab:param}  of Section \ref{sec:numerical}. 
\paragraph{Video90000\_{ts}} It shows an example of ballistic aggregation with $90000$ particles using the Time Stepping method with a volume fraction $V_f=0.2$ and the smallest free flight time step of Table \ref{tab:param}  of Section \ref{sec:numerical}. 
\paragraph{Video250000\_{ts}} It shows an example of ballistic aggregation with $250000$ particles using the Time Stepping method with a volume fraction $V_f=0.2$ and the smallest free flight time step of Table \ref{tab:param}  of Section \ref{sec:numerical}. 
\paragraph{Video1000000\_{ts}} It shows an example of ballistic aggregation with $10^6$ particles using the Time Stepping method with a volume fraction $V_f=0.2$ and the smallest free flight time step of Table \ref{tab:param}  of Section \ref{sec:numerical}.

%\newpage
\bibliography{Aggregation}
\bibliographystyle{plain}

\end{document}